\documentclass[aps,prb,amsmath,amssymb,floatfix,twocolumn,amsmath,superscriptaddress,
twocolumn,nofootinbib,tighten,letterpaper]{revtex4-2}
\usepackage{lmodern}
\usepackage[T1]{fontenc}
\usepackage[latin9]{inputenc}
\usepackage{float}
\usepackage[normalem]{ulem}
\usepackage{amsmath, dsfont}
\usepackage{amssymb}
\usepackage{mathtools}
\usepackage{mathdots}
\usepackage{graphicx}
\usepackage{wasysym}
\usepackage{color}
\usepackage[table]{xcolor}
\usepackage{bm}
\usepackage{hyperref}
\hypersetup{pdfpagemode=UseNone}
\usepackage[many]{tcolorbox}
\newsavebox\MBox

\allowdisplaybreaks

\newcommand{\bsub}{\begin{subequations}}
\newcommand{\esub}{\end{subequations}}

\newcommand{\tr}{{\rm \,tr\,}}

\newcommand\calphare{\cellcolor{green!10}}

\newcommand\caplus{\cellcolor{yellow!40}}
\newcommand\caminus{\cellcolor{yellow!20}}

\newcommand{\mfg}{\mathfrak g}

\newcommand{\mfn}{\mathfrak n}
\newcommand{\mfp}{\mathfrak p}
\newcommand{\mfk}{\mathfrak k}
\newcommand{\mfa}{\mathfrak a}

 \graphicspath{{Images/}}

\begin{document}

\title{Generalized multifractality in 2D disordered systems of chiral symmetry classes}

\author{Jonas F.~Karcher}
\affiliation{{Pennsylvania State University, Department of Physics, University Park, Pennsylvania 16802, USA}}
\affiliation{{Institute for Quantum Materials and Technologies, Karlsruhe Institute of Technology, 76021 Karlsruhe, Germany}}
\affiliation{{Institut f\"ur Theorie der Kondensierten Materie, Karlsruhe Institute of Technology, 76128 Karlsruhe, Germany}}

\author{Ilya A.~Gruzberg}
\affiliation{Ohio State University, Department of Physics, 191 West Woodruff Ave, Columbus OH, 43210, USA}

\author{Alexander D.~Mirlin}
\affiliation{{Institute for Quantum Materials and Technologies, Karlsruhe Institute of Technology, 76021 Karlsruhe, Germany}}
\affiliation{{Institut f\"ur Theorie der Kondensierten Materie, Karlsruhe Institute of Technology, 76128 Karlsruhe, Germany}}

\date{January 24, 2023}

\begin{abstract}

We study generalized multifractality that characterizes eigenstate fluctuations and correlations in disordered systems of chiral symmetry classes AIII, BDI, and CII. By using the non-linear sigma-model field theory, we construct pure-scaling composite operators and eigenfunction observables that satisfy Abelian fusion rules. The observables are labelled by two multi-indices $\lambda$, $\lambda'$ referring to two sublattices, at variance with other symmetry classes, where a single multi-index $\lambda$ (that can be viewed as a generalized version of a Young diagram) is needed.  Further, we analyze Weyl symmetries of multifractal exponents, which are also peculiar in chiral classes, in view of a distinct root system associated with the sigma-model symmetric space. The analytical results are supported and complemented by numerical simulations that are performed for a 2D lattice Hamiltonian of class AIII, both in the metallic phase and at the Anderson transition. Both in the metallic phase and at the transition, the numerically obtained exponents satisfy Weyl symmetry relations, confirming that the sigma-model is the right theory of the problem. Furthermore, in the metallic phase, we observe the generalized parabolicity (proportionality to eigenvalues of the quadratic Casimir operator), as expected in the one-loop approximation. On the other hand, the generalized parabolicity is strongly violated at the metal-insulator transition, implying violation of local conformal invariance.
\end{abstract}

\maketitle

\section{Introduction}
\label{sec:intro}

Disorder is ubiquitous in nature, which stimulates interest in the physics of Anderson localization~\cite{50_years_of_localization}. Of great importance in this context are Anderson transitions between localized and delocalized phases, which include metal-insulator transitions as well as transitions between topologically different localized phases~\cite{evers08}.

Field theories of Anderson localization are non-linear sigma models characterized by non-Abelian continuous symmetries, and Anderson transitions bear a certain similarity with conventional second-order phase transitions known from statistical mechanics. At the same time, the physics of Anderson transitions turns out to be distinct in many respects, which is related to the fact that the corresponding sigma-model field theories involve either the replica limit, $n \to 0$, or, alternatively, supersymmetry. One of consequences is that, at variance with the Mermin-Wagner theorem that forbids ordering in spatial dimensionalities $ d \le 2$, there is a wealth of Anderson transitions in two-dimensional (2D) geometries. The richness of the problem is enhanced by a rather large number (ten) of symmetry classes~\cite{altland1997nonstandard, zirnbauer1996riemannian, heinzner2005symmetry} and by topological phenomena giving rise to topological insulators and superconductors~\cite{chiu2016classification}.

Another remarkable property of Anderson transitions is the multifractality of critical states: different moments $\langle |\psi|^{2q} \rangle$ of the wave function amplitude (or, equivalently, of the local density of states) scale with distinct exponents that depend on $q$ in a non-linear way~\cite{evers08, rodriguez2011multifractal}. Furthermore, the notion of multifractality has been promoted to ``generalized multifractality''~\cite{karcher2021generalized} characterizing fluctuations and correlations of multiple eigenstates at criticality. In the field-theory (sigma-model) language, the generalized multifractality involves a full set of pure-scaling gradientless composite operators. These composite operators can be translated to the language of pure-scaling observables characterizing eigenfunctions of the Hamiltonian. This program has been started by Ref.~\cite{gruzberg2013classification}, where the construction was developed for the symmetry class A; the results were verified in Ref.~\cite{karcher2021generalized} by a numerical investigation of the integer quantum Hall transition. Subsequently, the generalized multifractality was explored in class C in Ref.~\cite{karcher2021generalized, karcher2022generalized}, with a focus on the spin quantum Hall transition.  Most recently, an analytical and numerical study of the generalized multifractality was carried out for symmetry classes AII, D, and DIII, which are characterized by weak antilocalization and exhibit metal-insulator transitions in two dimensions~\cite{karcher2022generalized-2}.

The scaling of generalized multifractality observables is characterized by an infinite set of exponents $x_\lambda$, where $\lambda$ is a multi-index, $\lambda= (q_1, \ldots, q_n)$, and $q_j$ can be in general arbitrary complex numbers. It was shown in Refs.~\cite{mirlin2006exact, gruzberg2011symmetries, gruzberg2013classification} that the multifractal spectra $x_\lambda$ in five symmetry classes (Wigner-Dyson classes A, AI, and AII, as well as Bogoliubov-de Gennes classes C and CI) satisfy certain exact symmetry relations (``Weyl symmetry'') that imply identity of scaling exponents characterizing seemingly unrelated observables.  (This extends the earlier result~\cite{mirlin2006exact} for conventional multifractality in Wigner-Dyson classes.) Furthermore, it was pointed out in Ref.~\cite{gruzberg2013classification} that the Weyl symmetry should hold also in Bogoliubov-de Gennes classes D and DIII as long as jumps between the disjoint components of the corresponding sigma-model manifolds are suppressed.

Numerical results in Refs.~\cite{karcher2021generalized, karcher2022generalized, karcher2022generalized-2} for the generalized multifractality  in the classes A, C, AII, D, and DIII confirm the construction of pure-scaling observables as well as the Weyl symmetry (in the cases when it should hold according to the analytical derivation). Furthermore, the numerical results in the metallic phases confirm the generalized parabolicity (proportionality to eigenvalues of the quadratic Casimir operator) that follows from the one-loop approximation. Since all these analytical predictions are based on the sigma-model formalism, this numerics has provided a clear confirmation of the use of the sigma models as field theories of Anderson localization. Another important result of the numerics is a strong violation of generalized parabolicity at the SQH transition and at 2D metal-insulator transitions in classes AII, D, and DIII. (In the case of SQH transition, this is also proven analytically in Ref.~\cite{karcher2022generalized} where exact results for a set of the generalized multifractal exponents are found.) As shown in Ref.~\cite{karcher2021generalized}, this implies a violation of local conformal invariance at these 2D transitions. This striking result puts strong constraints on fixed-point theories of the above 2D Anderson transitions; in particular, it excludes models of the Wess-Zumino-Novikov-Witten class.

The goal of the present paper is to study the generalized multifractality in the chiral classes AIII, BDI, and CII. These symmetry classes possess a variety of properties that distinguish them from other symmetry classes. On the level of the field theory, these properties result from an additional abelian [U(1)] degree of freedom  characterizing the sigma-model target space. As a consequence, random-matrix models of these classes may have an arbitrary number of strictly zero modes~\cite{ivanov2002supersymmetric}. As a closely related hallmark of chiral classes, quasi-one-dimensional models of these classes are  $\mathbb{Z}$ topological insulators, i.e., they undergo multiple transitions between topologically distinct insulating phases. At these transitions,  the density of states exhibits a strong singularity (first identified by Dyson in 1953)~\cite{dyson1953dynamics, mckenzie1996exact, titov2001fokker}, and the localization length diverges~\cite{theodorou1976extended, eggarter1978singular, fisher1995critical, balents1997delocalization, brouwer1998delocalization, mudry1999random, brouwer2000nonuniversality, altland2001spectral, altland2015topology}.  These critical points have infinite-randomness character, and the corresponding critical eigenfunctions show very strong fluctuations~\cite{balents1997delocalization, karcher2019disorder}.

It was understood long ago that 2D systems in the chiral symmetry classes also have very peculiar properties. Early numerical investigations demonstrated a resilience of chiral-class systems to localization, which led to a suggestion that 2D and 3D systems in the chiral classes always remain delocalized~\cite{antoniou1977absence}. The sigma-model field theories for chiral classes were derived by Gade and Wegner, who have also performed the renormalization-group (RG) analysis of these theories~\cite{gade1991the, gade1993anderson}. These sigma-model theories and the RG flows have a number of remarkable features. First, the sigma models involve an additional term (known as the Gade term) associated with the U(1) degree of freedom mentioned above. Second, the renormalization of the conductivity is absent to all orders in perturbation theory, which provides an apparent support to the absence of localization transition. Instead, this RG implies an extended metallic phase, with conductivity taking an arbitrary value, i.e., a line of  metallic infrared fixed points. Furthermore, there is a logarithmic flow of the Gade term, implying that a power-law singularity of the density of states at zero energy (chiral-symmetry point) is slowly enhanced along the flow to an infrared  fixed point and becomes very strong in the asymptotic infrared limit.

The Gade-Wegner sigma models, and some extensions of them, were subsequently rederived and analyzed in the context of various microscopic models of chiral symmetry classes~\cite{altland1999field, fabrizio2000anderson, fukui1999critical, guruswami2000super, dellanna2006anomalous, koenig2012metal-insulator, ostrovsky2014density}. It was understood that the ultimate infrared behavior in the metallic phase is of infinite-randomness nature and is associated with a ``freezing'' of the multifractality spectrum, which also leads to a refinement of the density-of-states asymptotics~\cite{motrunich2002particle-hole, mudry2003density, yamada2004random, dellanna2006anomalous}. On the numerical side, most of the studies found critical properties of the metallic phase characterized by non-universal exponents for various observables (such as the density of states, the finite-energy localization length, and multifractality)~\cite{hatsugai1997disordered, cerovski2000bond-disordered, eilmes2001exponents, eilmes2004exponents, markos2007critical, schweitzer2008disorder-driven},
at variance with what is expected in the infinite-randomness infrared limit. This is, however, by no means surprising, since the Gade-Wegner flow towards the line of infrared fixed points is logarithmically slow. Thus, for many models and parameter ranges, the infrared limiting behavior cannot be reached
on any realistic length scale. At the same time, several works~\cite{markos2010logarithmic, schweitzer2012scaling, markos2012disordered} reported evidence of the asymptotic behavior in some of the observables.

Contrary to earlier proposals mentioned above, more recent numerical investigations of suitably designed 2D models  have provided evidence of Anderson metal-insulator transitions in chiral classes~\cite{motrunich2002particle-hole, bocquet2003network}. Very recently, the phase diagram and the critical behavior at the metal-insulator transition in chiral unitary class AIII was studied in Ref.~\cite{karcher2022metal}. On the analytical side, a theory of Anderson transitions in 2D systems in chiral symmetry classes was developed in Ref.~\cite{koenig2012metal-insulator}. The central idea of Ref.~\cite{koenig2012metal-insulator} is that the sigma-model manifolds for chiral classes are not simply connected due to the U(1) degree of freedom and therefore allow for topological excitations---vortices. These vortices, characterized by a fugacity $y$, should be included in the RG analysis of the sigma-model, in analogy with the famous theory of the Berezinskii-Kosterlitz-Thouless (BKT) transition in the XY model in two dimensions. Derivation of the RG flow in  Ref.~\cite{koenig2012metal-insulator} indeed yielded a metal-insulator transition. At the same time, an important difference with the BKT transition was found: the fixed point for the chiral-class transition is at a finite fugacity, $y>0$, which should be contrasted to the BKT fixed point value $y=0$. Thus, while one may expect that the theory of Ref.~\cite{koenig2012metal-insulator} describes correctly the RG flow at the qualitative level, (as indeed supported by Ref.~\cite{karcher2022metal}), it is not parametrically controllable in what concerns quantitative characteristics of the transition (in particular, critical exponents). This additionally emphasizes importance of numerical studies of these transitions.

In this paper, we explore the generalized multifractality in systems of chiral symmetry classes by a combination of analytical and numerical approaches. Key results of the work are as follows:
\begin{enumerate}
	
	\item
	By using the Iwasava decomposition, we derive the pure-scaling observables in terms of sigma-model composite operators and in terms of eigenfunction observables. In agreement with our earlier arguments based on the sigma-model RG and on the physical considerations~\cite{karcher2022generalized-2}, the construction of observables follows the ``spinless'' pattern for classes AIII and BDI and the ``spinful'' pattern for class CII. At the same time, there is a crucial difference between the chiral classes and all other symmetry classes: for the chiral classes, the observables are labeled by a pair of multi-indices $\lambda, \lambda'$, with $\lambda$ corresponding to one sublattice and $\lambda'$ to another one. The developed construction satisfies Abelian fusion rules, both for the sigma-model composite operators and for the eigenfunction observables.
	
	\item   We derive Weyl symmetry relations for the chiral classes, which are very peculiar for several reasons. First, the observables are labeled by two multi-indices $\lambda, \lambda'$ (one for each sublattice), see above. Second, the scaling exponents $x_{\lambda, \lambda'}$ contain U(1) contributions that originate from the Gade term. Third, the root systems associated with the sigma-model manifolds of chiral classes are of type $A_n$, so that the corresponding Weyl group contains permutations but not reflections. The resulting symmetry can be written as $x_{\lambda, \lambda'} = x_{w(\lambda, \lambda')}$, where the Weyl group element $w$ involves permutations of all components of \emph{both} multi-indices $\lambda$ and $\lambda'$. We show, however, that for the subclass of ``balanced'' observables with $\lambda' = \lambda$ we have $x_{\lambda, \lambda} = x_{w(\lambda), w(\lambda)}$, where $w(\lambda)$ is obtained from $\lambda$ by an arbitrary combination of Weyl reflections and permutations.
	
	\item
	Using a tight-binding model of class AIII, we provide a numerical verification of the construction of pure-scaling observables. Furthermore, our simulations yield numerical values of the generalized multifractal scaling exponents. The numerical analysis is carried out both in the metallic phase and at the metal-insulator transition.
	
	\item
	In the metallic phase, we find that the Weyl symmetry, as predicted analytically, is fulfilled. Furthermore, for weak disorder, the generalized multifractal exponents are in excellent agreement with the one-loop predictions of the sigma-model RG (and thus with the generalized parabolicity). With increasing disorder, sizeable deviations from generalized parabolicity are observed, as expected analytically form higher-loop contributions.
	
	\item
	The analytically predicted Weyl symmetry relations hold also at the metal-insulator transition. At the same time, the generalized parabolicity gets strongly violated, which implies the violation of the local conformal invariance.
	
\end{enumerate}

\section{Sigma models for 2D disordered systems of chiral symmetry classes }
\label{sec:sigma}

The sigma-model field theories for 2D disordered systems of chiral symmetry classes were derived by Gade and Wegner~\cite{gade1991the, gade1993anderson}. In the fermionic replica formalism, the sigma-model manifolds for the chiral unitary class AIII, chiral orthogonal class BDI, and chiral symplectic class CII are $\text{U}(n)$, $\text{U}(2n)/\text{Sp}(2n)$, and $\text{U}(n)/\text{O}(n)$, respectively. The sigma-model action reads
\begin{align}
	S[U] &= - \! \int \! d^2 r \!  \Big[ \frac{\sigma}{8\pi s} \text{tr} (U^{-1} \nabla U)^2 + \frac{\kappa}{8\pi s} (\text{tr} U^{-1} \nabla U)^2
	\nonumber \\
	& \quad + i \frac{\pi \rho_0}{2s} \, \varepsilon \, \text{tr} (U + U^{-1}) \Big],
	\label{eq:gade-action}
\end{align}
where $s=1$ for class AIII and $s=2$ for classes BDI and CII.  Here $\sigma$ is the conductivity in units of $e^2/ \pi h$; the second term (known as the Gade term) couples only to the U(1) degree of freedom and is specific for chiral classes. In the last term, $\varepsilon$ is a running coupling whose bare value is the energy $E$ (which breaks the chiral symmetry), and $\rho_0$ is the bare density of states. Within the replica formalism, one should take the limit $n \to 0$ in the end of the calculation. An alternative formalism involves bosonic replicas, with non-compact symmetric spaces as target spaces of the sigma-model, see Table \ref{table:sigma-models}. Finally, the third option is to use supersymmetry, in which case the sigma-model target space is a product of the compact (fermionic) and non-compact (bosonic) spaces ``dressed'' by anticommuting variables (and no $n \to 0$ replica limit is needed). While the supersymmetric formalism is more accurate from the mathematical point of view, it is frequently sufficient to use one of the replica approaches.

%%%%%%%%%%%%%%%
\begin{table}
	\centering
	%\vskip 5mm
	\begin{tabular}{|c|c|c|}
		\hline
		Symmetry
		% & NL$\sigma$M
		& Compact (fermionic) & Non-compact (bosonic)
		\\
		class
		%& (n-c$|$c)
		& space & space
		\\
		\hline
		AIII
		%& A$|$A
		& $\text{U}(n) \vphantom{\biggr|}$
		& $\text{GL}(n,\mathbb{C})/\text{U}(n) \vphantom{\biggr|}$
		\\
		\hline
		BDI
		%& AI$|$AII
		& $\text{U}(2n)/\text{Sp}(2n) \vphantom{\biggr|}$
		& $\text{GL}(n,\mathbb{R})/\text{O}(n) \vphantom{\biggr|}$
		\\
		\hline
		CII
		%& AII$|$AI
		& $\text{U}(n)/\text{O}(n) \vphantom{\biggr|}$
		& $\begin{array}{c} \text{GL}(n,\mathbb{H})/\text{Sp}(2n) \\
			\equiv \text{U}^{*}(2n)/\text{Sp}(2n) \end{array}$
		\\
		\hline
	\end{tabular}\hfill \\
	\caption{Target spaces of sigma models for three chiral symmetry classes.}
	\label{table:sigma-models}
\end{table}
%%%%%%%%%%%%%%%%%%

Let us briefly recall the implications of the RG flow that follow from Eq.~\eqref{eq:gade-action}. For definiteness, we focus in this discussion on the chiral unitary class AIII (for which we will also perform numerical simulations presented below). The RG results for the other two chiral classes are very similar.

Perturbative RG equations for the couplings $\sigma$, $\kappa$, and $\varepsilon$ in class AIII read~\cite{gade1993anderson, koenig2012metal-insulator, ostrovsky2014density}
\begin{align}
	\dfrac{\partial \sigma}{\partial \ln L} &= 0,
	\label{eq:RG-sigma} \\[0.1cm]
	\dfrac{\partial \kappa}{\partial \ln L} &= 1,
	\label{eq:RG-kappa} \\[0.1cm]
	\dfrac{\partial \ln \varepsilon}{\partial \ln L} &= 2 + \frac{\kappa}{\sigma^2}.
	\label{eq:RG-E}
\end{align}
Remarkably, Eq.~\eqref{eq:RG-sigma} (the absence of renormalization of the conductivity) is exact to all orders in perturbation theory in all three chiral classes~\cite{gade1991the}. Moreover, in class AIII, Eq.~\eqref{eq:RG-kappa} is also perturbatively exact~\cite{guruswami2000super}.

In the right-hand side of Eq.~\eqref{eq:RG-E}, the first term is the normal dimension; the second term represents the one-loop anomalous dimension responsible for a non-trivial scaling of the density of states. Substituting the solution of Eq.~\eqref{eq:RG-kappa},
\begin{equation}
	\kappa (L) = \kappa_0 + \ln L,
	\label{eq:kappa-L}
\end{equation}
into Eq.~\eqref{eq:RG-E} and integrating the latter, we get
\begin{equation}
	\ln \varepsilon(L) = \ln E + B_0 \ln L + \frac{1}{2\sigma^2} \ln^2L,
	\label{eq:epsilon-L}
\end{equation}
where $\kappa_0$ is the bare (ultraviolet) value of the coupling $\kappa$ and $B_0 = 2 + \kappa_0 / \sigma^2$.
The RG transformation is performed until the running coupling $\varepsilon(L)$ ceases to be small, reaching the ultraviolet energy scale $\Delta$ of the problem (usually set by the bandwidth). At this scale, which we denote as $L_c(E)$, a crossover from class AIII to class A happens. Thus, the localization length $\xi(E)$ is given by
$\xi(E) \sim L_c(E) \exp(\sigma^2)$. Furthermore, the density of states $\rho(E)$  is determined by $L_c(E)$ as follows:
\begin{equation}
	\rho(E) \sim \frac{1}{E L_c^2(E)}.
	\label{eq:rho-E-general}
\end{equation}
Solving Eq.~\eqref{eq:epsilon-L} with $\varepsilon = \Delta$, one gets
\begin{equation}
	\ln L_c (E) = \sigma^2\left[ \sqrt{B_0^2 + \frac{2}{\sigma^2} 
		\Big|\ln \frac{E}{\Delta} \Big|} - B_0 \right].
	\label{eq:LcE-general}
\end{equation}

For asymptotically low energies,
\begin{equation}
	\left | \ln \frac{E}{\Delta} \right | \gg \frac{B_0 \sigma^2}{2},
	\label{eq:low-energy-condition}
\end{equation}
Eq.~\eqref{eq:LcE-general} yields $\ln L_c(E) \approx \sigma \sqrt{2  |\ln (E/\Delta) | }$ and thus a very strong singularity of the density of states,
\begin{equation}
	\rho(E) \sim \frac{1}{E} \exp \left\{ - 2 \sqrt{2} \, \sigma  |\ln (E/\Delta) |^{1/2} \right\}.
	\label{eq:rho-E-asympt}
\end{equation}
As was shown in Refs.~\cite{motrunich2002particle-hole, mudry2003density, yamada2004random, dellanna2006anomalous}, this regime is accompanied by freezing of the multifractal spectrum, which leads to a modification of Eq.~\eqref{eq:rho-E-asympt}: the exponent 1/2 in the second (subleading) factor is changed to 2/3.

In the intermediate-energy range,
\begin{equation}
	\left | \ln \frac{E}{\Delta} \right | \ll \frac{B_0 \sigma^2}{2},
	\label{eq:high-energy-condition}
\end{equation}
one has from Eq.~\eqref{eq:LcE-general}  $L_c \sim B_0^{-1} |\ln (E/\Delta) |$ and a non-universal power-law scaling of the density of states,
\begin{equation}
	\rho(E) \sim E^{-1 + 2/B_0}.
	\label{eq:rho-E-power-law}
\end{equation}

To shed more light on the crossover between the asymptotic regimes \eqref{eq:rho-E-power-law} and \eqref{eq:rho-E-asympt}, it is useful to consider a running differential exponent of $\rho(E)$:
%\begin{eqnarray}
\begin{align}
	\frac{d\ln \rho(E)}{d \ln E} &= - 1 -2 \frac{d \ln L_c(E)}{d \ln E} = -  1 + \frac{2}{B},   \nonumber \\
	B &\equiv B(L_c(E)).
	\label{eq:rho-E-running-exp}
\end{align}
%\end{eqnarray}
Here $B(L) = 2 + \kappa(L) / \sigma^2$ has the meaning of a dynamical exponent.
It should be emphasized that the bare coupling $\kappa_0$ is of order unity even in a good metal with $\sigma \gg 1$~\cite{altland1999field,fabrizio2000anderson,fukui1999critical}.
In this situation, the bare value $B_0$ of the dynamical exponent is close to 2, and the density of states $\rho(E)$ exhibits a relatively weak dependence on energy in a broad range of not too low energies. With decreasing energy, the running differential exponent~\eqref{eq:rho-E-running-exp} varies logarithmically according to Eq.~\eqref{eq:kappa-L}. In the asymptotic limit, $B(L)$ becomes large, ultimately leading to the strong divergence~\eqref{eq:rho-E-asympt} of the density of states (with the subleading term modified by freezing as mentioned above). This happens, however, only at extremely low energies, $|\ln (E/\Delta) | \gg 2 \sigma^2$. Furthermore, the existence of the asymptotic regime~\eqref{eq:rho-E-asympt} requires very large system sizes, $\ln L \gg 2\sigma^2$.  Clearly, such values of $L$ become unreachable already for moderately large $\sigma$. It is thus not surprising that most numerical studies of the metallic phase of 2D systems in chiral classes found a power-law behavior of the local density of states with non-universal exponents of the localization length and the density of states~\cite{hatsugai1997disordered, cerovski2000bond-disordered, eilmes2001exponents, eilmes2004exponents, markos2007critical, schweitzer2008disorder-driven}. Numerical observations of the ultimate infrared asymptotic behavior required much care in the choice of models and of observables to be studied, as well as special computational efforts for reaching very low energies~\cite{markos2010logarithmic, schweitzer2012scaling, markos2012disordered}.

The perturbative  RG, Eqs.~\eqref{eq:RG-sigma}--\eqref{eq:RG-E}, describes a flow towards a line of metallic fixed points labelled by the value of $\sigma$. At the same time, it fails to describe a transition to the localized phase, since Eq.~\eqref{eq:RG-sigma} is exact within this framework. This appears to be in contradiction with numerical studies~\cite{motrunich2002particle-hole, bocquet2003network,karcher2022metal} that indicated existence of a metal-insulator transition in 2D chiral-class systems. A non-perturbative extension of the Gade-Wegner theory that does describe the metal-insulator transition was developed in Ref.~\cite{koenig2012metal-insulator}. The key ingredient added by this work is topological configurations---vortices, which lead to the violation of Eq.~\eqref{eq:RG-sigma}, i.e., to renormalization of conductivity, and thus may induce localization. This leads to a more complex RG flow in the space of three couplings ($\sigma$, $\kappa$, and the vortex fugacity $y$) and to a critical surface in this parameter space separating the metallic and the insulating phases. The metal-insulator-transition fixed point of the modified RG equations derived in Ref.~\cite{koenig2012metal-insulator} is at $\sigma=0$, $\kappa =8$, and $y= \frac14$. Furthermore, there is a very slow RG flow of $\sigma$  on the critical surface towards this fixed point, implying an apparent non-universality in some characteristics of the transition when studied in systems of realistic size.  However, as was also pointed out in Ref.~\cite{koenig2012metal-insulator}, the quantitative predictions should be taken with caution, since the equations are parametrically controlled at $y \ll 1$, while the obtained fixed point is characterized by a finite value of $y$.  Thus, quantitative characteristics of the transition may differ substantially and need to be determined numerically. We refer the reader to a separate publication~\cite{karcher2022metal} for a numerical investigation of key characteristics of this metal-insulator transition (such as the critical exponent of the localization length and the critical conductance) and for a comparison of numerical results with analytical expectations.

\section{Generalized multifractality observables in systems of chiral classes}
\label{sec:obs}

\subsection{Observables in the sigma-model language}
\label{sec:ops}

As was shown by Gade and Wegner in Ref. \cite{gade1991the} and by Gade in Ref.~\cite{gade1993anderson}, gradientless pure-scaling operators in chiral classes are labeled by two multi-indices $\lambda = (q_1, \ldots, q_m)$ and $\lambda' = (q'_1, \ldots, q'_{m'})$, which are highest weights of the corresponding representations. Here $\lambda$ characterizes the dependence on $U$, and $\lambda'$ the dependence on $U^\dagger \equiv U^{-1}$. For polynomial composite operators considered in Refs. \cite{gade1991the, gade1993anderson}, $\lambda$ and $\lambda'$ correspond to conventional Young diagrams, with integer positive $q_i$ and $q_i'$ being the lengths of $i$-th rows of the two diagrams.

More generally, $q_i$ and $q_i'$ may be fractional, negative, and even complex, in analogy with construction and classification of composite operators in other symmetry classes  \cite{gruzberg2013classification,karcher2021generalized, karcher2022generalized,karcher2022generalized-2}. The pure-scaling composite operators for chiral classes can be chosen in a product form,
\begin{align}
	\mathcal{P}_{\lambda,\lambda'}[U] &= \mathcal{P}_{\lambda}[U] \mathcal{P}_{\lambda'}[U^{-1}].
	\label{eq:ops}
\end{align}
Here $ \mathcal{P}_{\lambda}[U]$ and $ \mathcal{P}_{\lambda'}[U^{-1}]$ have the same form as pure-scaling operators in the corresponding Wigner-Dyson class.
We refer the reader to Appendix B of Ref.~\cite{karcher2022generalized-2} for a discussion of invariant pure-scaling operators in all ten symmetry classes and of relations between different classes. For the chiral orthogonal class BDI, an explicit construction of polynomial pure-scaling operators can also be found in Ref. \cite{dellanna2006anomalous}.

In Refs.~\cite{gruzberg2013classification,karcher2021generalized, karcher2022generalized,karcher2022generalized-2}, two complementary approaches to construction and analysis of pure-scaling operators have been discussed. One of them is based on the analysis of one-loop RG equations and yields polynomial eigenoperators as well as associated one-loop results for scaling exponents. These one-loop formulas for the exponents can be extended to generic pure-scaling operators by analytic continuation. The second approach is based on the Iwasawa decomposition and allows one to construct generic pure-scaling operators satisfying Abelian fusion rules. In this paper we use both approaches, with details of the RG analysis of composite operators in chiral classes, as well as simple examples of such operators, are presented in Appendix \ref{appendix-rg}. A detailed exposition of the Iwasawa approach for chiral classes is then given in Appendix~\ref{app:Iwasawa}.

As shown in appendix \ref{sec:renorm-comp-op}, the scaling dimensions $x_{\lambda,\lambda'}$ of the operators $\mathcal{P}_{\lambda,\lambda'}$ in class AIII are given, in the one-loop order, by
\begin{align}
	x_{\lambda,\lambda'} &= -\frac{1}{\sigma} (z_\lambda + z_{\lambda'}) + (|\lambda|-|\lambda'|)^2 x_\nu + \mathcal{O}(\sigma^{-4}),
	\label{x-one-loop}
	%\\
	%\Delta_{\lambda,\lambda'} &= x_{\lambda,\lambda'} - (|\lambda|+|\lambda'|)x_\nu,
	%\label{one-loop-dimensions}
\end{align}
where 
\begin{align}
	|\lambda| &= \sum_j q_j, 
	&
	|\lambda'| &= \sum_j q'_j,
\end{align}
and $x_\nu$ is the scaling dimension of the average density of states. In class AIII, an exact formula  for $x_\nu$ holds:
\begin{align}
	x_\nu= - \frac{\kappa}{\sigma^2}.
\end{align}
Furthermore,
\begin{align}
	z_\lambda = \lambda\cdot(\lambda+\rho) \equiv \sum_{i} q_i(q_i + c_i)
	\label{z-Laplacian}
\end{align}
is the eigenvalue of the class-A Laplacian, where $\rho = (c_1, \ldots, c_n)$ is the half-sum of positive roots, with $c_j=1-2j$.  In classes BDI and CII, formulas analogous to Eq.~\eqref{x-one-loop} hold, with some modifications, see Eqs.~\eqref{eq:one-loop-x-AIII}--\eqref{eq:one-loop-x-BDI} in Appendix~\ref{sec:renorm-comp-op}.

In Sec.~\ref{sec:wave} we consider a correspondence between the pure-scaling operators $\mathcal{P}_{\lambda,\lambda'}[U]$ of the sigma-model and pure-scaling observables expressed in terms of eigenfunctions of a Hamiltonian.

\subsection{Observables in the eigenfunction language}
\label{sec:wave}

Here, we analyze  the connection between the sigma-model composite operators and wave-function observables. The bosonic replica formalism is used; a generalization to fermions or supersymmetry is straightforward. We closely follow the steps in Refs. \cite{karcher2022generalized, karcher2022generalized-2} and focus on class AIII, brifly commenting on classes BDI and CII in the end.

In a generic chiral system, we can write the Hamiltonian as
\begin{align}
	H &= \begin{pmatrix}
		0 & Z\\
		Z^\dagger& 0
	\end{pmatrix}_\sigma,
\end{align}
in the basis where the chiral operation is given by $\sigma_z$. A conventional realization of the chiral symmetry is a bipartite lattice. The chiral symmetry prohibits hopping inside the sublattices $A$ and $B$, permitting only hopping between the sublattices. Thus, we refer to the $2 \times 2$ space $\sigma$ as the sublattice space.

We sketch now the derivation of the sigma model, which follows the usual steps \cite{zirnbauer1996riemannian}. We start by defining the action in the presence of a disorder field $V$ in the chiral Hamiltonian $H$:
\begin{align}
	&S[\psi, {\psi}^\dagger, V] =  \int d^2r  \,\bar{\Psi}({ H}  +  i\tau_3 \eta)\Psi \nonumber\\
	&= \int d^2 r  \begin{pmatrix} \psi^\dagger & i \psi^\dagger\sigma_z \end{pmatrix}_\tau
	\begin{pmatrix} H  + i\eta & \\ & H  -i\eta \end{pmatrix}_\tau
	\begin{pmatrix} \psi \\ i\sigma_z\psi \end{pmatrix}_\tau, \nonumber \\ &
\end{align}
where a convenient choice of the bosonic integration variables $\Psi$ was used:
\begin{align}
	\Psi &= \begin{pmatrix} \psi \\ i\sigma_z\psi \end{pmatrix}_\tau,
	& \bar{\Psi} = \begin{pmatrix} \psi^\dagger & i \psi^\dagger\sigma_z \end{pmatrix}_\tau.
\end{align}
Here we introduced an additional auxiliary (``retarded-advanced'') space $\tau$. The field $\psi$ is a vector in replica and $\sigma$ spaces. (In the case of classes BDI and CII, we would need one more $2 \times 2$ space to take into account the time-reversal symmetry.)

To perform disorder average, we need to compute Gaussian integrals over the matrices $V$. These integrals are fully determined by the second moment~\cite{zirnbauer1996riemannian}:
\begin{align}
	\int \! d\mu(V) \, {\rm tr} (AV) \, {\rm tr} (BV) &= \lambda \: {\rm tr} (AB - A\sigma_zB\sigma_z).
	\label{eq:Vavg}
\end{align}
The matrices $A,B$ acting in sublattice space  are of the form $A_{\sigma,\sigma'} \equiv\sum_{a,\tau} \Psi_{\sigma, a,\tau}({\bf r}) \bar{\Psi}_{\sigma',a,\tau}({\bf r})$, where $a$ is the replica index. Using Eq.~\eqref{eq:Vavg} to average over disorder, we obtain, as usual, a quartic term in the action:
\begin{align}
	S_{\rm int} &= \lambda \int dr \!\! \sum_{ab,\tau\tau',\sigma\sigma'} \!\!
	{\Psi}_{\sigma, a,\tau} \bar{\Psi}_{\sigma',a,\tau}{\Psi}_{\sigma', b,\tau'} \bar{\Psi}_{\sigma, b,\tau'}.
\end{align}
Next, we decouple this ``interaction'' term using a Hubbard-Stratonovich field $Q$, which has a matrix structure in the replica space and the $\tau$-space. The coupling of the $Q$ field to the $\psi, {\psi}^\dagger$ fields reads
\begin{align}
	\sum_{\sigma} \mathrm{tr}\left[\begin{pmatrix}
		\psi_\sigma {\psi}^\dagger_\sigma &  i (-1)^\sigma\psi_\sigma {\psi}^\dagger_\sigma \\
		i (-1)^\sigma \psi_\sigma {\psi}^\dagger_\sigma &  - \psi_\sigma {\psi}^\dagger_\sigma
	\end{pmatrix}
	\begin{pmatrix}
		Q^{RR} & Q^{RA}\\
		Q^{RA} & -Q^{RR}
	\end{pmatrix} \right],
	\label{eq:coup}
\end{align}
where we have explicitly displayed the structure in the retarded-advanced ($\tau$) space. (For the brevity of notation, replica indices are suppressed here.) This means that the wave functions on sublattices $A, B$ couple to different linear combinations of the $Q$ field:
\begin{align}
	{\psi}_A {\psi}^\dagger_A &\leftrightarrow Q^{RR}+ i Q^{RA}, &
	{\psi}_B {\psi}^\dagger_B \leftrightarrow Q^{RR}- i Q^{RA}.
	\label{eq:sublat}
\end{align}
We verify that these distinct combinations of $Q$ live in different representations of the unitary group by means of the Iwasawa decomposition discussed in Sec.~\ref{sec:iwa}
and in more detail in Appendix \ref{app:Iwasawa}. Another way to see this is to note that these combinations can be used as building blocks of generic $K$-invariant sigma-model operators. Specifically, not only traces of products of $Q\Lambda$ are gauge-invariant as in non-chiral classes but also traces of products of $Q\Lambda^\pm$ with $\Lambda^\pm  = \tau_3\pm i\tau_1$.

After restriction to slow variations on the saddle-point manifold, the matrix field $Q$ fulfills the nonlinear constraint $Q^2 = \mathds{1}$ characteristic for sigma models. Thus, we have, using the structure of the $Q$-field in the retarded-advanced space (see Eq.~\eqref{eq:coup}),
\begin{align}
	&(Q^{RR}+ i Q^{RA}) (Q^{RR}- i Q^{RA})
	\nonumber\\
	&= (Q^{RR})^2 + (Q^{RA})^2 + i[Q^{RA}, Q^{RR}] = \mathds{1}.
	\label{eq:proof-Q-U}
\end{align}
Indeed, the first two terms in the second line of Eq.~\eqref{eq:proof-Q-U} sum up to $(Q^2)^{RR}=\mathds{1}$, whereas the commutator equals $(Q^2)^{RA}=0$. This allows us to identify 
\begin{align}
	U &\equiv Q^{RR} + i Q^{RA}, 
	& U^{-1} &\equiv Q^{RR}- i Q^{RA}, 
\end{align}
in the notations of Sec.~\ref{sec:sigma} and Sec.~\ref{sec:ops}. In view of Eq.~\eqref{eq:sublat}, we thus see that $U$  (and correspondingly the multi-index $\lambda$ in Eq.~\eqref{eq:ops}) is associated with the sublattice A, while $U^{-1}$ (and correspondingly the multi-index $\lambda'$) with the sublattice B.

In terms of the $Q$-field, the action~\eqref{eq:gade-action} reads:
\begin{align}
	S[Q] &=  \! \int \! d^2 r \!  \Big[ \frac{\sigma}{16\pi s} \text{tr} (\nabla Q)^2 - \frac{\kappa}{32\pi s} (\text{tr} Q \nabla Q)^2
	\nonumber \\
	& \quad -  i \frac{\pi \rho_0}{2s} \, \varepsilon \, \text{tr}\, Q \Big].
	\label{eq:gade-action2}
\end{align}
In this form, the action is presented in Refs.~\cite{fabrizio2000anderson, dellanna2006anomalous}.

An analysis along the lines of Refs.~\cite{karcher2022generalized, karcher2022generalized-2} allows us to translate now the sigma-model composite operators to the eigenfunction language. The building blocks of the eigenfunction construction are Slater determinants (spinless for class AIII)
\begin{align}
	P_{(1^k), (0)}[\psi] &= \mathrm{det} \left(\psi_i ({\bf r}_j, A)\right)_{k\times k},\nonumber\\
	P_{(0), (1^k)}[\psi] &= \mathrm{det} \left(\psi_i ({\bf r}_j, B)\right)_{k\times k}.
	\label{eq:det}
\end{align}
where we use the notation $q^k$ for a multi-index $(q, \ldots, q)$ with $k$ equal elements. The distinct feature of the chiral symmetry classes is that these determinants should be defined separately for each of the sublattices A and B.
A generic sigma-model pure-scaling operator~\eqref{eq:ops} now corresponds to
\begin{align}
	P_{\lambda,\lambda'}[\psi] &= P_{(1), (0)}^{q_1-q_2}[\psi]  \ldots P_{(1^{m-1}), (0)}^{q_{m-1} - q_m}[\psi] P_{(1^m), (0)}^{q_m}[\psi]
	\nonumber \\
	& \times P_{(0), (1)}^{q'_1-q'_2}[\psi]  \ldots P_{(0), (1^{m'-1})}^{q'_{m'-1} - q'_{m'}}[\psi] P_{(0), (1^{m'})}^{q_{m'}}[\psi].
	\label{eq:eigenfunction-scaling-observ}
\end{align}
A precise correspondence (worked out for class A in Ref.~\cite{gruzberg2013classification} and extended to other classes in Refs.~\cite{karcher2022generalized, karcher2022generalized-2}) establishes the pure-scaling nature of the observables~\eqref{eq:eigenfunction-scaling-observ} by mapping them to the $N$-radial functions $\phi_{\lambda, \lambda'}(Q)$ of the sigma model described in the next section. The correspondence implies that the eigenfunction observables~\eqref{eq:eigenfunction-scaling-observ} share the Abelian fusion property with the sigma-model operators $\phi_{\lambda, \lambda'}(Q)$, see Eq.~\eqref{eq:phi-chiral-fusion}:
\begin{align}
	P_{\lambda_1, \lambda'_1}[\psi] P_{\lambda_2, \lambda'_2}[\psi]
	= P_{\lambda_1 + \lambda_2, \lambda'_1 + \lambda'_2}[\psi]
	\label{eq:ab}.
\end{align}

The scaling of these generalized multifractal wavefunction observables (averaged over disorder realizations) with the system size $L$ is determined by the generalized-multufractality exponents $\Delta_{\lambda, \lambda'}$,
\begin{equation}
	L^{2q+2q'} \langle P_{\lambda, \lambda'}[\psi]\rangle \sim L^{-\Delta_{\lambda, \lambda'}},
\end{equation}
where $q = |\lambda|$ and $q' = |\lambda'|$. The factor  $L^{2q+2q'}$ takes care of the normal dimension of $\langle P_{\lambda, \lambda'}[\psi]\rangle$ (i.e. that in a perfect metal), so that $\Delta_{\lambda, \lambda'}$ is the anomalous dimension. The scaling dimension $\Delta_{\lambda, \lambda'}$ of an eigenfunction observable is related to the corresponding field-theoretical composite-operator dimension $x_{\lambda, \lambda'}$ via
\begin{align}
	x_{\lambda,\lambda'} &= \Delta_{\lambda, \lambda'} + (|\lambda| + |\lambda'|)x_{\nu}.
	\label{x-Delta}
\end{align}

The same construction of pure-scaling eigenfunction observables, Eqs.~\eqref{eq:eigenfunction-scaling-observ} and~\eqref{eq:det}, applies also to the chiral orthogonal class BDI, since it is also spinless. On the other hand, the chiral symplectic class CII is spinful (it possesses time reversal invariance satisfying $\mathcal{T}^2=-1$), so that  Eq.~\eqref{eq:det} is replaced by a spinful version of Slater determinants, see Ref.~\cite{karcher2022generalized-2}.

\subsection{The Iwasawa construction}
\label{sec:iwa}

In this section, we briefly discuss the Iwasawa construction; details can be found in Appendix \ref{app:Iwasawa}. We begin by recalling general properties of the Iwasawa decomposition that has been developed earlier for other symmetry classes and then describe features that are specific to chiral classes.

Within the sigma model field theory, observables characterizing the generalized multifractality are represented by gradientless composite operators $\mathcal{P}(Q)$. Here the sigma model field $Q \in G/K$ is a matrix, $Q = g\Lambda g^{-1}$, where $\Lambda$ is a matrix that commutes with all $k \in K$ (a standard choice is $\Lambda = \text{diag}(I_{m}, -I_{m})$, where the identity blocks are in the retarded-advanced space, and the integer $m$ depends on the symmetry class), and $g \in G$. Since $Q$ does not change when $g$ is multiplied on the right ($g \to gk$) by any element $k \in K$, the set of matrices $Q$ realizes the symmetric space $G/K$.

The pure-scaling sigma model observables $\mathcal{P}_{\lambda}(Q)$ can be constructed in different ways. One important choice is provided by the Iwasawa decomposition~\cite{Helgason-Differential-1978, Helgason-Groups-1984}. (In the supersymmetric approach we need a generalization to Lie supergroups that was worked out in Ref.~\cite{Alldridge-The-Harish-Chandra-2012}.) The Iwasawa decomposition was explicitly performed for class A in Ref.~\cite{gruzberg2013classification}, for class C in Ref.~\cite{karcher2021generalized}, and for classes AII, D, and DIII in Ref.~\cite{karcher2022generalized-2}. Here is a brief description of the method in the classical setting. In this context the label $\lambda$ is the highest weight of an irreducible representation of the group $G$.

Any connected non-compact semisimple Lie group $G$ has a global Iwasawa decomposition $G = NAK$, where $N$ is a nilpotent group, $A$ is an Abelian group, and $K$ is the maximal compact subgoup of $G$. This factorization provides a very useful parametrization of the target space $G/K$. An element $a \in A$ is fully specified by $n$ real numbers $x_i$, which play the role of radial coordinates on $G/K$. In terms of the radial coordinates, the pure-scaling operators $\phi_\lambda(Q)$ are simply ``plane waves'':
\begin{equation}
	\phi_\lambda(Q) = \phi_\lambda(x) = e^{-2 \sum_i q_i x_i}.
	\label{eq:phi-lambda}
\end{equation}
(Note that, in the supersymmetric formulation, $x_i$ are the Iwasawa radial coordinates in the boson sector.)

To construct the pure-scaling operators explicitly as combinations of matrix elements of $Q$, we use the key fact that there exists a choice of basis in which elements of $a \in A$ are diagonal matrices, while elements of $n \in N$ are upper triangular with units on the diagonal. This has immediate consequences for the matrix $Q \Lambda$: since elements of $K$ commute with $\Lambda$, the Iwasawa decomposition $g = nak$ leads to $Q \Lambda = n a^2 \Lambda n^{-1} \Lambda$, which is a product of an upper triangular, a diagonal, and a lower triangular matrices. In this form the lower principal minors of the advanced-advanced block of $Q \Lambda$ are simply products of diagonal elements of $a^2$, which are exponentials of the radial coordinates $x_i$ on $G/K$. These minors are basic building blocks, which can be raised to arbitrary powers and multiplied to produce the most general exponential functions~\eqref{eq:phi-lambda}. A great advantage of this choice is that the functions $\phi_\lambda(x)$ are positive highest-weight vectors. Thus, they satisfy Abelian fusion and can be raised to any power:
\begin{align}
	\phi_{\lambda_1}(x) \phi_{\lambda_2}(x) &= \phi_{\lambda_1 + \lambda_2}(x),
	&
	(\phi_\lambda(x))^c &= \phi_{c\lambda}(x).
	\label{eq:phi-lambda-fusion}
\end{align}

Let us now focus on the features that are specific to the chiral classes (see Appendix \ref{app:Iwasawa} for detail). First of all, as was discussed in Sec.~\ref{sec:wave},
microscopic models in these classes have two sublattices, and this leads to a block-diagonal structure of the sigma model field, where the two blocks can be used to construct generalized MF scaling observables for each sublattice using the Iwasawa construction described above. The most general multifractal observables are then products over the two sublattices, and they depend on two sets of radial variables, $x_1,\ldots, x_n$ and $x'_1,\ldots, x'_n$, one for each sublattice, and also labeled by two weights $\lambda$ and $\lambda'$: $\phi_{\lambda, \lambda'}(x,x')$. The corresponding scaling dimensions $x_{\lambda, \lambda'}$ also depend on two weights. The general scaling observables satisfy the Abelian fusion
\begin{align}
	\phi_{\lambda_1, \lambda'_1}(x,x') \phi_{\lambda_2, \lambda'_2}(x,x')
	= \phi_{\lambda_1 + \lambda_2, \lambda'_1 + \lambda'_2}(x,x').
	\label{eq:phi-chiral-fusion}
\end{align}
As we discussed above in Sec.~\ref{sec:wave}, the observables $\phi_{\lambda, \lambda'}$ are in direct correspondence with the pure-scaling eigenfunction observables satisfying analogous properties, including Abelian fusion, Eq.~\eqref{eq:ab}.

Secondly, another distinctive feature of the chiral classes is that the groups $G$ and $K$ involved in the construction are not \emph{semisimple}. The non-semisimplicity allows for the presence of the Gade term in the sigma model action (the second term in Eqs.~\eqref{eq:gade-action} and~\eqref{eq:gade-action2}). This term then leads to  contributions in the scaling dimensions of generalized multifractal observables and thus should be taken into account in the analysis of the Weyl symmetry of the spectra of generalized multifractality. Results of the analysis of the Weyl symmetries performed in Appendix~\ref{app:Iwasawa} are presented in
Sec.~\ref{sec:weyl}.

It is worth mentioning here that there is an alternative method of construction of pure scaling operators in the sigma model that employs the Cartan decomposition $G=KAK$, and leads to the so-called $K$-invariant  (or $K$-radial) eigenfunctions $\mathcal{P}_{\lambda,\lambda'}(Q)$. These functions naturally appear in the perturbative RG construction described in detail in Appendix~\ref{appendix-rg}, see, for example, Eq.~\eqref{K-radial-P}. While the $K$-radial scaling operators $\mathcal{P}_{\lambda,\lambda'}(Q)$ do not satisfy Abelian fusion, they belong to the same $G$-representation as the $N$-radial functions $\phi_{\lambda, \lambda'}(Q)$, and have the same RG eigenvalues (the scaling dimensions) $x_{\lambda, \lambda'}$.

\subsection{Weyl symmetries: specifics for chiral classes}
\label{sec:weyl}

The generalized multifractal scaling dimensions $x_\lambda$ in all symmetry classes satisfy a set of symmetry relations that follow from the symmetry of their weights $\lambda$ under the action of a Weyl group $W$. The Weyl group $W$ depends on the symmetry class, and may involve two types of actions on components $q_i$ of the weights: reflections
\begin{align}
	q_i \rightarrow -c_i - q_i,
\end{align}
and permutations
\begin{align}
	q_i &\rightarrow q_j+(c_j-c_i)/2,
	&
	q_j &\rightarrow q_i+(c_i-c_j)/2.
	\label{A-reflections}
\end{align}
In both cases, the numbers $c_i$ are components of the so-called Weyl vector $\rho$ (which was already introduced in Sec.~\ref{sec:ops}) equal to the half-sum of positive restricted roots. For the chiral classes these are
\begin{align}
	c_j &= 1 - 2j,  & \text{class AIII},  \\
	c_j &= \frac12 - j,   & \text{class BDI},  \\
	c_j &= 2 - 4j,  & \text{class CII}.
\end{align}

In all symmetry classes studied by us earlier in Refs.~\cite{gruzberg2013classification, karcher2021generalized, karcher2022generalized-2}, the relevant Weyl groups included both reflections and permutations. Let us call such a group ``type A'' and denote it by $W_\text{A}$. On the other hand, in the chiral classes the Weyl groups include only permutations. At first sight, this substantially reduces  implications  of the Weyl symmetry. It is important, however, that the Weyl groups in chiral classes include  permutations
between all components of \emph{both} weights:
\begin{align}
	q_i &\to q_j + (c_j - c_i)/2,
	&
	q_j &\to q_i + (c_i - c_j)/2,
	%\label{exchange-1}
	\nonumber \\
	q'_i &\to q'_j + (c_j - c_i)/2,
	&
	q'_j &\to q'_i + (c_i - c_j)/2,
	%\label{exchange-2}
	\nonumber \\
	q_i &\to - q'_j - (c_i + c_j)/2,
	&
	q'_j &\to - q_i - (c_i + c_j)/2.
	%\label{exchange-3}
	\label{AIII-exchanges}
\end{align}
Let us call such a group ``type AIII'' and denote it by $W_\text{AIII}$. Thus, the most general observables $\phi_{\lambda, \lambda'}(x,x')$ characterized by two different weights $\lambda \neq \lambda'$ satisfy the type AIII symmetry relations:
\begin{align}
	x_{\lambda, \lambda'} &= x_{w(\lambda, \lambda')},
	&
	w &\in W_\text{AIII}.
	\label{Weyl-AIII}
\end{align}

Notice that if the weights $\lambda$ and $\lambda'$ contain equal entries at the position $i$: $q_i = q'_i$, then the last type of exchange in Eq.~\eqref{AIII-exchanges} is equivalent to simultaneous reflections~\eqref{A-reflections} of both $q_i$ and $q'_i$! This leads to the following result: for ``balanced'' observables characterized by two equal weights $\lambda' = \lambda$, the dimensions $x_{\lambda, \lambda}$ enjoy both the permutation and reflection Weyl symmetries (i.e., those of class A) in the sense that
\begin{align}
	x_{\lambda, \lambda} &= x_{w(\lambda), w(\lambda)},
	&
	w &\in W_\text{A},
	\label{Weyl-A}
\end{align}
which are among the class-AIII symmetries~\eqref{Weyl-AIII}.

The natural equivalence of the two sublattices also implies that the scaling dimensions should be symmetric under the exchange of the two weights that label them:
\begin{align}
	x_{\lambda, \lambda'} = x_{\lambda', \lambda}.
	\label{sublattice-symmetry}
\end{align}
This sublattice symmetry combined with a perturbative analysis  also leads to the above results~\eqref{Weyl-AIII} and~\eqref{Weyl-A}, see Appendix~\ref{app:Iwasawa} for details.

\section{Model}
\label{sec:model}
%%%%%%%%%%%%%%%%%%
\begin{figure}
	\centering
	\includegraphics[width=0.4\textwidth]{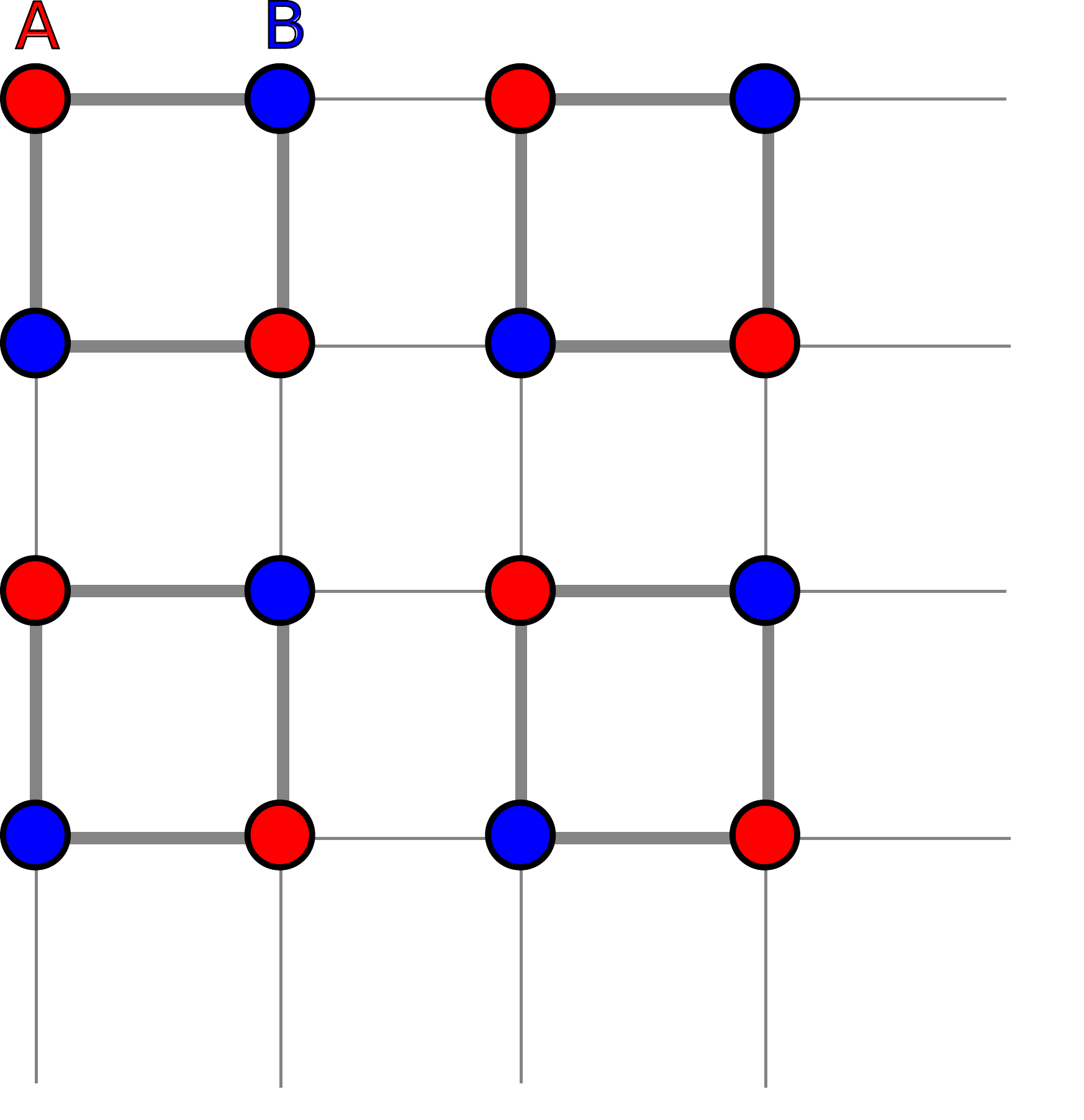}
	\caption{Schematic presentation of the tight-binding model \eqref{eq:stag} used for numerical investigation of the generalized multifractality. The sites of two sublattices A and B are shown by red and blue colors, respectively. In view of a non-zero staggering $\delta$, there are strong (thin lines) and weak (thick lines) bonds.
		%{\bf \color{blue} Corrections to the figure: (i) We do not perform transfer matrix analysis in this paper, so that showing the corresponding slices by pale colors is not needed; (ii) Notations for hoppings  $t_n^{AB}$ on the figure are totally different from the notations in Eq.~\eqref{eq:stag}. Either they should be totally removed on the figure or replaced by correct notations.}
	}
	\label{fig:lattice_aiii}
\end{figure}
%%%%%%%%%%%%%%%%%%%

The model that we study numerically is defined by a bipartite tight-binding Hamiltonian  defined on a square lattice:
\begin{align}
	H &= \sum_{i,j}  \left[ c_{i,j}^\dagger t^{(x)}_{i,j} c_{i+1,j} + c_{i,j}^\dagger t^{(y)}_{i,j} c_{i,j+1} + \text{h.c.} \right].
	\label{eq:ham_aiii}
\end{align}
Clearly, this Hamiltonian possesses chiral symmetry since the only non-zero matrix elements are nearest-neighbor hopping terms, which are off-diagonal in the sublattice space. The hoppings $t^{(x)}$ or $t^{(y)}$ are complex, so that the time-reversal symmetry is broken, and the model belongs to the unitary chiral symmetry class AIII. We leave numerical analyses of generalized multifractality in models in the chiral orthogonal (BDI) and the chiral symplectic (CII) classes to future studies.

We choose the following form of the hopping matrix elements
\begin{align}
	t^{(x)}_{i,j} &= \begin{cases}
		e^{-\delta} (1 + v_{i,j}), & i\textrm{ even},\\
		1 + v_{i,j} \:, & i\textrm{ odd},
	\end{cases} \nonumber\\
	t^{(y)}_{i,j} &= \begin{cases}
		e^{-\delta} (1 + w_{i,j}), & j\textrm{ even},\\
		1 + w_{i,j} \:, & j\textrm{ odd}.
	\end{cases}
	\label{eq:stag}
\end{align}
The disorder is introduced via complex terms $v_{i,j}$ and $w_{i,j}$, whose real and imaginary parts are independent random variables with a box distribution on $[-W/2, W/2]$. Further, $\delta$ is a real parameter that controls the staggering. For $\delta = 0$ there is no staggering, while for $\delta\rightarrow \pm\infty$ the staggering is maximal and the system breaks up into $2 \times 2$ plaquettes.  The model is illustrated in Fig.~\ref{fig:lattice_aiii}, where the sites of two sublattices A and B  are shown by red and blue dots, respectively.  Because of staggering, $\delta \ne 0$, there are strong and weak bonds, which are presented by thick and thin lines, respectively.

The metal-insulator transition in this model was explored numerically in Ref.~\cite{karcher2022metal}. In that work, we determined the metal-insulator transition line $W_c(\delta)$ in the parameter plane spanned by the disorder $W$ and the staggering $\delta$ (see Fig.~3 in Ref.~\cite{karcher2022metal}), and investigated the behavior of various observables. The results were in agreement with analytical predictions of Ref.~\cite{koenig2012metal-insulator} (within the $\sigma$-model formalism) of the RG flow and the metal-insulator transition driven by vortices. Among other observables, we studied in Ref.~\cite{koenig2012metal-insulator} the ``conventional'' multifractality (the scaling of eigenfunction moments) at the critical line. In Sec.~\ref{sec:metal} below, we extend this analysis to the generalized multifractality at the metal-insulator transition line $W_c(\delta)$.  It is worth noting that, in the absence of staggering ($\delta = 0$), we do not find a transition up to the strongest disorder studied ($W=20$). It is thus likely that for $\delta=0$ the model remains in the metallic phase for all values of disorder $W$.

In Sec.~\ref{sec:metal}, we carry out the generalized-parabolicity analysis for the metallic phase, $W < W_c(\delta)$. In Ref.~\cite{koenig2012metal-insulator}, the model defined above was used to study numerically the ``conventional'' multifractality in this phase. Here, we choose for this purpose a slightly modified model, with imaginary parts of  $v_{i,j}$ and $w_{i,j}$ having a box distribution on $[0,1]$ (the real parts are still characterized by the box distribution on $[-W/2,W/2]$ as above). While both models have similar properties, finite-size effects in the metallic phase turn out to be somewhat weaker in the latter model.

\section{Numerical study of generalized multifractality:  Metallic phase}
\label{sec:metal}
We perform numerical studies of the generalized multifractality by using the tight-binding model defined in Sec.~\ref{sec:model}. Employing sparse-matrix libraries, we extract a few wavefunctions around zero energy for systems of linear sizes $L$ between $24$ and $1024$.
For each realization, we calculate the generalized-multifractality observables $P_{\lambda, \lambda'}[\psi]$ as defined in Sec.~\ref{sec:wave}. The spatial arguments ${\bf r}_i$ of all involved wave functions are chosen within a distance $r \ll L$ from their ``center of mass''; the smallest $r$ is of order of the lattice spacing.

In the numerical studies in the present paper, we focus on two classes of the generalized-multifractality observables:  one-sublattice observables $P_{\lambda, (0)}[\psi]$ and balanced observables $P_{\lambda, \lambda}[\psi]$.  For each simulation, we perform ensemble averaging over $10^5$ configurations of disorder and over $L^2$ points in the sample.
The scaling of $\langle P_{\lambda, \lambda'}[\psi] \rangle $ with the system size $L$ yields the generalized-multifractality exponents $\Delta_{\lambda, \lambda'}$. The analysis proceeds along the lines of Refs.~\cite{karcher2021generalized, karcher2022generalized, karcher2022generalized-2}, where more details on the numerical scaling analysis of the generalized multifractality are provided. Our first goal here is to verify correctness of the analytical construction of the pure-scaling observables $P_{\lambda, \lambda'}[\psi]$.

Within one-loop approximation, the sigma-model RG predicts (see Eq.~\eqref{x-one-loop} above as well as  Eq.~\eqref{eq:x1l} in Appendix~\ref{sec:renorm-comp-op}) a generalized parabolic form of the spectrum  $x_{\lambda, \lambda'}$ with two independent parameters $b$ and $x_\nu$:
\begin{align}
	x_{\lambda,\lambda'} &\simeq -b (z_\lambda + z_{\lambda'}) +  (|\lambda|-|\lambda'|)^2 x_\nu.
	\label{eq:one-loop-dim-repeated}
\end{align}
We recall (see Sec.~\ref{sec:ops}) that $z_\lambda$ are eigenvalues of the class-A Laplacian, $b=1/\sigma$, and that the exponents $x_{\lambda,\lambda'}$ and $\Delta_{\lambda, \lambda'}$ are related via $x_{\lambda,\lambda'} = \Delta_{\lambda, \lambda'} +  (|\lambda| + |\lambda'|) x_\nu$ (see Eq.~\eqref{x-Delta}). Further, the exponent $x_\nu$ controls the scaling of the density of states: 
\begin{align}
	\nu(L) &\sim L^{-x_\nu}, 
	&
	\nu(E) &\sim E^{\alpha_\nu}, 
	&
	\alpha_\nu &= \frac{x_\nu}{2-x_\nu}. 
\end{align}
The one-loop approximation is parametrically controlled when the dimensionless conductivity is large, $\sigma \gg 1$. Corrections to Eq.~(\ref{eq:one-loop-dim-repeated}) predicted by the sigma-model are of the order $\sigma^{-4}$; they break the generalized parabolicity.  It is thus expected from analytical consideration that for a ``bad metal'', with a not too large $\sigma$, deviations from parabolicity should become sizeable. Verification of these predictions (parabolicity deeply in the metallic phase and its violation for smaller $\sigma$, i.e., stronger disorder) is another goal in this part of the work.

A further important prediction of the field theory is the Weyl symmetry, which is discussed in detail in Sec.~\ref{sec:weyl}. It implies multiple relations between the generalized multifractality exponents.  In particular, the balanced exponents $x_{\lambda,\lambda}$ obey invariance under the conventional Weyl reflections and permutations as in class A. Relations that are relevant for our analysis include 
\begin{align}
	x_{(1),(1)} &= 0, 
	&
	x_{(2,1),(2,1)} &= 0, 
	&
	x_{(2^2),(2^2)} &= 0. 
\end{align}
Furthermore, the Weyl invariance predicts that the exponents $x_{(q_1^m), (q_1^m)} $ should exhibit the symmetry $q_1 \longleftrightarrow m - q_1$.  The numerics presented here allow us to verify these relations.

\begin{figure}
	\centering
	\includegraphics[width=0.48\linewidth]{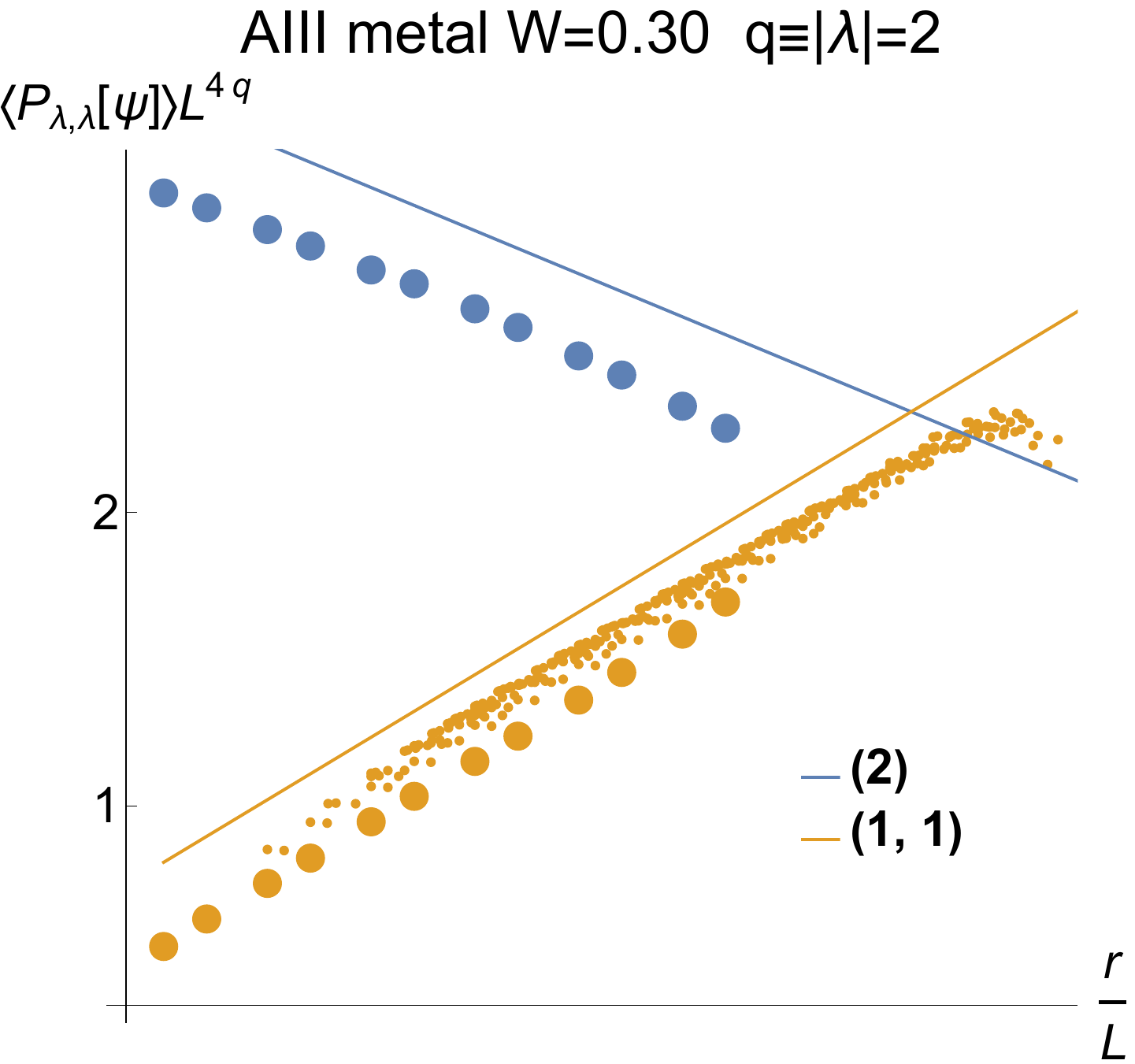}
	\includegraphics[width=0.48\linewidth]{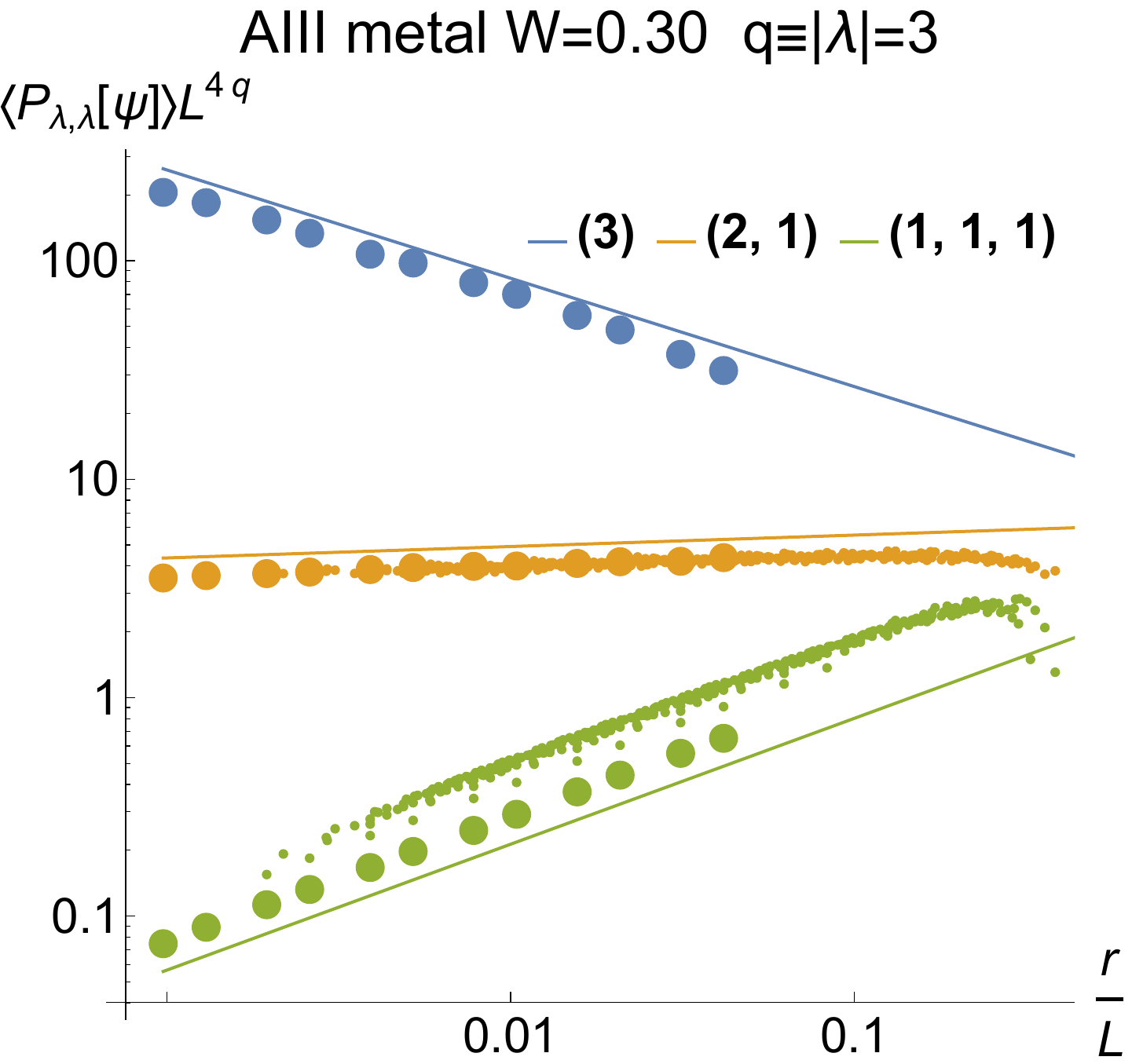}
	\\[0.3cm]
	\includegraphics[width=0.48\linewidth]{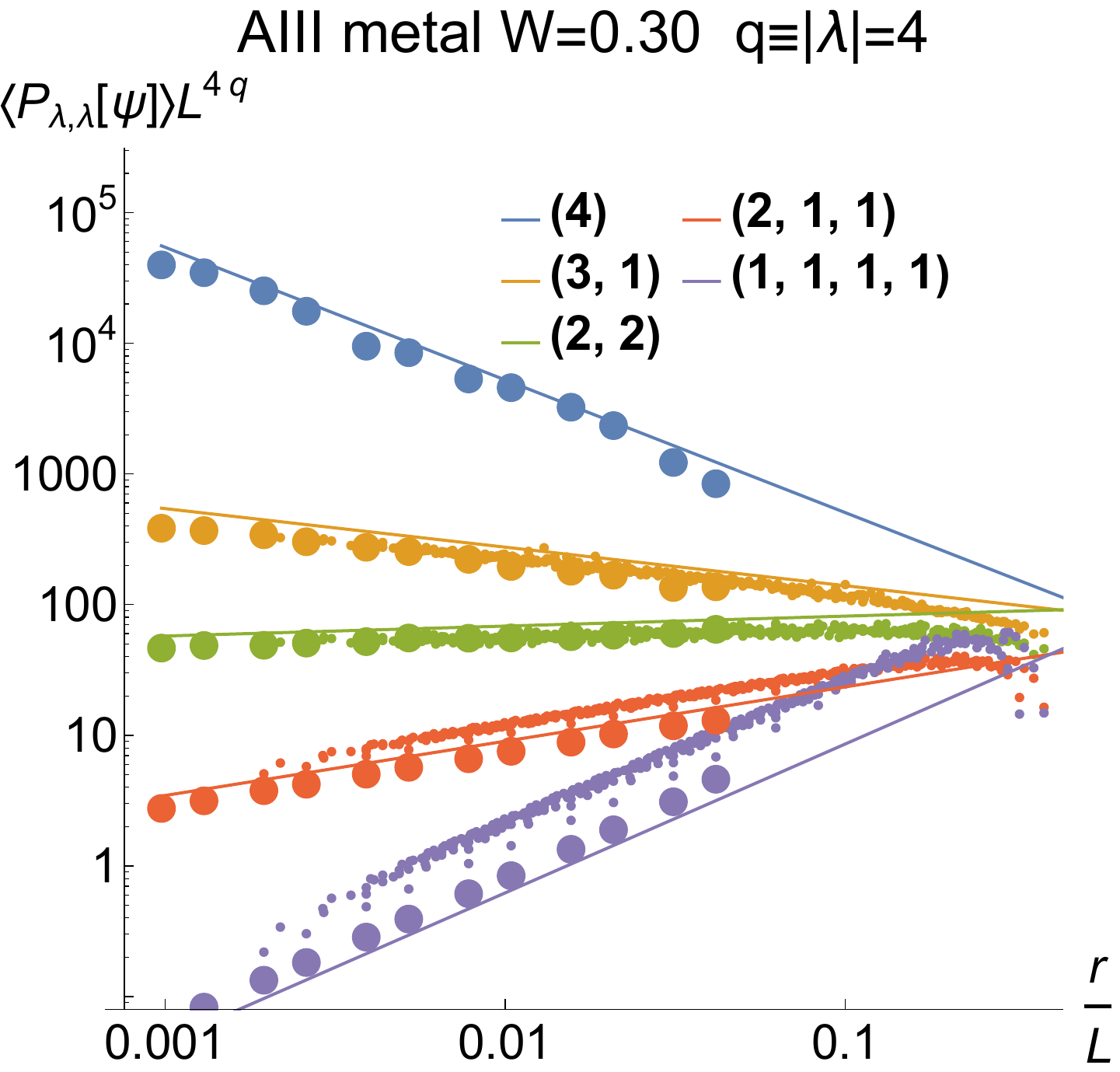}
	\includegraphics[width=0.48\linewidth]{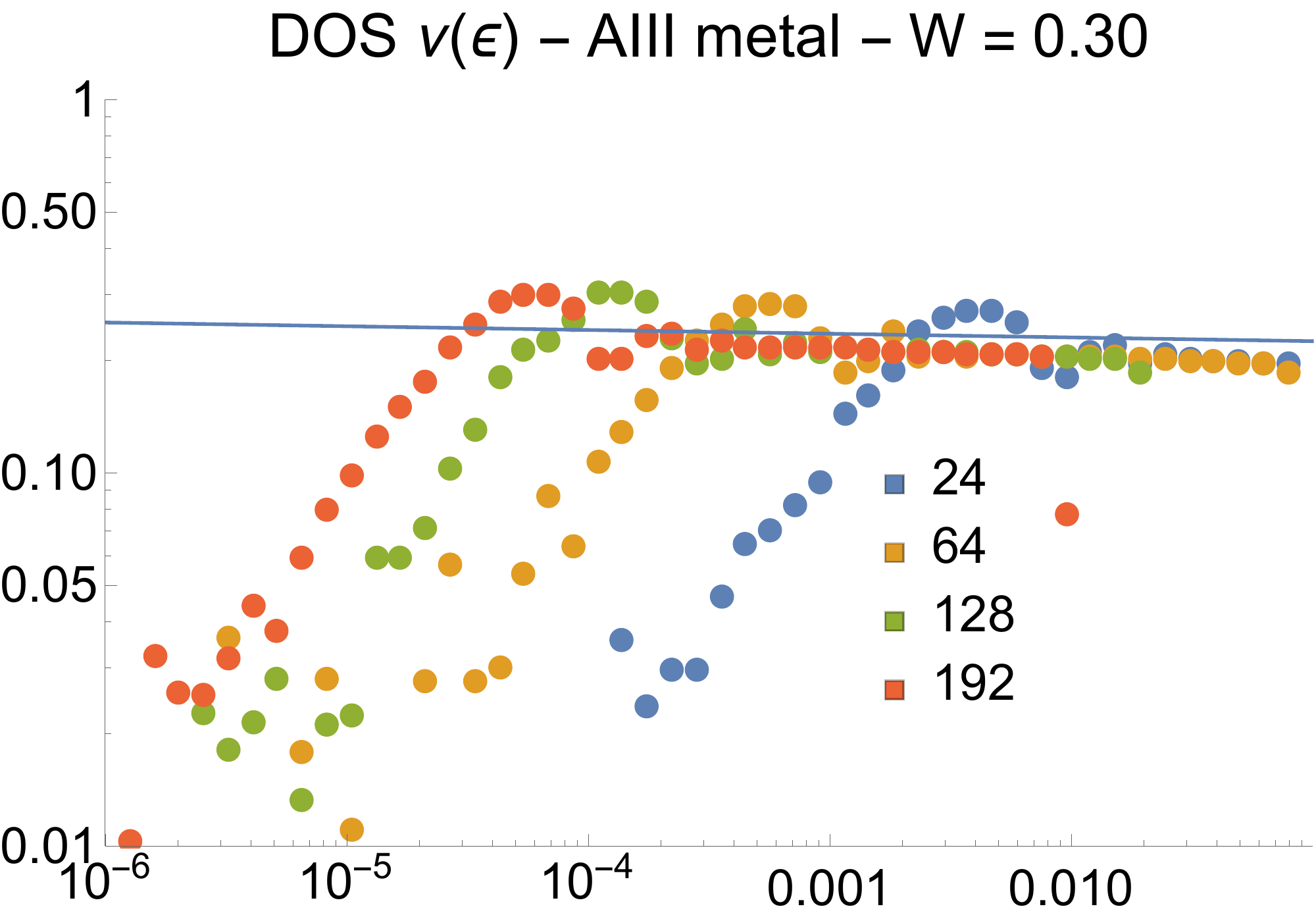}
	\caption{Generalized multifractality in the metallic phase: disorder $W=0.3$, no staggering ($\delta=0$).  Scaling of averaged balanced eigenfunction observables   with $r/L$ is shown in three panels (upper row and bottom left) for polynomial observables with $q \equiv |\lambda| = 2$, 3, and 4, respectively. Data points for the smallest $r \sim 1$ are shown by large dots. The fitted slopes (straight lines) determine the exponents $\Delta_{\lambda, \lambda}$. Bottom right panel: density of states $\nu(E)$ for different system sizes $L=24,\ldots, 192$. There is a very weak power-law dependence $\nu(E) \sim E^{\alpha}$ with $\alpha \sim - 0.01$ (straight line).
	}
	\label{fig:scaling_aiii}
\end{figure}

The Weyl relation $x_{(1),(1)}=0$ implies that 
\begin{align}
	\Delta_{(1),(1)} = -2x_\nu. 
	\label{Delta11-nu}
\end{align}
We have checked that our results---with the exponent $x_\nu$ obtained from the energy scaling of the density of states $\nu(E)$ (see, e.g., the lower right panel of Fig.~\ref{fig:scaling_aiii})---are in full consistency with this relation. Furthermore, it turns out that this relation allows us to determine $x_\nu$ more accurately than from the $\nu(E)$ scaling. Thus, we use the numerically obtained $\Delta_{(1),(1)}$ exponent to calculate $x_\nu$, which is then in turn used to translate other exponents $\Delta_{\lambda, \lambda'}$  into $x_{\lambda, \lambda'}$.

%%%%%%%%%%%%%%%%%%%%
%%%%%%%%%%%%%%%%%%%%

In Fig.~\ref{fig:scaling_aiii}, we present numerical results for the scaling of balanced ($\lambda' = \lambda$) eigenfunction observables $\langle P_{\lambda, \lambda}[\psi] \rangle $ deeply in the metallic phase ($W=0.3$, no staggering). Specifically, polynomial observables with $q \equiv |\lambda| = 2$, 3, and 4 are displayed.  The data are presented on log-log scale, so that straight lines correspond to power-law scaling with $r/L$, and their slopes to scaling exponents $\Delta_{\lambda, \lambda}$. The results confirm that the analytically derived composite objects $P_{\lambda, \lambda}[\psi]$, see Sec.~\ref{sec:obs}, are indeed pure-scaling observables. Large dots represent the data for the smallest $r \sim 1$. Straight lines are power-law fits to these data.
The bottom right panel of Fig.~\ref{fig:scaling_aiii} shows the density of states $\nu(E)$ for system sizes in the range $L=24,\ldots, 192$. The data exhibit a power-law behavior $\nu(E) \sim E^{\alpha}$ with a very small negative exponent, $\alpha \sim - 0.01$.  As discussed above, a more accurate way to extract the density-of-state exponent is to use the relation~\eqref{Delta11-nu}:  $\Delta_{(1),(1)} = -2x_\nu$. For the disorder $W=0.3$, this yields $x_\nu = -0.008$. This value is used to convert the exponents $\Delta_{\lambda, \lambda}$ to $x_{\lambda, \lambda}$, which are presented in Table~\ref{tab:lchiral}. As seen from this Table (and discussed in more detail below), the $W=0.3$ exponents satisfy the generalized parabolicity with an excellent accuracy, in full agreement with analytical predictions for weak disorder.

\begin{figure}
	\includegraphics[width=0.85\linewidth]{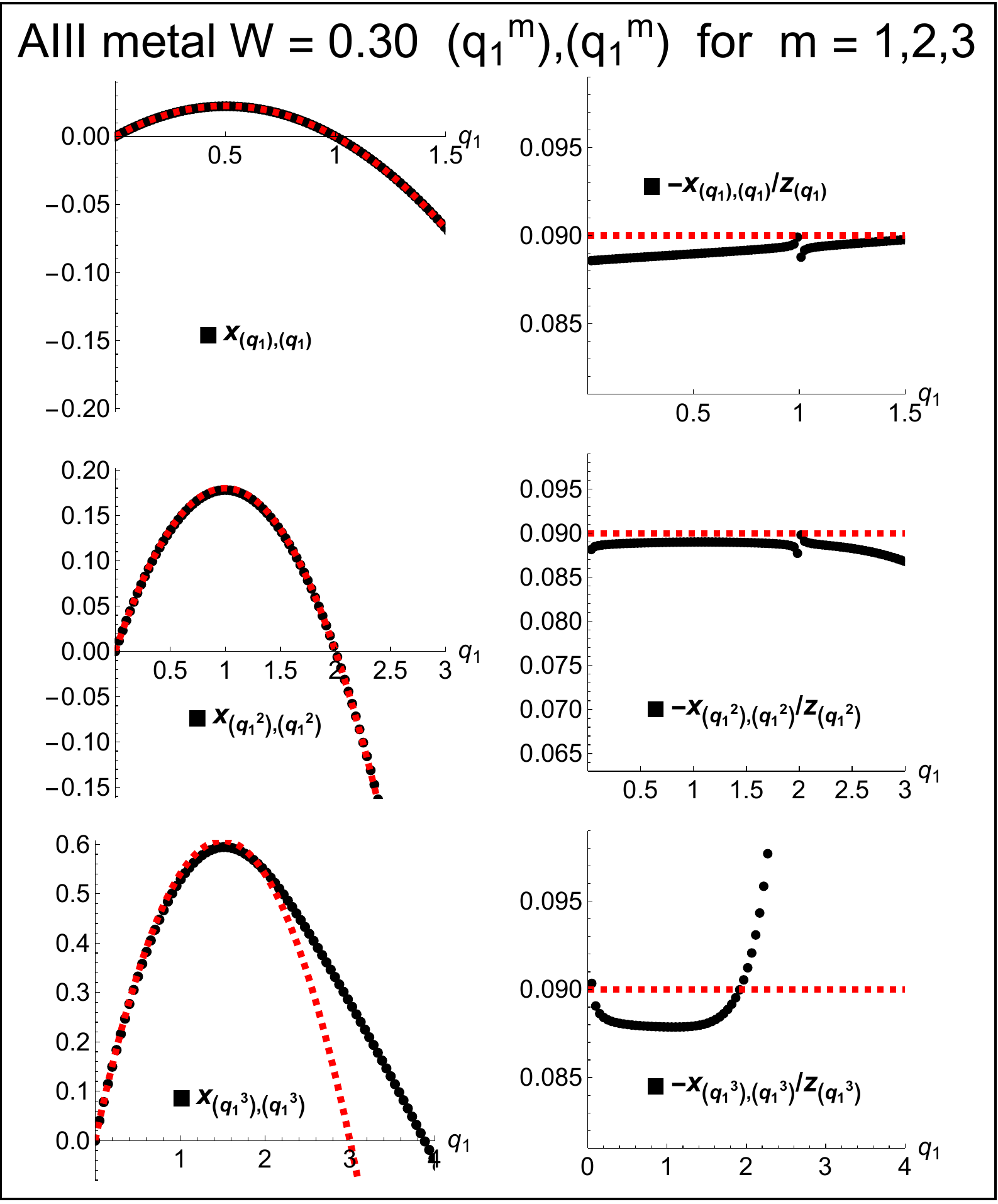}
	\caption{Generalized multifractality in the metallic phase: disorder $W=0.3$, no staggering ($\delta=0$).  Left panels: exponents $x_{(q_1^m),(q_1^m)}$ for  $m=1$ (top), $m=2$ (middle), and $m=3$ (bottom). Black symbols are numerical data; red dotted lines represent parabolic approximation $x_{\lambda,\lambda} = -2bz_\lambda$ with  $b=0.045$. The parabolicity holds with a high accuracy as expected for week disorder (i.e., large conductivity, $\sigma = 1/b \simeq 22$). Deviations at large $q_1$ in the bottom panel are due to insufficient averaging. Right panels: same data shown as $- x_{(q_1^m),(q_1^m)} / z_{(q_1^m)}$.
	}
	\label{fig:scaling_aiii2}
\end{figure}

In Fig.~\ref{fig:scaling_aiii2}, the generalized parabolicity at weak disorder ($W=0.3$) is probed in a different way. We show there the balanced exponents $x_{(q_1^m), (q_1^m)}$ for $m=1$, $m=2$, and $m=3$. The parabolic approximation for them (see Eqs.~\eqref{x-one-loop} and~\eqref{z-Laplacian}) reads
\begin{align}
	x_{(q_1^m), (q_1^m)} \simeq -2b z_{(q_1^m)} \equiv 2 m b q_1(m-q_1). 
\end{align}
This approximation holds with a very good accuracy as expected. Strong deviations of numerical data from a parabola that are observed for the $m=3$ case for $q > 2$ are explained by the fact that the ensemble averaging becomes insufficient for high-order correlators.

%%%%%%%%%%%%%%%%%%%
%%%%%%%%%%%%%%%%%%

%%%%%%%%%%%%%%%%%%%%
\begin{table*}
	\centering
	\begin{tabular}{cc|cccc|cccc}
		& rep. $\lambda$ & & $x_{\lambda,\lambda}^{\rm num}/b_W$ & & & & $(x_{\lambda,(0)}^{\rm num} - |\lambda|^2 x_{\nu})/b_W$ & &\\
		& & $W=0.30$ & $W=1.00$ & $W=3.00$ & $-2z_\lambda$ &$W=0.30$ & $W=1.00$ & $W=3.00$ & $-z_\lambda$\\
		\hline
		\hline
		&(2) & $-4.046\pm 0.030$ & $-3.985\pm 0.028$ & $-4.589\pm 0.100$ & $-4$ & \
		$-1.990\pm 0.006$ & $-1.951\pm 0.007$ & $-1.601\pm 0.015$ & $-2$\\[3pt]
		& $(1^2)$
		%(1, 1) 
		& $4.017\pm 0.011$ & $3.877\pm 0.014$ & $3.990\pm 0.023$ & $4$ &\
		$1.979\pm 0.006$ & $1.940\pm 0.006$ & $2.307\pm 0.014$ & $2$\\[3pt]
		&&&&&&\\[-5pt]
		\hline
		&(3) & $-12.280\pm 0.164$ & $-12.190\pm 0.233$ & $-11.490\pm 0.374$ & $-12$ & \
		$-5.995\pm 0.022$ & $-5.826\pm 0.029$ & $-3.161\pm 0.080$ & $-6$\\[3pt]
		&(2, 1) & $0.040\pm 0.046$ & $-0.083\pm 0.054$ & $0.097\pm 0.180$ & $0$ & \
		$-0.005\pm 0.014$ & $0.024\pm 0.020$ & $1.833\pm 0.105$ & $0$\\[3pt]
		& $(1^3)$
		%(1, 1, 1) 
		& $11.740\pm 0.019$ & $11.540\pm 0.021$ & $11.320\pm \
		0.081$ & $12$ & $5.886\pm 0.015$ & $5.833\pm 0.016$ & $7.582\pm 0.065$ & $6$\
		\\[3pt]
		&&&&&&\\[-5pt]
		\hline
		&(4) & $-24.510\pm 0.787$ & $-23.480\pm 0.848$ & $-18.520\pm 0.649$ & $-24$ & \
		$-12.050\pm 0.056$ & $-11.440\pm 0.120$ & $-2.990\pm 0.175$ & $-12$ \\[3pt]
		&(3, 1) & $-8.194\pm 0.261$ & $-8.019\pm 0.295$ & $-5.831\pm 0.567$ & $-8$ & \
		$-4.000\pm 0.030$ & $-3.715\pm 0.065$ & $2.458\pm 0.294$ & $-4$\\[3pt]
		& $(2^2)$
		%(2, 2) 
		& $0.204\pm 0.106$ & $-0.256\pm 0.226$ & $1.358\pm 0.276$ & $0$ & \
		$0.032\pm 0.027$ & $0.071\pm 0.062$ & $5.170\pm 0.282$ & $0$\\[3pt]
		& $(2,1^2)$
		%(2, 1, 1) 
		& $7.752\pm 0.059$ & $7.634\pm 0.107$ & $7.422\pm 0.289$ & $8$ & \
		$3.927\pm 0.027$ & $4.025\pm 0.037$ & $9.180\pm 0.182$ & $4$ \\[3pt]
		& $(1^4)$
		%(1, 1, 1, 1) 
		& $23.940\pm 0.032$ & $23.050\pm 0.036$ & $21.830\pm \
		0.181$ & $24$ & $12.040\pm 0.027$ & $11.770\pm 0.031$ & $16.440\pm 0.201$ & $12$\
		\\[3pt]
		&&&&&&\\[-5pt]
		\hline
		\hline
	\end{tabular}
	\caption{Numerically determined scaling exponents of generalized multifractality for polynomial eigenstate observables with $q\equiv |\lambda|\leq 4$ in the metallic phase (disorder $W=0.3$, 1.0, and 3.0; no staggering). In the left part of the table, balanced exponents $x_{\lambda,\lambda}$ are presented. The right half of the table reports one-sublattice exponents $x_{\lambda,(0)}$. To translate eigenfunction exponents $\Delta_{\lambda,\lambda'}$ into field-theoretical operator dimensions $x_{\lambda,\lambda'}$, the values  $x_\nu^{W=0.30} = -0.008$, $x_\nu^{W=1.00} = -0.037$ and $x_\nu^{W=3.00} = -0.295$ of the density-of-states exponent were used, as obtained from the relation $\Delta_{(1),(1)} = -2x_\nu$. To emphasize the generalized parabolicity and deviations from it, the exponents are divided by $b_W$ obtained from the parabolic approximation
		to $x_{(q),(q)}$ in the range $q\in[0,2]$. The obtained values of $b_W$ are $b_{W=0.30}=0.045$, $b_{W=1.00}=0.083$ and $b_{W=3.00}=0.24$.
	}
	%	\label{tab:sC}
	\label{tab:lchiral}
\end{table*}
%%%%%%%%%%%%%%%%%%%%%%

In Table~\ref{tab:lchiral}, we collect numerical results for  scaling exponents   $x_{\lambda,\lambda}$ and
$x_{\lambda,(0)}$ corresponding to polynomial observables with $q\equiv |\lambda|\leq 4$ in the metallic phase. Specifically, we consider three values of disorder, $W=0.3$
(as in Fig.~\ref{fig:scaling_aiii}), $W=1.0$, and $W=3.0$, with no staggering, $\delta=0$, in all the cases. We also show in the table statistical error bars (one standard deviation).
Within the generalized-parabolicity approximation (that holds in the one-loop order), we have
\begin{align}
	x_{\lambda, \lambda} &= -2bz_\lambda, 
	&
	x_{\lambda,(0)} &= -bz_\lambda + |\lambda|^2 x_\nu.
	\label{eq:num-metal-one-loop-exp}
\end{align}
To demonstrate the accuracy of the parabolic approximation and to quantify the deviations, we show in Table~\ref{tab:lchiral} the ratios $x_{\lambda, \lambda} / b_W$ and
$(x_{\lambda,(0)} - |\lambda|^2 x_\nu)/ b_W$. The first of them is equal to $-2z_\lambda$ and the second one to $-z_\lambda$ in the parabolic approximation; the corresponding values are also included in the table for convenience of comparison. The parameters $b_W$ are obtained by fitting the exponents $x_{(q),(q)}$ to the corresponding parabolic approximation $2bq(1-q)$ within the range $q\in[0,2]$; the results are 
\begin{align}
	b_{W=0.30} &= 0.045, 
	&
	b_{W=1.00} &= 0.083, 
	&
	b_{W=3.00} = 0.24. 
\end{align}
By inspecting the table, we see that, whereas deviations from the generalized parabolicity are very small for weak disorder ($W=0.3$), they become progressively stronger when the disorder increases, as expected.

As we have shown analytically, two of the exponents presented in the table---$x_{(2,1),(2,1)}$ and $x_{(2^2),(2^2)}$---should be in fact exactly zero. We observe that the Weyl-sym\-met\-ry relations $x_{(2,1),(2,1)}=0$ and $x_{(2^2),(2^2)}=0$ indeed perfectly hold for our numerical values when statistical error bars (which are rather small) are taken into account. The only exception is the exponent $x_{(2^2),(2^2)}$ for the strongest disorder $W=3.0$, in which case the deviation from zero is five times larger than the statistical standard deviation. These indicates some systematic errors in this case: presumably, the averaging is not fully sufficient, or finite-size corrections to scaling intervene, or both. We note that the exponent $x_{(2^2),(2^2)}$ corresponds to an observable which is of the 16-th order in wave-function amplitudes, so that emergence of such a deviation (still quite small) in the case of a rather strong disorder is not so surprising.

It is worth recalling that the sigma-model coupling $\kappa$ flows to infinity in the metallic phase, see Sec.~\ref{sec:sigma}. This implies a flow to an infinite-randomness fixed point, with $x_\nu \to - \infty$ (and, correspondingly, $\alpha_\nu \to -1$) for the density-of-states exponent. The values of $x_\nu$ that we find numerically are, however, rather small: 
\begin{align}
	x_\nu^{W=0.30} &= -0.008, 
	\nonumber \\
	x_\nu^{W=1.00} &= -0.037, 
	\nonumber \\
	x_\nu^{W=3.00} &= -0.295. 
\end{align}
The explanation for this was discussed in Sec.~\ref{sec:sigma}:  reaching the ultimate infrared behavior in our model requires astronomically large system sizes $L$. The exponents $x_\nu$ that we find correspond to the accessible range of $L$; they are in fact slowly changing with $L$. This applies also to most of the generalized-multifractality exponents. A notable exception is provided by the class of balanced exponents $x_{\lambda,\lambda}$, which are not influenced by the renormalization of $x_\nu$ [see Eq.~\eqref{eq:num-metal-one-loop-exp}] and thus should stay finite in the limit $L \to \infty$.

\section{Numerical study of generalized multifractality:  Metal-insulator transition}
\label{sec:mit}

We turn now to a numerical investigation of the generalized multifractality at the metal-insulator transition, which is carried out analogously to the study of the metallic phase in Sec.~\ref{sec:metal}. We perform this numerical analysis at five points $(W, \delta_c(W))$ on the critical line in the plane $(W, \delta)$ with $W= 0.3$, 0.5, 1.0, 2.0, and 3.0.  The critical values of staggering $\delta_c(W)$ at which the transition occurs were determined in Ref.~\cite{karcher2022metal}, where transport properties and conventional multifractality of our model \eqref{eq:ham_aiii} were studied. They are 
\begin{align}
	\delta_c(0.3) &= 1.64, 
	&
	\delta_c(0.5) &= 1.22, 
	&
	\delta_c(1.0) &= 0.73, 
	\nonumber \\
	\delta_c(2.0) &= 0.33,
	&
	\delta_c(3.0) &= 0.22.
\end{align}

\begin{figure*}
	\centering
	\includegraphics[width=0.4075\textwidth]{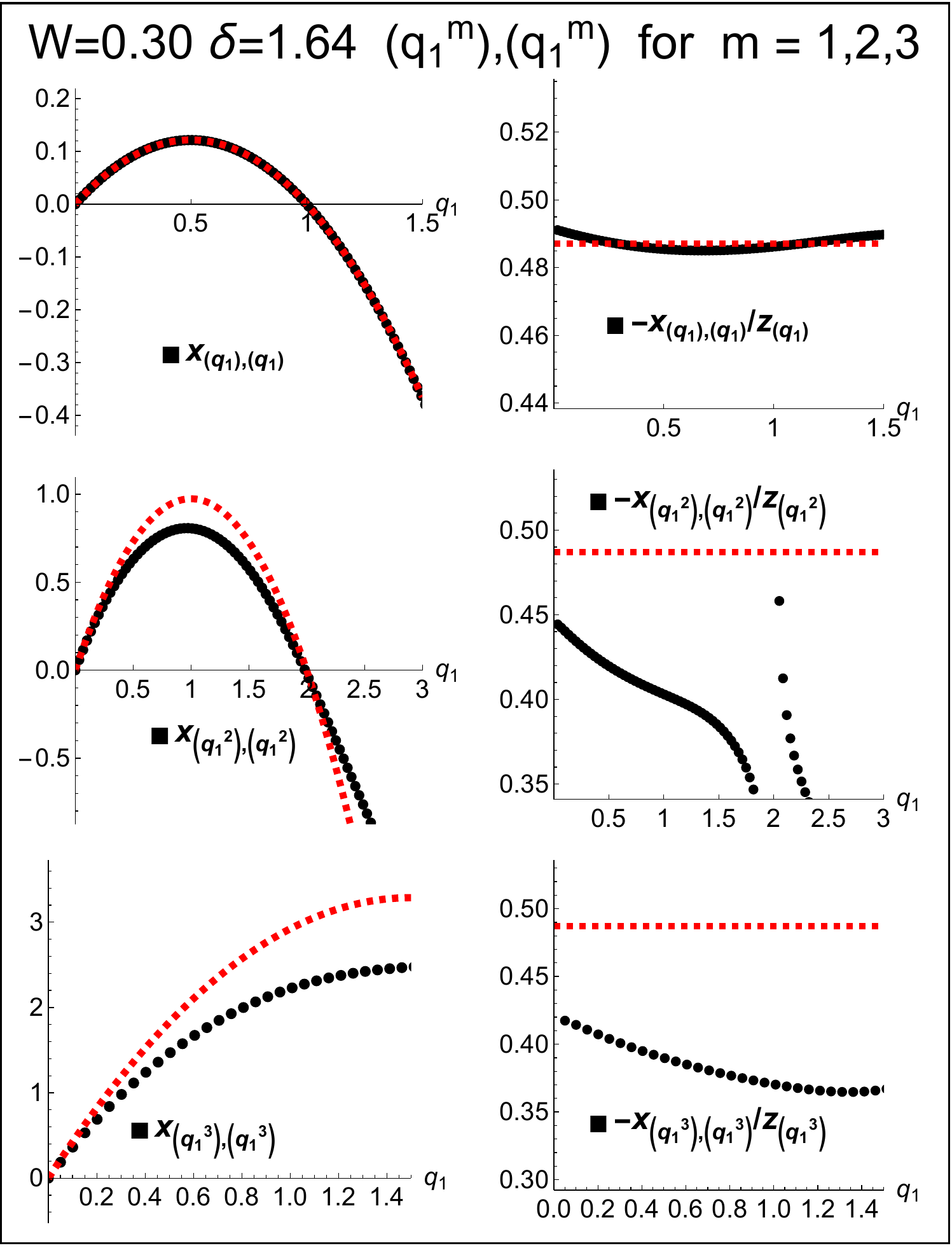}
	\includegraphics[width=0.3925\textwidth]{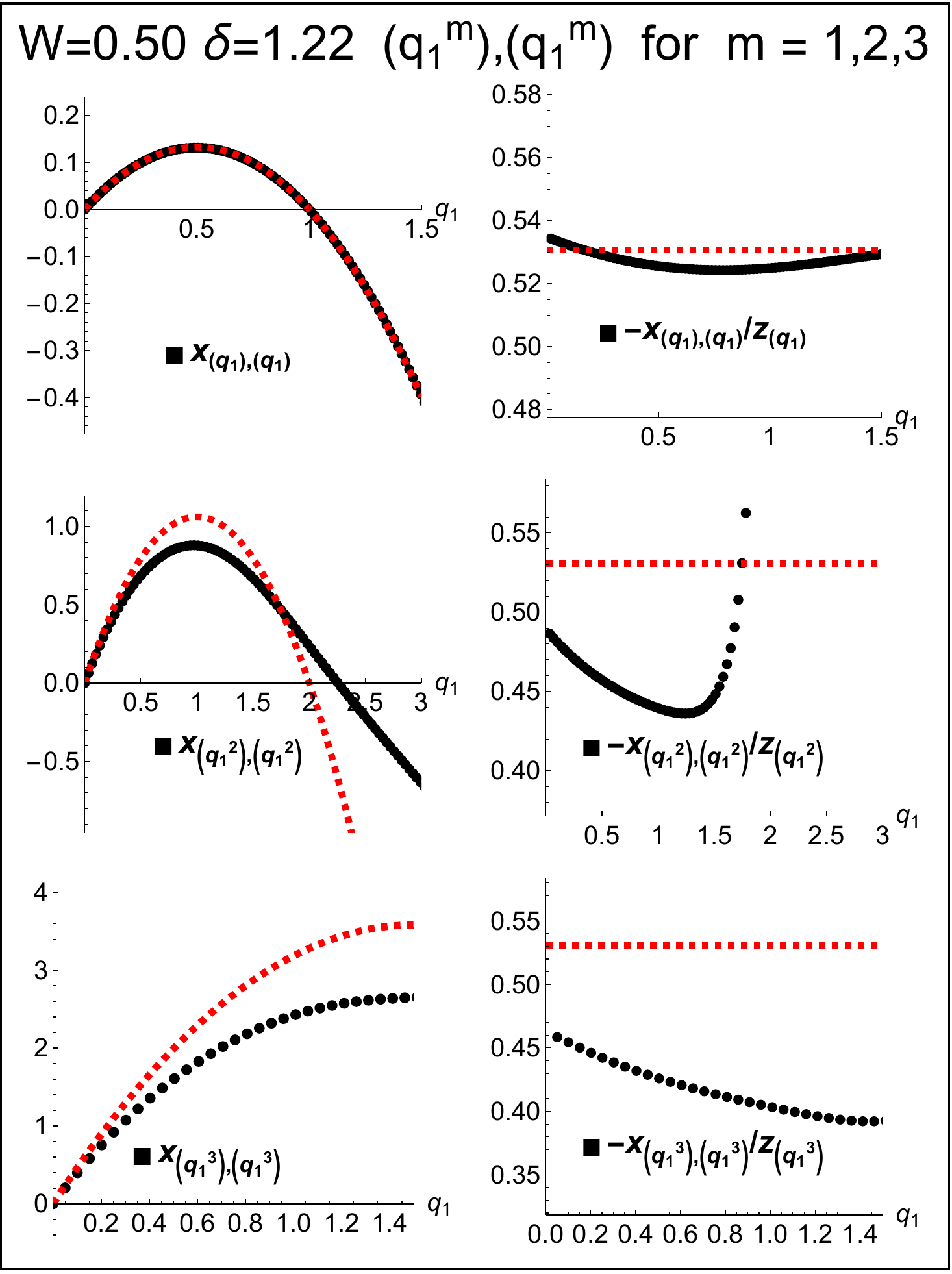}
	\caption{Generalized multifractality at the metal-insulator transition: disorder $W=0.3$ and staggering $\delta = \delta_c(0.3) = 1.64$ (left side),  and $W=0.5$,  $\delta = \delta_c(0.5) = 1.22$ (right side). For each of the disorder values, left panels present exponents $x_{(q_1^m),(q_1^m)}$ for  $m=1$ (top), $m=2$ (middle), and $m=3$ (bottom), whereas right panels show the same data in the form $- x_{(q_1^m),(q_1^m)} / z_{(q_1^m)}$, as in Fig.~\ref{fig:scaling_aiii2}.
		Black symbols are numerical data; red dotted lines represent parabolic approximation $x_{\lambda,\lambda} = -2bz_\lambda$ with $b = 0.24$ for $W=0.3$ and $b = 0.27$ for $W=0.5$. 
	}
	\label{fig:aiii-mit}
\end{figure*}

We focus here on balanced observables $P_{\lambda,\lambda}$, which allows us to use the Weyl symmetry as an indicator of the accuracy of the numerics. In Fig.~\ref{fig:aiii-mit}, we present the behavior of the exponents $x_{(q_1^m), (q_1^m)}$ as functions of $q_1$ for $W=0.3$ and $W=0.5$. For each of these disorder values, the presentation of the data is fully analogous to that in Fig.~\ref{fig:scaling_aiii2}: the left panels show $x_{(q_1^m), (q_1^m)}$ for $m=1$, 2, and 3, and the right panels show the ratio $- x_{(q_1^m), (q_1^m)}/ z_{(q_1^m)}$. The red dashed lines show the parabolic approximation, $x_{\lambda,\lambda} = - 2b z_\lambda$. For the $x_{(q_1), (q_1)}$ exponent, the parabolicity holds with a good accuracy. In fact, there are deviations, as the panels $- x_{(q_1), (q_1)} / z_{(1)}$ show, but they are quite small, on the level of a few percent. On the other hand, when one considers the results for $x_{(q_1^2), (q_1^2)}$ and $x_{(q_1^3), (q_1^3)}$, strong deviations from the generalized parabolicity become obvious.  We also note that the data are consistent with the Weyl symmetry, which implies that the curves $x_{(q_1^m), (q_1^m)}$ possess a symmetry $q_1 \longleftrightarrow m - q_1$ and thus have a maximum at $q_1 = m/2$.

%%%%%%%%%%%%%%%%
%%%%%%%%%%%%%%%

In Table \ref{tab:mit}, we show the critical exponents $x_{\lambda, \lambda}$ at the metal-insulator transition, $\delta = \delta_c(W)$, for different disorder strengths $W$.
In analogy with Table~\ref{tab:lchiral}, we present results for polynomial observables with $q \equiv |\lambda| = 2$, 3, and 4, and indicate statistical errors (which grow with $q$ for obvious reasons).

Let us first consider the values of the exponents with $q =2$ and 3  (i.e., the first five lines of the table). A special role is played here by the exponent $x_{(2,1),(2,1)}$, since the Weyl symmetry implies an exact equality $x_{(2,1),(2,1)}=0$. This equality indeed holds to an excellent precision (well within the small statistical error bars) for all values of $W$, thus providing a clear confirmation of a high accuracy of our numerics. Another important observation is a strong violation of the generalized parabolicity, in full agrement with the above discussion of  Fig.~\ref{fig:aiii-mit}. This is evident already after an inspection of values of $q=2$ exponents, which have a particularly high accuracy. Indeed, for a spectrum satisfying generalized parabolicity, we would have $x_{(2),(2)} = - x_{(1^2),(1^2)}$. It is immediately seen from the two upper lines of the table that this equality (and thus the generalized parabolicity) is strongly violated.

One more striking result is an apparent non-universality of the critical behavior: the exponents $x_{\lambda,\lambda}$ change when we move along the critical line in the phase diagram. This corroborates results of Ref.~\cite{karcher2022metal} on (the apparent) non-universality of other observables characterizing the transition. The reason for this was discussed in Ref.~\cite{karcher2022metal}: the sigma-model RG equations of Ref.~\cite{koenig2012metal-insulator} predict a very slow flow along the critical  line towards the ultimate fixed point (which is expected to have a zero conductivity and very strong multifractality). It is worth recalling that a somewhat similar situation of an apparent non-universality because of a slow RG flow towards an infrared fixed point has been discussed above in the context of the metallic phase.

We proceed now to a discussion of the $q=4$ part of Table \ref{tab:mit}. Here, there is also an exponent that should be exactly zero due to Weyl symmetry: $x_{(2^2),(2^2)}=0$. At variance with the perfectly fulfilled Weyl-symmetry relation $x_{(2,1),(2,1)}=0$, we observe substantial deviations of $x_{(2^2),(2^2)}$ from zero (several times larger than statistical error) for $W=0.5$ and stronger disorder. This shows an importance of systematic errors (due to insufficient averaging and/or finite-size corrections) and provides an estimate of the magnitude of such errors for $q=4$ exponents.  Note that these errors are still quite small in comparison with a typical magnitude of $q=4$ exponents. A similar effect appeared for our strongest disorder ($W=3.0$) in the metallic phase, see Sec.~\ref{sec:metal}. As discussed there, the emergence of such deviations for $q=4$ observables (involving observables that are of the 16-th order in wave-function amplitudes) does not come as a big surprise.

%%%%%%%%%%%%%%%%%%
\begin{table*}
	\centering
	\begin{tabular}{cc|cccccc}
		& rep. $\lambda$ & & & $x_{\lambda,\lambda}^{\rm num}$ & & & \\
		&  & $W=0.3$ & $W=0.5$ & $W=1.0$ & $W=2.0$ & $W=3.0$ &\\[3pt]
		\hline
		\hline
		&(2) & $-0.972\pm 0.010$ & $-1.054\pm 0.017$ & $-1.515\pm 0.033$ & \
		$-1.628\pm 0.045$ & $-1.657\pm 0.082$ & \\[3pt]
		& $(1^2)$
		%(1, 1) 
		& $0.813\pm 0.003$ & $0.883\pm 0.003$ & $1.165\pm 0.006$ & \
		$1.247\pm 0.008$ & $1.289\pm 0.009$ & \\[3pt]
		&&&&&&&\\[-5pt]
		\hline
		&(3) & $-2.590\pm 0.047$ & $-2.780\pm 0.070$ & $-3.671\pm 0.103$ & \
		$-3.786\pm 0.118$ & $-3.602\pm 0.187$ & \\[3pt]
		&(2, 1) & $-0.021\pm 0.027$ & $0.001\pm 0.022$ & $0.014\pm 0.048$ & \
		$0.002\pm 0.071$ & $-0.024\pm 0.070$ & \\[3pt]
		& $(1^3)$
		%(1, 1, 1) 
		& $2.223\pm 0.007$ & $2.423\pm 0.007$ & $3.107\pm 0.021$ & \
		$3.347\pm 0.025$ & $3.526\pm 0.031$ & \\[3pt]
		&&&&&&&\\[-5pt]
		\hline
		&(4) & $-4.277\pm 0.099$ & $-4.553\pm 0.123$ & $-5.782\pm 0.175$ & \
		$-5.888\pm 0.184$ & $-5.472\pm 0.277$ & \\[3pt]
		&(3, 1) & $-1.464\pm 0.078$ & $-1.486\pm 0.064$ & $-1.774\pm 0.120$ & \
		$-1.788\pm 0.164$ & $-1.909\pm 0.177$ & \\[3pt]
		& $(2^2)$
		%(2, 2) 
		& $0.048\pm 0.057$ & $0.202\pm 0.036$ & $0.313\pm 0.089$ & \
		$0.485\pm 0.077$ & $0.345\pm 0.124$ & \\[3pt]
		& $(2,1^2)$
		%(2, 1, 1) 
		& $1.468\pm 0.022$ & $1.663\pm 0.032$ & $2.196\pm 0.070$ & \
		$2.564\pm 0.051$ & $2.577\pm 0.069$ & \\[3pt]
		& $(1^4)$
		%(1, 1, 1, 1) 
		& $4.195\pm 0.013$ & $4.564\pm 0.012$ & $5.796\pm \
		0.044$ & $6.337\pm 0.078$ & $6.817\pm 0.084$ & \\[3pt]
		&&&&&&&\\[-5pt]
		\hline
	\end{tabular}
	\caption{Numerically determined scaling exponents of generalized multifractality $x_{\lambda,\lambda}$ for balanced polynomial observables $P_{\lambda,\lambda}$ with $q\equiv |\lambda|\leq 4$ on the metal-insulator-transition line $(W, \delta_c(W))$. The disorder values are $W= 0.3$, 0.5, 1.0, 2.0, and 3.0.   For each disorder, the staggering $\delta$ is fixed to the corresponding critical value, $\delta_c(W)$ given by $\delta_c(0.3) = 1.64$, $\delta_c(0.5) = 1.22$, $\delta_c(1.0) = 0.73$, $\delta_c(2.0) = 0.33$, and $\delta_c(3.0) = 0.22$.
		Statistical errors are indicated.
	}
	\label{tab:mit}
\end{table*}
%%%%%%%%%%%%%%%%%%%%%%%

\section{Summary and outlook}
\label{sec:summary}

In this paper, we have explored generalized multifractality in systems of chiral symmetry classes. On the analytical side, we have studied all three chiral classes: chiral unitary AIII, chiral orthogonal BDI, and chiral symplectic CII. We have supported and complemented the analytical results by numerical simulation of a class-AIII model on a bipartite square lattice. These simulations were performed for several points $(W, \delta_c(W))$ on the critical line of metal-insulator transition in the parameter plane $(W,\delta)$ spanned by disorder $W$ and staggering $\delta$.  Furthermore, we carried out the numerical analysis for several points $(W,0)$ in the metallic phase. Our main results are as follows:

\begin{itemize}
	
	\item We have developed a construction of pure-scaling observables of generalized multifractality, both in the field-theory (sigma-model) language and in terms of Hamiltonian eigenfunctions, and verified it numerically. While it is largely analogous to the corresponding construction for other symmetry classes~\cite{gruzberg2013classification,karcher2021generalized, karcher2022generalized, karcher2022generalized-2}, there is one major difference. Specifically, for chiral classes, the observables are labeled by two multi-indices (weights), $\lambda = (q_1, \ldots, q_m)$ and $\lambda' = (q'_1, \ldots, q'_{m'})$,  corresponding to two sublattices.
	
	\item We have explored the impact of the Weyl symmetry for the generalized-multifractality exponents $x_{\lambda,\lambda'}$ in chiral classes. At variance with other symmetry classes, the Weyl group for the chiral classes contains only permutations but not reflections. This substantially reduces the impact of the Weyl symmetry if one considers exponents $x_{\lambda,(0)}$  corresponding to one-sublattice observables. At the same time, we show that, for generic observables, the allowed symmetry transformations include also permutations between the sublattices. This implies, in particular, that exponents for balanced ($\lambda = \lambda'$) observables satisfy symmetries $x_{\lambda, \lambda} = x_{w(\lambda), w(\lambda)}$, with $w$ belonging to the conventional  Weyl group (like, e.g., in class A), including both permutations and reflections.
	
	\item Deeply in the metallic phase (large conductivity, $\sigma \gg 1$), the sigma-model RG predicts the generalized-multifractality spectrum satisfying the generalized parabolicity (up to corrections of order $\sigma^{-4}$). At variance with other symmetry classes, such a generalized parabolic spectrum in chiral classes is parametrized by two (rather than by one) parameters. The second parameter is associated with an additional U(1) degree of freedom, which is a peculiarity of the sigma models for the chiral classes. Our numerical results for the metallic phase fully confirm these predictions of the generalized parabolicity for weak disorder. At the same time, with increasing disorder (i.e., for smaller $\sigma$), we observe clear deviations from the generalized parabolicity, in consistency with analytical expectations.  On the other hand, the Weyl symmetry holds for balanced observables even for stronger disorder---again in agreement with analytical predictions.
	
	\item Numerical results for balanced exponents $x_{\lambda, \lambda}$ at the metal-insulator transition also exhibit the Weyl symmetry. At the same time, we find that the generalized parabolicity is strongly violated. This implies, according to Ref.~\cite{karcher2021generalized}, that the local conformal invariance is violated (at least partly) at these 2D transitions, in analogy with several 2D Anderson-localization critical points studied earlier~\cite{karcher2021generalized, karcher2022generalized, karcher2022generalized-2}.
	
	The numerically determined exponents $x_{\lambda, \lambda}$ exhibit an apparent non-universality along the critical line. This is consistent with the analytical prediction of a very slow RG flow along the critical line towards the ultimate fixed point~\cite{koenig2012metal-insulator, karcher2022metal}, and with numerical results on the apparent non-universality of other observables at criticality in the same model~\cite{karcher2022metal}.
	
\end{itemize}

Before closing the paper, we briefly discuss some prospects for future investigations of chiral-class systems. The analytical results derived here allow one to extend the numerical analysis of the generalized multifractality in chiral classes (performed in this work for a 2D class-AIII  model on a square lattice) in several directions. First, numerical investigations of different models that would provide access to the ultimate infrared critical behavior is of much interest. Second, we foresee an extension of our numerical studies to models of the other two chiral classes (BDI and CII). Third, investigations of the generalized multifractality at chiral Anderson transitions in three dimensions would be important. (Such transitions were recently discussed in Refs.~\cite{luo2020critical, wang2021universality}.) Fourth, the generalized multifractality may serve as a sensitive tool of ``stacked criticality'' that is conjectured to emerge on surfaces of class-AIII topological superconductors~\cite{sbierski2020spectrum-wide}. Fifth, we envision that our results can be extended to non-Hermitian Anderson transitions that currently attract much interest~\cite{hatano1996localization, kawabata2021nonunitary, luo2021universality, luo2022unifying} and are closely related~\cite{feinberg1997non, gong2018topological, luo2022unifying} to models of chiral symmetry classes. Sixth, the status of conformal invariance and its violation requires a better understanding. In this work, we have demonstrated a strong violation of the generalized parabolicity and thus of local conformal invariance. At the same time, we found in Ref.~\cite{karcher2022metal} that the invariance with respect to a particular global conformal transformation---the exponential map~\cite{Obuse-Conformal-2010}---holds within the numerical accuracy. It remains to be understood whether Anderson transitions possess a partial conformal symmetry. The situation in this respect is similar to that for 2D critical points in other disordered systems. Finally, one may attempt an extension of our results (both analytical and numerical) to interacting models; some analytical results in this direction are available for other symmetry classes~\cite{burmistrov2016mesoscopic, babkin2022generalized}.

\section*{Acknowledgement}
J.F.K. and A.D.M. acknowledge support by the Deutsche Forschungsgemeinschaft (DFG) via the grant MI 658/14-1.

\pagebreak
\appendix

\section{RG for sigma models of chiral symmetry classes and one-loop renormalization of composite operators}
\label{appendix-rg}

Target manifolds $G/K$ for the fermionic-replica sigma models of chiral classes are
$\mathrm{U}(n)$ for the class AIII, $\mathrm{U}(n)/\mathrm{O}(n)$ for the class CII, and $\mathrm{U}(n)/\mathrm{Sp}(n)$ with even $n$ for the class BDI , see
Table \ref{table:sigma-models}. Thus, in class AIII (i.e., in the absence of time-reversal symmetry), the sigma-model field $U$ is a generic unitary matrix.
In classes CII and BDI, there are additional constraints due to time-reversal symmetry. Specifically,  in class CII the unitary matrix $U$ is symmetric, $\bar{U} \equiv U^T=U$, while for class BDI, the matrix field $U$ satisfies $\bar{U} \equiv CU^TC=U$ with $C=\sigma_y$ acting in time-reversal space. (In the latter case, the size of the matrix is doubled to incorporate the time-reversal symmetry.)

The action is given by Eq. \eqref{eq:gade-action} of the main text,
\begin{eqnarray}
	S[U] &=& - \! \int \! d^2 r \!  \left[ \frac{\sigma}{8\pi s} \text{Tr} (U^{-1} \nabla U)^2 + \frac{\kappa}{8\pi s} (\text{Tr} U^{-1} \nabla U)^2 \right.
	\nonumber \\
	& + & \left.  i \frac{\pi \rho_0}{2s} \, \varepsilon \, \text{Tr} (U + U^{-1}) \right ],
	\label{eq:ac}
\end{eqnarray}
%\begin{align}
%S &= -\int \dfrac{d^2r}{8\pi s}\left[\sigma \;\mathrm{tr} (U^{-1}\nabla U U^{-1}\nabla U) \right. \nonumber \\
%& \left. + c\; \mathrm{tr} (U^{-1} \nabla U) \mathrm{tr} (U^{-1} \nabla U) \right] +i \frac{\pi \rho_0}{2s} \int d^2r \mathrm{tr}\left( U + U^{-1} \right)
%\end{align}
where $s=1$ for the class AIII and $s=2$ for the classes BDI and CII, $\sigma$ is the conductivity in units of $e^2/ \pi h$,  $\rho_0$ is the bare density of states, and $\varepsilon$ is a running coupling whose bare value is the energy $E$.

\subsection{Background-field RG formalism for the chiral classes}
\label{sec:background-field}

We begin by reviewing the RG formalism and the renormalization of coupling constants of the sigma-model action, Eq.~\eqref{eq:ac}. In this part of the presentation, we follow
Ref.~\cite{koenig2012metal-insulator}. After this, in Sec.~\ref{sec:renorm-comp-op}, we extend the one-loop RG analysis to composite operators describing the generalized multifractality.

Following Ref.~\cite{koenig2012metal-insulator}, we parametrize the sigma-model field as
\begin{align}
	U=\bar{V} \tilde{U} V \,, \label{eq:decomp}
\end{align}
where $\bar{V}$ and $V$ are slow fields whereas $\tilde{U}$ is a fast field. In class AIII, the slow fields $\bar{V}$ and $V$ are independent unitaries. In classes BDI or CII, $\bar{V}$ and $V$ are related by the respective time-reversal operation
\begin{align}
	\begin{cases}
		\bar{V}= V^T, & \mathrm{CII} ,\\
		\bar{V} = C V^T C, & \mathrm{BDI},
	\end{cases}
\end{align}
as the notation suggests.

\subsubsection{Fast-mode action}

For all three chiral classes, the $D$-dimensional sigma-model target space is spanned by $D-1$ traceless hermitean generators $T^a$ and the unit matrix $\mathds{1}$. In class AIII,  the target space is the full unitary group $U(n)$, and its dimension is $D=n^2$. In the class CII, the time-reversal symmetry constraint eliminates the $U(n)$ generators that are antisymmetric, yielding the dimension $D=\frac12 n(n+1)$, whereas  in BDI  the symmetry constraint leaves $D=\frac12 n(n-1)$ degrees of freedom.

The fast field $\tilde{U}$ can be parametrized as
\begin{align}
	\tilde{U} &= e^{i W}, & W &= n^{-\frac12}w_0 \mathds{1} + W_1\,, & W_1 = \sum_a w_a  T^a.
	\label{eq:tilde-U-param}
\end{align}
Here, $w_0$ parametrizes the diagonal $U(1)$ subgroup, while $T^a$ are traceless generators fulfilling $\mathrm{Tr} (T^a T^b) = \delta^{ab}$.
Taken together, $T^a$ and $n^{-1/2} \mathds{1}$ form an orthonormal basis in the tangent space to the target manifold.

As a consequence of the tracelessness of the generators $T^a$, a term singular in the replica limit appears in their completeness identity:
\begin{align}
	\sum_a T^a_{ij} T^a_{kl} &=-n^{-1}\delta_{ij}\delta_{kl} +
	\begin{cases}
		\delta_{il}\delta_{jk}, & \mathrm{AIII},\\
		\frac12(\delta_{il}\delta_{jk} + \delta_{ik}\delta_{jl}), & \mathrm{CII},\\
		\frac12(\delta_{il}\delta_{jk} + C_{ik}C_{jl}), & \mathrm{BDI}.
	\end{cases} \label{eq:div1}
\end{align}
As we will see, the divergent terms cancel with terms from $w_0 n^{-\frac12}$ contractions when we consider the RG of physical operators.

With this parametrization, the gaussian fast-field action  becomes (discarding the energy term):
\begin{align}
	&S_L[W] = \int \dfrac{d^2r}{8\pi s} \left \{ (\sigma_0 + n\kappa_0)(\nabla w_0)^2+ \sigma_0L^{-2}(w_0)^2  \vphantom{\sum_a} \right. \nonumber \\
	& \left. + \sigma_0\sum_a\left[ (\nabla w_a)^2 + L^{-2}(w_a)^2\right] \right\}.
\end{align}
Here $L$ is the running RG scale and a mass $\sim 1/L$ ensures that $W$ contains only fast modes \cite{koenig2012metal-insulator}.
The corresponding propagators are:
\begin{align}
	\langle w_0(q) w_0(-q) \rangle_f  &= \dfrac{4\pi s}{(\sigma_0+n\kappa_0)q^2+\sigma_0L^{-2}},\\
	\langle w_a(q) w_a(-q) \rangle_f  &= \dfrac{4\pi s}{\sigma_0q^2+\sigma_0L^{-2}}.
\end{align}
This leads to the one-loop integrals:
\begin{align}
	&I_f^{(0)} = \sum_q \langle w_0(q) w_0(-q) \rangle_f\nonumber\\&= \int ^{a^{-1}}\dfrac{d^2q}{(2\pi)^2} \dfrac{4\pi s}{(\sigma_0 + n \kappa_0)q^2+\sigma_0 L^{-2}}
	\nonumber \\
	&  = \dfrac{2s}{\sigma_0 + n \kappa_0}\ln\left(\dfrac{L}{a}\right),
	\label{eq:If0}
	\\[0.3cm]
	&I_f^{(a)} = \sum_q \langle w_a(q) w_a(-q) \rangle_f \nonumber\\ &= \int ^{a^{-1}}\dfrac{d^2q}{(2\pi)^2} \dfrac{4\pi s}{\sigma_0q^2+\sigma_0 L^{-2}} = \dfrac{2s}{\sigma_0}\ln\left(\dfrac{L}{a}\right),
	\label{eq:Ifa}
\end{align}
where $a^{-1}$ is the ultraviolet cutoff in the momentum space.
In the $w_0$ contraction, this fast mode integral comes generically with a factor $n^{-1}$ [due to a factor $n^{-1/2}$ in Eq.~\eqref{eq:tilde-U-param}]:
\begin{align}
	n^{-1}\sum_q\langle w_0(q) w_0(-q) \rangle_f &= \dfrac{2s}{\sigma_0}\left[n^{-1} - \dfrac{\kappa_0}{\sigma_0} +\ldots \right]\ln\left(\dfrac{L}{a}\right) \label{eq:div2}.
\end{align}
The first term in square brackets on the right-hand side of Eq.~\eqref{eq:div2}
diverges in the replica limit $n \to 0$.  We expect, however, that all RG functions have a finite replica limit. Indeed, as we will see below, this singular term cancels with singular terms from $w_a$ contractions.

\subsubsection{Renormalization of $\sigma$ and $\kappa$}

As a first step in application of the RG formalism, we recall the renormalization of the coupling constants $\sigma$ and $\kappa$ as carried out in Ref.~\cite{koenig2012metal-insulator}. Upon integration over fast fields $\tilde{U}$, the effective action $S[\bar{V},V]$ depends only on the gauge-invariant combination $U_s = \bar{V} V$ of the slow fields.
The action in terms of $U_s$ takes the same form \eqref{eq:ac} but with renormalized couplings. Evaluating these renormalized couplings, one gets the RG equations \cite{koenig2012metal-insulator}
\begin{align}
	\dfrac{\partial\sigma }{\partial \ln L} &= -n + n\: O(1/\sigma), & \dfrac{\partial\kappa }{\partial \ln L} &= 1 +  O (1/\sigma).
\end{align}
In the replica limit $n \to 0$, they take the form \eqref{eq:RG-sigma}, \eqref{eq:RG-kappa}.  As also emphasized in Sec.~\ref{sec:sigma},  the absence of renormalization of $\sigma$ in the $n\to 0$ limit holds to all orders of the perturbation theory in all three chiral classes~\cite{gade1991the}. Also, $\partial \kappa / \partial \ln L = 1$ is perturbatively exact in the class AIII~\cite{guruswami2000super}. These perturbatively exact statement are violated when one takes into account vortices~\cite{koenig2012metal-insulator}.

\subsection{Renormalization of composite operators}
\label{sec:renorm-comp-op}

After having reviewed the RG formalism and the renormalization of the couplings $\sigma$ and $\kappa$, we turn to the renormalization of composite operators representing the generalized multifractality. We will perform this RG analysis within the one-loop approximation for all three chiral classes. The results of this RG analysis in this section are presented in Sec. \ref{sec:obs} of the main text.

\subsubsection{General analysis}
\label{app:obs}
%As was shown by Gade and Wegner in Refs. \cite{gade1991the} and \cite{gade1993anderson}, gradientless pure-scaling operators in chiral classes are labeled by two multi-indices $\lambda = (q_1, \ldots, q_m)$ and $\lambda' = (q'_1, \ldots, q'_{m'})$, which are highest weights of the corresponding representations. Here $\lambda$ characterizes the dependence on $U$, and $\lambda'$ the dependence on $U^\dagger \equiv U^{-1}$. For polynomial composite operators, $\lambda$ and $\lambda'$ correspond to conventional Young diagrams, with integer, positive $q_i$ and $q_i'$ being the  lengths of $i$-th row. More generally, $q_i$ and $q_i'$ may be fractional, negative, and even complex.  {\color{blue} \bf Add a reference to the subsection and appendix where the Iwasawa is discussed.}
It is well established \cite{gade1991the, gade1993anderson, dellanna2006anomalous}, that gradientless pure-scaling operators in chiral classes are labeled by two multi-indices $\lambda = (q_1, \ldots, q_m)$ and $\lambda' = (q'_1, \ldots, q'_{m'})$. Here $\lambda$ characterizes the dependence on $U$, and $\lambda'$ the dependence on $U^\dagger \equiv U^{-1}$.

%We can choose the pure-scaling composite in a product form,
%\begin{align}
%\mathcal{P}_{\lambda,\lambda'}[U] &= \mathcal{P}_{\lambda}[U] \mathcal{P}_{\lambda'}[U^{-1}] \,.
%\end{align}
%Here $ \mathcal{P}_{\lambda}[U]$ and $ \mathcal{P}_{\lambda'}[U^{-1}]$ have the same form as pure-scaling operators in the corresponding Wigner-Dyson class.
%We refer the reader to Appendix B of Ref.~\cite{karcher2022generalized-2} for a discussion of invariant pure-scaling operators in all ten symmetry classes and of relations between different classes. For the chiral orthogonal class BDI, an explicit construction of pure scaling operators can also be found in Ref. \cite{dellanna2006anomalous}.

For the RG, we choose the pure-scaling composite in a product form like Eq. \eqref{eq:ops} in Sec. \ref{sec:obs} of the main text. As a simple example for this form, consider polynomial operators of total degree $|\lambda|+|\lambda'|=2$ for class AIII. (Here $|\lambda| = \sum_i q_i $ and $|\lambda'| = \sum_i q'_i $.)   It is easy to see that there are  five such irreducible representations.  Choosing the operators invariant with respect to $U \mapsto kUk^{-1}$, we get the following pure-scaling operators:
\begin{align}
	\mathcal{P}_{(2),(0)}[U] &= \mathrm{tr}(U^2)-\mathrm{tr} (U)\mathrm{tr} (U),
	\nonumber\\
	\mathcal{P}_{(0),(2)}[U] &= \mathrm{tr}(U^{-2})-\mathrm{tr} (U^{-1})\mathrm{tr} (U^{-1}),\nonumber \\
	\mathcal{P}_{(1,1),(0)}[U] &= \mathrm{tr}(U^2)+\mathrm{tr} (U)\mathrm{tr} (U),
	\nonumber\\
	\mathcal{P}_{(0),(1,1)}[U] &= \mathrm{tr}(U^{-2})+\mathrm{tr} (U^{-1})\mathrm{tr} (U^{-1}),
	\nonumber\\
	\mathcal{P}_{(1),(1)}[U] &= \mathrm{tr} (U)\mathrm{tr} (U^{-1}).
	\label{K-radial-P}
\end{align}

In order to find scaling dimensions of the operators $\mathcal{P}_{\lambda,\lambda'}$, we employ the background field method described in Sec.~\ref{sec:background-field}.
Specifically, as detailed in Sec.~\ref{sec:background-field}, we split $U$ in fast and slow field, $U=\bar{V} \tilde{U} V$, see Eq.~\eqref{eq:decomp}, with the slow degrees of freedom $U_s=\bar{V} V$ and all the fast modes being contained in $\tilde{U} = e^{iw_0n^{-\frac12} + iW_1}$.

Scaling dimensions $x_{\lambda,\lambda'}$ of the composite operators $\mathcal{P}_{\lambda,\lambda'}$ are defined as
\begin{align}
	\dfrac{\partial }{\partial \ln L}\ln\langle \mathcal{P}_{\lambda,\lambda'}[\bar{V} \tilde{U} V] \rangle_f &= -x_{\lambda,\lambda'} \,.
	\label{eq:xqdef1}
\end{align}
At one-loop level, the dimension $x_{\lambda,\lambda'}$ by inspecting the leading logarithimic correction resulting from the renormalization:
\begin{align}
	\langle \mathcal{P}_{\lambda,\lambda'}[\bar{V} \tilde{U} V] \rangle_f &= \left[ 1 -x^{(1)}_{\lambda,\lambda'}\ln\left(\dfrac{L}{a}\right)\right] \mathcal{P}_{\lambda,\lambda'}[\bar{V} V]. \label{eq:xq1}
\end{align}
Here the superscript ``(1)'' refers to the one-loop order.
We thus need to evaluate the fast-mode averages in the left-hand side of Eq.~\eqref{eq:xq1} to find the scaling dimensions.
Let us begin with considering the contribution of the $U(1)$ part of the fast field that is proportional to the identity matrix and therefore  factors out exactly:
\begin{align}
	& \mathcal{P}_{\lambda,\lambda'}[\bar{V} \tilde{U} V]  = e^{iw_0n^{-\frac12}(|\lambda|-|\lambda'|)}
	\nonumber\\
	& \qquad \times \mathcal{P}_{\lambda}[\bar{V} e^{iW_1} V] \mathcal{P}_{\lambda'}[V^{-1} e^{- iW_1} (\bar{V})^{-1}].
\end{align}
Evaluating the average of the U(1) fast-mode factor, we get
\begin{align}
	&\langle e^{iw_0n^{-\frac12}(|\lambda|-|\lambda'|)} \rangle_f = 1 - \frac1{2n} (|\lambda|-|\lambda'|)^2 \sum_q\langle w_0(q) w_0(-q) \rangle \nonumber\\
	&= 1-(|\lambda|-|\lambda'|)^2\dfrac{s}{\sigma_0}\left[n^{-1} - \dfrac{\kappa_0}{\sigma_0} +\ldots \right]\ln\left(\dfrac{L}{a}\right) \label{eq:div3}.
\end{align}
There is a contribution here that is divergent in replica limit $n \to 0$. As was pointed out in Sec.~\ref{sec:background-field}, such contributions originating
from the divergent (at $n \to 0$) term in Eq.~\eqref{eq:div2} generically cancel with singular terms originating from the structure factors \eqref{eq:div1}.
Indeed, the divergent contribution from the contractions of the $w_a$ fields in $W_1$ reads
\begin{align}
	&-n^{-1}\left(\frac12 |\lambda|(|\lambda|-1) i^2  + |\lambda|\frac12  i^2 \right.\nonumber\\& \left.  +\frac12|\lambda'|(|\lambda'|-1) (-i)^2  +  |\lambda'| \frac12 (-i)^2 + |\lambda'||\lambda| i(-i) \right) I_f{(a)}\nonumber\\
	&= (|\lambda|-|\lambda'|)^2\dfrac{s}{\sigma_0}n^{-1}\ln\left(\dfrac{L}{a}\right),
\end{align}
which cancels exactly with the singular term in Eq. \eqref{eq:div3}.

As a consequence, we obtain for the fast mode average in the one-loop order
\begin{align}
	\langle\mathcal{P}_{\lambda,\lambda'}[\bar{V} \tilde{U} V]\rangle_f
	& = \left[\dfrac{s\kappa_0 (|\lambda|-|\lambda'|)^2}{\sigma_0^2}\ln\left(\dfrac{L}{a}\right)\mathcal{P}_{\lambda,\lambda'}[\bar{V} V]
	\right. \nonumber \\
	&  \left. + \langle \mathcal{P}_{\lambda,\lambda'}[\bar{V} e^{iW_1} V] \rangle_f^{{\rm (ns)}}\right].
	\label{eq:ren_ns}
\end{align}
%
%\onecolumngrid
%\begin{align}
%\langle\mathcal{P}_{\lambda,\lambda'}[\bar{V} \tilde{U} V,(\bar{V} \tilde{U} V)^{-1}]\rangle_f &= \left[\dfrac{s\kappa_0 (|\lambda|-|\lambda'|)^2}{\sigma_0^2}\ln\left(\dfrac{L}{a}\right)\mathcal{P}_{\lambda,\lambda'}[\bar{V} V] + \langle \mathcal{P}_{\lambda,\lambda'}[\bar{V} e^{iW_1} V,(\bar{V} e^{iW_1} V)^{-1}] \rangle_f^{(ns)}\right] \label{eq:ren_ns}.
%\end{align}
%\twocolumngrid
The superscript ``(ns)'' here indicates that the corresponding fast-mode average is understood as a non-singular part, i.e., without the singular ($1/n$) contribution of the $W_1$ contractions.

To evaluate the last term on the right-hand side of Eq.~\eqref{eq:ren_ns}, we recall that, in the one-loop order, RG eigenvalues are proportional to eigenvalues $z_\lambda$ of the Laplace operator on the corresponding symmetric space \cite{Friedan-Nonlinear-1980,Friedan-Nonlinear-1985,karcher2021generalized,karcher2022generalized-2}. This yields, in the one-loop order,
\begin{align}
	x_{\lambda,\lambda'}&= -b(z_\lambda +z_{\lambda'}) - \dfrac{s\kappa_0}{\sigma_0^2}(|\lambda|-|\lambda'|)^2 \,,
	\label{eq:x1l}
\end{align}
where $b$ is a prefactor (to be determined below) and $z_\lambda$ are Laplace-operator eigenvalues on the symmetric spaces generated by $W_1$ modes, i.e., SU(n) for class AIII, SU(n)/Sp(n) for class BDI, and SU(n)/SO(n) for class CII.  The explicit form of $z_\lambda$ is
\begin{equation}
	z_\lambda = \sum_j q_j (q_j + c_j) \,,
	\label{eq:z-lambda}
\end{equation}
where $c_j$ (with $j=1,2, \ldots$) are coefficients of the half-sum of positive roots for the corresponding symmetric space, which are given by \cite{gruzberg2013classification}
\begin{align}
	& c_j = 1 - 2j \,, \qquad  & \text{class AIII}\,,  \\
	& c_j = \frac12 - j \,, \qquad  & \text{class BDI}\,,  \\
	& c_j = 2 - 4j \,, \qquad  & \text{class CII}\,.
\end{align}

It remains to determine the prefactor $b$ in Eq.~\eqref{eq:x1l} for each of the three chiral classes. Since $b$ originates from the one-loop integral \eqref{eq:Ifa}, it is clear that $b \sim 1/\sigma_0$ up to a numerical factor.   To fix the constant, it is sufficient to evaluate the required average $\langle \mathcal{P}_{\lambda,\lambda'}[\bar{V} e^{iW_1} V] \rangle_f^{{\rm (ns)}}$ for one non-trivial choice of $\lambda, \lambda'$. For the classes BDI and CII, the simplest choice $\lambda, \lambda' = (1),(0)$ does the job. For the class AIII, we have $z_{(1)} = 0$, so that we choose $\lambda, \lambda' = (2),(0)$ to determine $b$.

\subsubsection{Class AIII}

For $\lambda, \lambda' = (2),(0)$, the last term in the square brackets in Eq.~\eqref{eq:ren_ns} is
\begin{align}
	\langle \mathrm{tr} (\bar{V} \tilde{U} V\bar{V} \tilde{U} V)-\mathrm{tr}(\bar{V} \tilde{U} V)\mathrm{tr}(\bar{V} \tilde{U} V) \rangle_f^{\rm (ns)}\,,
\end{align}
with $\tilde{U} = e^{iW_1}$.  We expand this to the second order in $W_1$ and evaluate the contractions. Terms with both factors $W_1$ coming from the same fast field $\tilde{U}$ vanish in the replica limit ($n\to 0$), since we have in class AIII
\begin{align}
	\langle \frac12(iW_1)^2 \rangle_f^{\rm (ns)} &= -n \dfrac{1}{\sigma_0} \ln\left(\dfrac{L}{a}\right) \mathds{1}. \label{eq:aiii_dos}
\end{align}
We are thus left with the following terms:
\begin{align}
	&\langle \mathrm{tr} (\bar{V} iW_1 V \bar{V} iW_1 V) \rangle_f^{\rm (ns)} = -\dfrac{2}{\sigma_0} \ln\left(\dfrac{L}{a}\right) \mathrm{tr} (\bar{V} V )\mathrm{tr} ( \bar{V} V), \nonumber\\
	&\langle \mathrm{tr} (\bar{V} iW_1 V)\mathrm{tr} ( \bar{V} iW_1 V) \rangle_f^{\rm (ns)}= -\dfrac{2}{\sigma_0} \ln\left(\dfrac{L}{a}\right) \mathrm{tr} (\bar{V} V \bar{V} V).
\end{align}
Combining them, we obtain
\begin{align}
	\langle \mathcal{P}_{(2),(0)}[\bar{V} e^{iW_1} V] \rangle_f^{{\rm (ns)}} &= \dfrac{2}{\sigma_0}\ln\left(\dfrac{L}{a}\right)\mathcal{P}_{(2),(0)}[\bar{V} V].
	\label{eq:aiii_ns}
\end{align}
Since $z_{(2)}=2$ in class AIII according to Eq.~\eqref{eq:z-lambda}, we find for the coefficient $b$ in Eq. \eqref{eq:x1l}
\begin{equation}
	b = \frac{1}{\sigma_0} \,, \qquad \text{class AIII} \,.
	\label{eq:b-AIII}
\end{equation}

%
%\begin{align}
%&\langle \mathrm{tr} (\bar{V} \tilde{U} V\bar{V} \tilde{U} V)-\mathrm{tr}(\bar{V} \tilde{U} V)\mathrm{tr}(\bar{V} \tilde{U} V) \rangle_f^{\rm (ns)} \nonumber\\
%&= \dfrac{2}{\sigma_0}\ln\left(\dfrac{L}{a}\right)\langle\mathcal{P}_{(2),(0)}[\bar{V} V] \label{eq:aiii_ns}
%\end{align}
%in replica limit, since terms with $W_1^2$ do not contribute by  Eq. \eqref{eq:aiii_dos}. Combining Eq. \eqref{eq:aiii_ns} and Eq. \eqref{eq:aiii_s}, this means that
%\begin{align}
%&\langle\mathcal{P}_{(2),(0)}[\bar{V} \tilde{U} V]\rangle_f = \nonumber\\ &\left[1+\dfrac{\kappa_0 2^2}{\sigma_0^2}\ln\left(\dfrac{L}{a}\right) + \dfrac{2}{\sigma_0}\ln\left(\dfrac{L}{a}\right)\right] \mathcal{P}_{(2),(0)}[\bar{V} V].
%\end{align}
%
%Consequently by Eq. \eqref{eq:ren_ns}, the corresponding scaling dimension of $\mathcal{P}_{(2),(0)}$ is
%\begin{align}
%x_{(2),(0)} = -\frac{2}{\sigma_0} - \dfrac{\kappa_0 2^2}{\sigma_0^2}
%\end{align}
%in replica limit $n\rightarrow 0$.

\subsubsection{Class CII}

For $\lambda, \lambda' = (1),(0)$, the calculation is very simple.  Expanding $ \langle \mathcal{P}_{(1),(0)}[\bar{V} e^{iW_1} V] \rangle_f^{{\rm (ns)}}
= \langle \mathrm{tr} [\bar{V} e^{iW_1} V] \rangle_f^{{\rm (ns)}}$ up to the second order in $W_1$ and calculating the contraction, we get
\begin{align}
	\langle \mathrm{tr} (\bar{V} \frac12(iW_1)^2 V) \rangle_f^{(\rm ns)} &= -\dfrac{1}{\sigma_0} \ln\left(\dfrac{L}{a}\right) \mathrm{tr} (\bar{V} V ).
	\label{eq:P10-average-CII}
\end{align}
Since $z_{(1)}=-1$ in class CII according to Eq.~\eqref{eq:z-lambda}, we find for the coefficient $b$ in Eq. \eqref{eq:x1l}
\begin{equation}
	b = \frac{1}{\sigma_0} \,, \qquad \text{class CII} \,.
	\label{eq:b-CII}
\end{equation}

%
%using Eq. \eqref{eq:ren_ns} and Eq. \eqref{eq:xqdef}, we can follow that
%\begin{align}
%x_{(1),(0)} +\dfrac{2\kappa_0 1^2}{\sigma_0^2} = \frac{1}{\sigma_0}.
%\end{align}
%Since $z_{(1)}=-1$, by Eq. \eqref{eq:x1l} the coefficient $b=1$ can be uniquely fixed:
%\begin{align}
%x_{\lambda,\lambda'} &= -\frac{1}{\sigma_0}(z_\lambda + z_{\lambda'})  -\dfrac{2\kappa_0}{\sigma_0^2}(|\lambda|-|\lambda'|)^2
%\end{align}
%in the general form.

\subsubsection{Class BDI}

The calculation for the class BDI proceeds in the same way as for CII. We have, in analogy with Eq.~\eqref{eq:P10-average-CII},
\begin{align}
	\langle \mathrm{tr} (\bar{V} \frac12(iW_1)^2 V) \rangle_f^{\rm (ns)} &= \dfrac{1}{\sigma_0} \ln\left(\dfrac{L}{a}\right) \mathrm{tr} (\bar{V} V ).
	\label{eq:P10-average-BDI}
\end{align}
Since $z_{(1)}= \frac12$ in class BDI according to Eq.~\eqref{eq:z-lambda}, we find for the coefficient $b$ in Eq. \eqref{eq:x1l}
\begin{equation}
	b = \frac{2}{\sigma_0} \,, \qquad \text{class BDI} \,.
	\label{eq:b-BDI}
\end{equation}

\subsubsection{One-loop scaling dimensions of composite operators: Summary}

Summarizing, we have obtained Eq.~\eqref{eq:x1l} for the one-loop scaling dimensions of composite operators, with the coefficient $b$ given by
Eqs.~\eqref{eq:b-AIII}, \eqref{eq:b-CII} , and \eqref{eq:b-BDI}.  The derivation was performed assuming that the sigma-model couplings are given by their bare value $\sigma_0$ and $\kappa_0$ (i.e., the system is close to their ultraviolet cutoff). Under RG, the couplings flow, and the same analysis applies with $\sigma_0, \kappa_0$ replaced by their renormalized values $\sigma, \kappa$. Thus, we have the following results for one-loop scaling dimensions in three chiral classes:
\begin{align}
	x_{\lambda,\lambda'} &= -\frac{1}{\sigma}(z_\lambda + z_{\lambda'})  -\dfrac{\kappa}{\sigma^2}(|\lambda|-|\lambda'|)^2, \ \  \text{class AIII},
	\label{eq:one-loop-x-AIII}
	\\
	x_{\lambda,\lambda'} &= -\frac{1}{\sigma}(z_\lambda + z_{\lambda'})  -\dfrac{2\kappa}{\sigma^2}(|\lambda|-|\lambda'|)^2, \ \  \text{class CII},
	\label{eq:one-loop-x-CII}
	\\
	x_{\lambda,\lambda'} &= -\frac{2}{\sigma}(z_\lambda + z_{\lambda'})  -\dfrac{2\kappa}{\sigma^2}(|\lambda|-|\lambda'|)^2, \ \  \text{class BDI}.
	\label{eq:one-loop-x-BDI}
\end{align}

\subsubsection{Energy scaling}

As is evident from the action \eqref{eq:ac}, the coupling $\varepsilon$  (whose bare value is the energy $E$) couples to the operator $\mathcal{P}_{(1),(0)} + \mathcal{P}_{(0),(1)}$.
According to Eq.~\eqref{eq:x1l}, this operator has the scaling dimension
\begin{align}
	x_{\nu}\equiv x_{(1),(0)} = x_{(0),(1)} = -bz_{(1)} - \dfrac{s\kappa}{\sigma^2}.
\end{align}
Explicitly, we have for each of the three chiral classes:
\begin{align}
	x_\nu&=\begin{cases}
		-\dfrac{\kappa}{\sigma^2}\,, & \textrm{AIII}\,, \\[0.4cm]
		\hspace{.1cm}
		\dfrac{1}{\sigma}-\dfrac{2\kappa}{\sigma^2}\,, & \textrm{CII} \,, \\[0.4cm]
		\hspace{.1cm}
		- \dfrac{1}{\sigma}-\dfrac{2\kappa}{\sigma^2}\,, & \textrm{BDI} \,.
	\end{cases}
\end{align}
We use the notation $x_\nu$ for this exponent since it determines the scaling of the density of states $\nu(E)$. Specifically, $\nu(E) \propto E^{\alpha_\nu}$ with $\alpha_\nu = x_\nu / (2-x_\nu)$.

\section{The Iwasawa construction}
\label{app:Iwasawa}

In this appendix, we describe the construction of pure-scaling $\sigma$-model observables %$\mathcal{P}_{\lambda, \lambda'}(Q)$ 
$\phi_{\lambda, \lambda'}(Q)$
based on the Iwasawa decomposition, see Refs.~\cite{Helgason-Differential-1978, onishchik1990} for rigorous definitions and details. The pure-scaling observables obtained in this way satisfy the abelian fusion. The construction explicitly demonstrates the difference between ``spinless'' and ``spinful'' symmetry classes. We first focus on the ``spinless'' class AIII, and emphasize the role of the U(1) sector on the scaling dimensions of the scaling operators and their Weyl symmetries. Results for the other two chiral classes are presented at the end of this appendix.

\subsection{Generalities}

The Iwasawa construction has already been presented for class A  by two of us and M. Zirnbauer in Ref.~\cite{gruzberg2013classification}, for class C by the present authors and N. Charles in Ref.~\cite{karcher2021generalized}, and for classes AII, D, and DIII in Ref.~\cite{karcher2022generalized-2}, so here we only provide basic steps. Further details relevant to the three chiral symmetry classes, AIII, CII, and BDI, studied in this paper, will be presented in the subsequent sections.

It is sufficient for our purposes in this paper to work within the bosonic replica formalism. This requires to take the limit $n \to 0$, where $n$ is the number of bosonic replicas. The replica limit will be often implicitly assumed in equations that follow. The bosonic $\sigma$-model target spaces have the form $M_B = G/K$ where $G$ is a real non-compact group and $K$ is its maximal compact subgroup. As we explained in Ref.~\cite{gruzberg2013classification}, the pure-scaling operators $\mathcal{P}_\lambda(Q)$ are joint eigenfunctions of the $G$-invariant differential operators on $G/K$, also known as the Laplace-Casimir operators. The Iwawasa decomposition allows us to construct the desired eigenfunctions as the $N$-radial spherical functions $\phi_\lambda(Q)$ on $G/K$.

An essential feature of the chiral classes that makes them different from other symmetry classes, is that the groups $G$ and $K$ are not semisimple. They have non-trivial centers (abelian subgroups) $Z(G)$ and $Z(K)$ that lead to the factorization
\begin{align}
	M_B &= \mathbb{R}_+ \times M_B^{(s)},
	&
	M_B^{(s)} &= G^{(s)}/K^{(s)},
	\label{MB-R-MBs}
\end{align}
where $G^{(s)} = G/Z(G)$ and $K^{(s)} = K/Z(K)$ are semisimple subgroups of $G$ and $K$. The abelian factors $\mathbb{R}_+$ are non-compact counterparts of the U$(1)$ factors present in the fermionic replicas. The factorization allows for the presence of two terms in the sigma-model action on $M_B$: the usual kinetic term, and the so-called Gade term, with two independent coupling constants. The factor $\mathbb{R}_+$ and the associated Gade term give a contribution to the scaling dimensions of generalized multifractal observables (gradientless operators) that affects the Weyl symmetry of the MF spectra. For brevity, we will refer to the relevant degrees of freedom as ``the U(1) sector''.

The classical Iwasawa construction applies to semisimple groups. However, we can largly ignore the issue of the semisimplicity and work out the Iwasawa decomposition for the full $M_B$. It will be easy to impose the semisimplicity (tracelessness) condition on the relevant Lie algebras at any moment. We will see the role of the U(1) sector in the process, and it will mostly present itself via the tracelessness condition on the Abelian subalgebra $\mfa$ which will lead to an equivalence relation between $N$-radial functions on $M_B^{(s)}$.

Another essential feature of the chiral classes is that the matrices $Q \in M_B$ have (after an appropriate transformation, see below) a block-diagonal form in the retarded-advanced space. The two diagonal blocks are related to the two sublattices in the microscopic models of the chiral symmetry classes. Then we can separately construct generalized multifractal observables for each sublattice from the corresponding block of the $Q$ matrix using two different weights (multi-indices) $\lambda$ and $\lambda'$.

We begin with the Cartan decomposition
\begin{align}
	\mfg= \mfk \oplus \mfp
	\label{Cartan-decomposition-real}
\end{align}
of the Lie algebra of $G$, $\mfg = \text{Lie}(G)$, into a maximal compact subalgebra $\mfk$ and the complementary subspace $\mfp$. The two parts of the Cartan decomposition are the $+1$ and $-1$ eigenspaces of a Cartan involution (a Lie algebra automorphism that squares to the identity) $\theta$. If we write an element $Z \in \mfg$ as $Z = X + Y$ where $X \in \mfk$ and $Y \in \mfp$, then
%\begin{align}
$\theta(X + Y) = X - Y$.
%\end{align}
The parts of the Cartan decomposition satisfy the commutation relations
\begin{align}
	[\mfk, \mfk] &\subseteq \mfk,
	&
	[\mfk, \mfp] &\subseteq \mfp,
	&
	[\mfp, \mfp] &\subseteq \mfk.
	\label{comm-rel-Cartan-real}
\end{align}

Then we choose a maximal Abelian subalgebra $\mfa \subset \mfp$ and consider the adjoint action of elements $H \in \mfa$ on $\mfg$. The eigenvectors $E_\alpha$ of this action satisfy
\begin{align}
	[H, E_\alpha] = \alpha(H) E_\alpha
\end{align}
and are called restricted root vectors, and the eigenvalues $\alpha$ are called restricted roots. The dimension $m_\alpha$ of the restricted root space
%\begin{align}
$\mfg_\alpha = \text{span}\, \{E_\alpha\}$
%\end{align}
is called the multiplicity of the restricted root $\alpha$, and can be bigger that 1. Restricted roots are linear functions on $\mfa$, and lie in the space $\mfa^*$ dual to $\mfa$. The dimension $n$ of both $\mfa$ and $\mfa^*$ is the {\it rank} of the symmetric space $G/K$. This is what we earlier called the number of bosonic replicas. Basis elements of $\mfa$ will be denoted by $H_k$, so that a generic element $H \in \mfa$ is $H = \sum_{k=1}^n h_k H_k$.
The dual basis in $\mfa^*$ is defined as elements $x_i$ such that $x_i(H) = h_i$ ($i = 1,\ldots, n$). In terms of this basis the restricted roots for all chiral classes will be
\begin{align}
	\pm \alpha_{ij} &= \pm(x_i - x_j), \qquad i < j.
	\label{roots}
\end{align}
These are ordinary roots with multiplicities $m_{o}$ known for all classes, and the system of these roots in $A_{n-1}$ in the standard Cartan notation. In what follows, we will compute these multiplicites for the three classes AIII, BDI, and CII that are the focus of this paper.

A system of positive restricted roots is defined by choosing some hyperplane through the origin of $\mfa^*$ which divides $\mfa^*$ in two halves, and then defining one of these halves as positive. We will choose $\alpha_{ij}$ as the positive roots. The Weyl vector $\rho$  is defined as the half-sum of positive restricted roots accounting for their multiplicities. In the replica limit $n \to 0$ this gives
\begin{align}
	\rho &= \lim_{n \to 0} \frac{1}{2} \sum_{\alpha > 0} m_\alpha \alpha = \sum_{i} c_i x_i,
	%\nonumber \\
	&
	c_i &= m_{o} \Big( \frac{1}{2} - i \Big).
	\label{Weyl-vector}
\end{align}
Positive restricted roots generate the nilpotent Lie algebra
%\begin{align}
$\mfn = \sum_{\alpha > 0} \mfg_\alpha$.
%\end{align}
The Iwasawa decomposition at the Lie algebra level is
\begin{align}
	\mfg= \mfk\oplus \mfa \oplus \mfn
	\label{Iwasawa-decomposition-algebra}
\end{align}

Exponentiation of Eq.~\eqref{Iwasawa-decomposition-algebra} gives the global form of the Iwasawa decomposition
\begin{align}
	G = NAK,
	\label{Iwasawa-decomposition-group}
\end{align}
which allows us to represent any element $g \in G$ in the form $g = nak$, with $n \in N = e^\mfn$, $a \in A = e^\mfa$, and $k \in K = e^{\mfk}$. This factorization is unique once the system of positive restricted roots is fixed, and provides a very useful parametrization of the target space $G/K$. An element $a \in A$ is fully specified by $n$ real numbers $x_i(\ln a)$, which play the role of radial coordinates on $G/K$. For simplicity, we will denote these radial coordinates simply by $x_i$. Thus $x_i$ may now have two different meanings: either its original meaning as a basis element in $\mfa^*$, or the new one as an $N$-radial function $x_i(\ln a)$ on $G/K$. It should be clear from the context which of the two meanings is being used.

Using the radial coordinates, the joint $N$-radial eigenfunctions of the Laplace-Casimir operators on $G/K$ take a very simple exponential form
\begin{align}
	\phi_\mu(Q) = e^{(\mu + \rho)(\ln a)},
	\label{plane-wave-1}
\end{align}
where $a$ is the $a$-factor in the Iwasawa decomposition of $g$ in $Q = g \Lambda g^{-1}$, and
%\begin{align}
$\mu = \sum_{i} \mu_i x_i$
%\end{align}
is a weight vector in $\mfa^*$ with arbitrary real or even complex components $\mu_i$. We will also use the notation
\begin{align}
	\lambda &= -\frac{\mu + \rho}{2}
	= (q_1,q_2,\ldots,q_n),
	&
	q_i &= -\frac{\mu_i + c_i}{2},
\end{align}
and the exponential functions~\eqref{plane-wave-1} become
\begin{align}
	\phi_\lambda(x)
	%\equiv \phi_{(q_1,q_2,\ldots,q_n)}(x)
	= \exp \Big(\!-\!2 \sum_{i} q_i x_i \Big).
	\label{plane-wave-2}
\end{align}

To construct the exponential $N$-radial eigenfunctions explicitly as combinations of matrix elements of $Q$, we use the key fact that there exists a choice of basis in which elements of $\mfa$ and $a \in A$ are diagonal matrices, while elements of $\mfn$ are strictly upper triangular, and elements $n \in N$ are upper triangular with units on the diagonal. This has immediate consequences for the matrix $Q \Lambda$: since elements of $K$ commute with $\Lambda$, the Iwasawa decomposition $g = nak$ leads to $Q \Lambda = n a^2 \Lambda n^{-1} \Lambda$, which is a product of an upper triangular, a diagonal, and a lower triangular matrices. In this form the lower principal minors of the advanced-advanced (AA)  block of $Q \Lambda$ are simply products of diagonal elements of $a^2$, which are exponentials of the radial coordinates $x_i$ on $G/K$. These minors are basic $N$-radial spherical functions on $G/K$ which can be raised to arbitrary powers and multiplied to produce the most general exponential functions~\eqref{plane-wave-1}. A great advantage of this construction is that is directly gives the general positive scaling operators that can be raised to arbitrary powers and satisfy the Abelian fusion rules.

The block-diagonal structure of the $Q$ matrix mentioned above makes it useful to completely separate the $x_i$ variables that are used to describe observables on the two sublattices. To this end, we will double the number of replicas, which enlarges the root system to $A_{2n-1}$, use $x_i$ and $q_i$ with $i = 1,\dots, n$ for one sublattice, and redefine
\begin{align}
	x'_i &\equiv x_{2n+1-i},
	&
	q'_i &\equiv q_{2n+1-i},
\end{align}
also with $i = 1,\dots, n$, for the other sublattice. These two sets of the radial variables are not overlapping. Then we consider $N$-radial eigenfunction of the form
\begin{align}
	\phi_{\lambda, \lambda'}(x,x')
	&= \exp \Big(\!-\!2 \sum_{i=1}^{n} q_i x_i \Big)
	\exp \Big(2 \sum_{i=1}^{n} q_i' x_i' \Big),
	\label{phi-lambda-bar-lambda}
\end{align}
where the subscripts now stand for two independent weights
\begin{align}
	\lambda &= (q_1,q_2,\ldots,q_n),
	&
	\lambda' &= (q'_1,q'_2,\ldots,q'_n),
\end{align}
one for each sublattice. For our purposes, it is important to consider multi-indices $\lambda$ and $\lambda'$ such that all components $q_i$ and $q'_i$ beyond the first $m$ (respectively, $m'$) are zero. We omit these zero components, using the notations
\begin{align}
	\lambda &= (q_1,q_2,\ldots,q_m),
	&
	\lambda' &= (q'_1,q'_2,\ldots,q'_{m'}).
\end{align}
The replica limit $n \to 0$ is taken at fixed $m$ and $m'$.

Let us now present elements of the Iwasawa construction that are the same for all symmetry classes. The groups $G$ and $K$ will act in the space
\begin{align}
	\mathbb{C}^{4n} &= \mathbb{C}^2 \otimes \mathbb{C}^2 \otimes \mathbb{C}^n,
	\label{tensor-product-space-3}
\end{align}
where the factors in the tensor product correspond in this order to advanced-retarded, spin, and replica spaces. We will use the standard Pauli matrices $\sigma_i$ including the identity matrix $\sigma_0$. These act in either of the two first factors in Eq.~\eqref{tensor-product-space-3}, and we introduce short-hand notations for various tensor products
\begin{align}
	\Sigma_{i} &\equiv \sigma_i \otimes I_n,
	\qquad
	\sigma_{jk} \equiv \sigma_j \otimes \sigma_k,
	\nonumber \\
	\Sigma_{jk} &\equiv \sigma_{jk} \otimes I_n
	= \sigma_j \otimes \sigma_k \otimes I_n.
	\label{Sigma-ij-definition}
\end{align}
For example $\Sigma_{00} = I_{4n}$, and $\Sigma_{30} = \Lambda$, the usual $\Lambda$ matrix from the sigma model. In  ``spinless'' symmetry classes, like AIII and BDI studied in this paper, we can omit the second factor in the space~\eqref{tensor-product-space-3}, resulting in
\begin{align}
	\mathbb{C}^{2n} &= \mathbb{C}^2 \otimes \mathbb{C}^n.
	\label{tensor-product-space-2}
\end{align}

We will use a standard notation for the matrix units: $E_{ij}$ is the matrix with $1$ in the $i$-th row and $j$-th column, all other entries being zero. The symmetric and anti-symmetric combinations of matrix units are denoted as
\begin{align}
	E^+_{ij} & = E_{ij} + E_{ji},
	& i &\leqslant j,
	&
	E^-_{ij} & = E_{ij} - E_{ji},
	& i &< j.
	\label{E^pm}
\end{align}

Another common element in the constructions below are basis rotations in the spaces~\eqref{tensor-product-space-3} or~\eqref{tensor-product-space-2} facilitated by the unitary matrix
\begin{align}
	R &=  (\sigma_0 + i \sigma_1)/\sqrt{2}.
	\label{U-Iwasawa-chiral}
\end{align}
The conjugation by $R$ permutes the Pauli matrices as follows:
\begin{align}
	R \sigma_1 R^{-1} &= \sigma_1,
	&
	R \sigma_2 R^{-1} &= -\sigma_3,
	&
	R \sigma_3 R^{-1} &= \sigma_2.
\end{align}

\subsection{Class AIII}

In this section we present details of the Iwasawa construction for class AIII which is the simplest of the three classes considered in this paper. In class AIII we have $M_B = \text{GL}(n, \mathbb{C})/\text{U}(n)$, which is not irreducible and can be factorized as
\begin{align}
	M_B &= \frac{\text{GL}(1, \mathbb{C})}{\text{U}(1)}
	\times \frac{\text{SL}(n, \mathbb{C})}{\text{SU}(n)}
	= \mathbb{R}_+ \times M_B^{(s)},
	\label{MB-R-MBs-AIII}
\end{align}
where $\mathbb{R}_+$ is the multiplicative group of positive real numbers. As we mentioned in the previous section, we will ignore the issues related to the presence of the abelian factor $\mathbb{R}_+$, which will be taken into account later.

We write the elements of $\mfg = \text{gl}(n, \mathbb{C})$ as matrices in the space~\eqref{tensor-product-space-2}
\begin{align}
	Z = \sigma_0 \otimes X + i \sigma_2 \otimes Y
	= \begin{pmatrix*}[r] X & Y \\ -Y & X \end{pmatrix*},
	\label{Z-in-gl(n,C)}
\end{align}
where all entries are $n \times n$ blocks in the RA space satisfying $X^\dagger = - X$ and $Y^\dagger = - Y$. This definition is equivalent to the standard one where the algebra $\mfg = \text{gl}(n, \mathbb{C})$ consists of all complex $n \times n$ matrices. In the semisimple case we also need to impose the conditions
\begin{align}
	\text{tr} \, X &= 0, & \text{tr} \, Y &= 0.
	\label{trace-condition}
\end{align}
The elements $Z \in \mfg$ satisfy the conditions
\begin{align}
	Z^\dagger \Sigma_{3} + \Sigma_{3} Z &= 0,
	&
	Z \Sigma_{2} - \Sigma_{2} Z = 0,
	\label{Z-conditions-AIII}
\end{align}
and their combinations. The subalgebra u$(n)$ is the one with $Y = 0$.

The Cartan involution is
\begin{align}
	\theta(Z) = - Z^\dagger = \Sigma_{3} Z \Sigma_{3},
\end{align}
and its eigenspaces are characterised as follows: $Z \in \mfk$ if $Y = 0$, and $Z \in \mfp$ if $X = 0$. We have two groups of generators in both $\mfk$ and $\mfp$:
\begin{align}
	X^{(0)}_{ij} &= \sigma_{0} \otimes E^-_{ij},
	&
	Y^{(0)}_{ij} &= \sigma_{2} \otimes E^+_{ij},
	\label{0-generators-AIII}
	\\
	X^{(1)}_{ij} &= i \sigma_{0} \otimes E^+_{ij},
	&
	Y^{(1)}_{ij} &= i \sigma_{2} \otimes E^-_{ij}.
	\label{1-generators-AIII}
\end{align}

We choose the maximal Abelian subspace $\mfa \subset \mfp$ as
\begin{align}
	\mfa &= \text{span}\left\{ H_k = \sigma_{2} \otimes E_{kk}
	= Y^{(0)}_{kk}/2 \right\}.
\end{align}
The tracelessness condition for the semisimple case is imposed by the requirement that we only consider elements $H = \sum_k h_k H_k \in \mfa$ with $\sum_k h_k = 0$. Straightforward computations show that the system of restricted roots is $A_{n-1}$ given by Eq.~\eqref{roots} with $m_o = 2$. The positive restricted root vectors are
\begin{align}
	E_{\alpha_{ij}}^{(0)} &= X^{(0)}_{ij} + Y^{(0)}_{ij},
	&
	E_{\alpha_{ij}}^{(1)} &= X^{(1)}_{ij} + Y^{(1)}_{ij},
	\label{positive-RRV-AIII}
\end{align}
and the Weyl vector in the replica limit is
\begin{align}
	\rho &= \sum_{i}^n c_i x_i,
	&
	c_i &= 1 - 2i.
	\label{Weyl-vector-AIII}
\end{align}
We note that the values of the components $c_i$ obtained here are the same as in class A.

The unitary transformation that makes the generators of $\mfa$ diagonal and the generators of $\mfn$ strictly upper-triangular is accomplished with the help of the matrix
\begin{align}
	R_\text{AIII} &= R \otimes I_n,
	\label{U-AIII}
\end{align}
where the matrix $U$ was defined in Eq.~\eqref{U-Iwasawa-chiral}. We also need the permutation matrix $\Pi$ with elements $\Pi_{ij} = \delta_{\pi_1(i), j}$ where the permutation $\pi$ of the basis of the space~\eqref{tensor-product-space-2} is given by $\pi(i) = n + 1 - i$ for $i \in 1, \ldots, n$. The unitary transformation
\begin{align}
	\tilde{M} &= \Pi^{-1} R_\text{AIII} M R_\text{AIII}^{-1} \Pi
	\label{tilde-M-AIII}
\end{align}
leads to
\begin{align}
	\tilde{\Lambda} &= \sigma_{2} \otimes \mathcal{I}_n,
\end{align}
where $\mathcal{I}_n$ is the $n \times n$ matrix with units on the ``anti-diagonal'', that is, $(\mathcal{I}_n)_{ij} = \delta_{i, n + 1 - j}$. It is easy to show that in the new basis the positive restricted root vectors $\tilde{E}$ are strictly upper-triangular.

We can visualize the restricted root vectors for $n=3$ as a schematic matrix diagram by indicating the matrix positions where various generators have non-zero entries. For brevity we write $\alpha^{(i)} \equiv \tilde{E}^{(i)}_{\alpha_{kl}}$ with indices suppressed since they can be inferred from the matrix grid (uncolored cells have zero entries):
\begin{equation}
	\tilde{E} = \left(\begin{array}{ccc|ccc}
		\caminus -x_3 & \calphare \alpha^{(01)} & \calphare \alpha^{(01)} & & & \\
		& \caminus -x_2 & \calphare \alpha^{(01)} & & & \\
		& & \caminus -x_1 & & & \\
		\hline
		& & & \caplus x_1 & \calphare \alpha^{(01)} & \calphare \alpha^{(01)} \\
		& & & & \caplus x_2 & \calphare \alpha^{(01)} \\
		& & & & & \caplus x_3 \\
	\end{array}\right).
	\label{matrix-positive-RRV-AIII-triangular}
\end{equation}

The block-diagonal structure of the matrix in Eq.~\eqref{matrix-positive-RRV-AIII-triangular} transfers to the subgroups $A$ and $N$ that are necessary to construct the $Q$ matrix. As a result, the $\mathcal{Q} = Q\Lambda$ matrix itself is block-diagonal in the rotated basis. Indeed, we have
\begin{align}
	\tilde{\mathcal{Q}} = \tilde{n} \tilde{a}^2 \tilde{\Lambda} \tilde{n}^{-1} \tilde{\Lambda}.
\end{align}
The matrices $\tilde{n}$ and $\tilde{n}^{-1}$ are block-diagonal, and each block is upper-triangular with units on the diagonals, while $\tilde{a}$ is diagonal:
\begin{align}
	\tilde{a} = \text{diag} (e^{-x_n}, \ldots, e^{-x_1}, e^{x_1},\ldots, e^{x_n}).
\end{align}
Conjugation by $\tilde{\Lambda}$ converts $\tilde{n}^{-1}$ into $\tilde{\Lambda} \tilde{n}^{-1} \tilde{\Lambda}$ which is block-diagonal with {\it lower-triangular} blocks with units on the diagonal. This results in the following structure of the matrix $\tilde{\mathcal{Q}}$:
\begin{align}
	\tilde{\mathcal{Q}} &= \begin{pmatrix} \tilde{\mathcal{Q}}_{RR} & 0 \\
		0 & \tilde{\mathcal{Q}}_{AA} \end{pmatrix},
	\label{Q-tilde-blocks}
\end{align}
where the diagonal blocks in the RA space have the following structure of their lower-right $m \times m$ submatrices for any $m \leq n$:
\begin{align}
	\tilde{\mathcal{Q}}_{RR}^{(m)} &=
	\begin{pmatrix} 1 & \ldots & * \\ \vdots & \ddots & \vdots \\
		0 & \ldots & 1 \end{pmatrix}
	\begin{pmatrix}
		e^{-2x_{m}} & \ldots & 0 \\ \vdots & \ddots & \vdots \\
		0 & \ldots & e^{-2x_1} \end{pmatrix}
	\begin{pmatrix}
		1 & \ldots & 0 \\ \vdots & \ddots & \vdots \\ * & \ldots & 1 \end{pmatrix},
	%\label{Q-tilde-RR-m}
	\nonumber \\
	\tilde{\mathcal{Q}}_{AA}^{(m)} &=
	\begin{pmatrix} 1 & \ldots & * \\ \vdots & \ddots & \vdots \\
		0 & \ldots & 1 \end{pmatrix}
	\begin{pmatrix}
		e^{2x_{n - m + 1}} & \ldots & 0 \\ \vdots & \ddots & \vdots \\
		0 & \ldots & e^{2x_n} \end{pmatrix}
	\begin{pmatrix}
		1 & \ldots & 0 \\ \vdots & \ddots & \vdots \\ * & \ldots & 1 \end{pmatrix}.
	%\label{Q-tilde-AA-m}
	\label{Q-tilde-blocks-m}
\end{align}

The transformation of the basis used above is very natural since it is directly related to the way the sublattice structure is present in the $Q$ matrix. In the original basis we have
\begin{align}
	Q = \begin{pmatrix} Q_{RR} & Q_{RA} \\ Q_{RA} & -Q_{RR} \end{pmatrix}
	= \sigma_3 \otimes Q_{RR} + \sigma_1 \otimes Q_{RA}
\end{align}
(the restrictions on the blocks follow from the symmetries of the problem),
and it is easy to see that
\begin{align}
	R_\text{AIII} \mathcal{Q} R_\text{AIII}^{-1}
	&= \sigma_0 \otimes Q_{RR} + i \sigma_3 \otimes Q_{RA}
	\nonumber \\
	&= \begin{pmatrix} Q_{RR} + i Q_{RA} & 0 \\ 0 & Q_{RR} - iQ_{RA} \end{pmatrix}.
\end{align}
The two diagonal subblocks here are inverses of each other and couple to the two sublattices. Thus, we can use the subblock $\tilde{\mathcal{Q}}_{RR}$ to construct scaling observables for sublattice A, and the subblock $\tilde{\mathcal{Q}}_{AA}$ to construct scaling observables for sublattice B.

As we mentioned above, it is useful to separate the $x_i$ variables that are used to describe observables on the two sublattices. To this end, let us double the number of replicas, which enlarges the root system to $A_{2n-1}$, and use $x_i$ with $i = 1,\dots, n$ for one sublattice, and $x'_i \equiv x_{2n+1-i}$ also with $i = 1,\dots, n$ for the other sublattice. These two sets of the radial variables are not overlapping. Then
%we consider only $N$-radial eigenfunction of the form
%\begin{align}
%\phi_{\lambda, \lambda'}
%&= \exp \Big(\!-\!2 \sum_{i=1}^{n} q_i x_i \Big)
%\exp \Big(2 \sum_{i=1}^{n} q'_i \bar{x}_i \Big).
%\label{phi-lambda-bar-lambda}
%\end{align}
we define two types of elementary building blocks as determinants of the matrices $\tilde{\mathcal{Q}}_{RR}^{(m)}$ and $\tilde{\mathcal{Q}}_{AA}^{(m)}$:
\begin{align}
	d_{m}(x) &= \det \tilde{\mathcal{Q}}_{RR}^{(m)}
	%= \prod_{i = 1}^{m} e^{-2x_i}
	= \exp \Big(- 2 \sum_{i=1}^{m} x_i \Big),
	\nonumber \\
	d'_m(x') &= \det \tilde{\mathcal{Q}}_{AA}^{(m)}
	%= \prod_{i = 1}^{m} e^{2x_{n - i + 1}}
	= \exp \Big( 2 \sum_{i=1}^{m} x'_i \Big),
\end{align}
and form two $N$-radial eigenfunctions, each labeled by its own weight vector:
\begin{align}
	\phi_\lambda(x) &=
	\prod_{k=1}^{n} d_k^{q_k - q_{k+1}}(x)
	%d_1^{q_1 - q_2} d_2^{q_2 - q_3}
	%\ldots d_{n-1}^{q_{n-1} - q_n} d_{n}^{q_n}
	= \exp \Big(\!-\!2 \sum_{i=1}^{n} q_i x_i \Big),
	\nonumber \\
	\phi_{\lambda'}(x')
	&= \prod_{k=1}^{n} d_k^{\prime \,q'_k - q'_{k+1}}(x')
	%d_1^{\prime \,q'_1 - q'_2} d_2^{\prime \,q'_2 - q'_3}
	%\ldots d_{n-1}^{\prime \,q'_{n-1} - q'_n} d_{n}^{\prime \,q'_n}
	= \exp \Big(2 \sum_{i=1}^{n} q'_i x'_i \Big).
\end{align}
As in other symmetry classes, the components of the weights $\lambda$ and $\lambda'$ are arbitrary complex numbers, and we denote $q_{n+1} = 0$. Products of such functions labeled by the pair of weights
\begin{align}
	\phi_{\lambda, \lambda'}(x,x')
	&= \exp \Big(\!-\!2 \sum_{i=1}^{n} q_i x_i \Big)
	\exp \Big(2 \sum_{i=1}^{n} q_i' x_i' \Big)
	\label{d-product-AIII}
\end{align}
are the most general $N$-radial eigenfunctions with separated contributions of the two sublattices. They can also be written as $\phi_{\lambda, \lambda'} = e^{\rho + \mu}$, where
\begin{align}
	\rho &= \sum_{i=1}^n (c_i x_i + c'_i x'_i),
	&
	\mu &= \sum_{i=1}^{n} (\mu_i x_i + \mu'_i x'_i),
	\label{rho-mu-AIII}
	\\
	c_i &= 1 - 2i, & \mu_i &= - 2q_i - c_i,
	\nonumber \\
	c'_i &= - c_i = 2i - 1,
	&
	\mu'_i &= 2q'_i - c'_i = 2q'_i + c_i.
\end{align}

\subsubsection{The role of the U(1) sector}

Let us now consider the role of the U(1) sector. We use the non-compact (bosonic) repliacs, but many steps and equations will be similar to the ones in Ref.~\cite{gade1991the} by Gade und Wegner, which we will denote as GW in this appendix.

In the semisimple case the elements of $\mfa$ are restricted by the condition $\sum_k^{2n} h_k = 0$, which implies
$\forall H \in \mfa$
\begin{align}
	\sum_{i=1}^{n} x_i(H) + \sum_i^{n} x'_i(H) = \sum_i^{2n} h_i = 0.
\end{align}
Thus, the dual basis in $\mfa^*$ automatically satisfies
\begin{align}
	\sum_{i=1}^{n} (x_i + x'_i) = 0,
	\label{semisimplicity}
\end{align}
which is a constraint selecting the subspace $M_B^{(s)}$ in the coset $G/K$. Restricting our consideration to this subspace has several consequences.

First of all, the representation of an arbitrary element of $\mfa^*$ as $\mu = \sum_i^{n} (\mu_i x_i + \mu'_i x'_i)$ is not unique. In this notation all $\mu_i$ and $\mu'_i$ can be shifted by an arbitrary constant without changing the result. This defines an equivalence relations between weights: $\mu \sim \mu + c(1^{2n})$, and all members of the equivalence class $[\mu]$ of a weight $\mu$ give the same $N$-radial function $\phi_\mu = e^{\rho + \mu}$ on $M_B^{(s)}$. Let us adopt the following terminology. We will call
\begin{align}
	Q_\mu \equiv \sum_{i=1}^{n} (\mu_i + \mu'_i) = 2 \sum_{i=1}^{n} (q'_i - q_i)
	= 2(|\lambda'| - |\lambda|)
	\label{Q-lambda-bar-lambda}
\end{align}
the \emph{charge} of the weight $\mu$ and the corresponding $N$-radial function~\eqref{d-product-AIII}.
%The charge of the Weyl vector in the replica limit is
%\begin{align}
%Q_\rho = \sum_{i=1}^{n} (c_i + c'_i) = 0.
%\end{align}

Among all members of the equivalence class $[\mu]$ of a weight $\mu$ there is a unique element \begin{align}
	\mu^{(0)} &= \mu - \frac{Q_\mu}{2n} (1^{2n})
\end{align}
whose charge is zero: $Q_{\mu^{(0)}} = 0$. We will call such weights (and the corresponding $N$-radial functions) ``neutral''.
%Notice that if we are interested in the replica limit $n \to 0$, the shift by $Q_\mu/n$ is singular in the limit, if the charge $Q_\mu$ is finite for $n=0$. In addition, we normally consider weights $\mu$ with a finite number of nonzero parts $\mu_i$. If such a weight has nonzero charge $Q_\mu$, then its neutral equivalent $\mu^{(0)}$ has all of its $n$ parts nonzero. In the replica limit this number becomes infinity, and this is problematic.
An arbitrary $N$-radial exponential function on $M_B$ can be factorized as
\begin{align}
	\phi_\mu = e^{\rho + \mu} = e^{\rho + \mu^{(0)}}
	e^{(Q_\mu/2n) \sum_{i=1}^{n} (x_i + x'_i)}.
	\label{phi-mu-AIII}
\end{align}
This factorization is analogous to Eq.~(14) in GW. The first factor here is neutral, and is thus a function on $M_B^{(s)}$. The sum $\sum_i (x_i + x'_i)$ in the second factor plays the role of the generator of the group $\mathbb{R}_+$ in Eq.~\eqref{MB-R-MBs}. It is the analog of the phase $\phi$ in Eq.~(14) in GW.

The second consequence of Eq.~\eqref{semisimplicity} is that the radial variables $x_i$, $x'_i$ are not independent on the space $M_B^{(s)}$. Its dimension is $2n-1$ as opposed to the dimension of $M_B$, which is $2n$. Then all commuting differential operators on $M_B$ should separate into parts, one for $M_B^{(s)}$ and one for the ``U(1)'' part $\mathbb{R}_+$, and the same is true for their eigenvalues and the scaling dimensions of the scaling operators. This is seen in Eq.~(15) in GW, which for our purposes can be written in terms of the scaling dimensions as
\begin{align}
	x_{\lambda, \lambda'} = x_{\lambda, \lambda'}^{(s)} + x_{\lambda, \lambda'}^\text{U(1)}.
	\label{x-mu-splitting}
\end{align}
Gade and Wegner have argued that
\begin{align}
	x_{\lambda, \lambda'}^\text{U(1)} = \alpha(n) Q_\mu^2
	= 4 \alpha(n) (|\lambda| - |\lambda'|)^2
\end{align}
exactly to all orders in perturbation theory. The coefficient $\alpha(n)$ depends on the coupling constant of the Gade term and on the number of replicas $n$, see Eq.~(16) in GW. The splitting~\eqref{x-mu-splitting} of the scaling dimension is singular in the sense that both terms in the right-hand side diverge in the replica limit. The divergences, however, cancel out and the result for $x_{\lambda, \lambda'}$ is finite as $n \to 0$. It is easy to see how this works in the perturbative RG at weak coupling, where the scaling dimensions at one loop are proportional to the eigenvalues of the quadratic Casimir-Laplace operators.

Let us denote $\partial_i \equiv \partial/\partial x_i$, $\partial'_i \equiv \partial/\partial x'_i$. Then on the space $M_B$ the quadratic Casimir-Laplace operator is
\begin{align}
	\Delta = \frac{1}{4} e^\rho \sum_{i=1}^{n} (\partial^2_i + \partial^{\prime 2}_i) \circ e^{-\rho}.
	\label{quadratic-Casimir}
\end{align}
Its eigenvalues corresponding to the eigenfunctions~\eqref{d-product-AIII} are
\begin{align}
	z_{\lambda, \lambda'} = \frac{1}{4} \sum_{i=1}^{n} (\mu_i^2 + \mu^{\prime 2}_i)
	= z^{\text{A}}_{\lambda} + z^{\text{A}}_{\lambda'},
	\label{quadratic-Casimir-AIII}
\end{align}
where we denote
\begin{align}
	z^{\text{A}}_{\lambda} &= \sum_{i=1}^{n} q_i(q_i + c_i),
	&
	z^{\text{A}}_{\lambda'} &= \sum_{i=1}^{n} q'_i(q'_i + c_i).
	\label{quadratic-Casimir-A}
\end{align}
These are eigenvalues of the quadratic Casimir-Laplace operator that appear in class A, as indicated by the superscript.

On the ``U(1)'' part $\mathbb{R}_+$ there is only one basic differential operator: the total ``momentum'' operator
\begin{align}
	D = e^\rho \sum_{i=1}^{n} (\partial_i + \partial'_i) \circ e^{-\rho}
	= \sum_{i=1}^{n} (\partial_i + \partial'_i),
\end{align}
whose eigenvalue on $e^{(Q_\mu/2n) \sum_i (x_i + x'_i)}$ is $Q_\mu$. The function~\eqref{phi-mu-AIII} on $M_B$ is also an eigenfunction of the operator $D$ with the eigenvalue $Q_\mu$. Therefore, the operator $D$ nullifies all neutral $\phi_\mu$. It is now easy to see that the combination
\begin{align}
	\Delta^{(s)} = \Delta - \frac{1}{2n} D^2
\end{align}
nullifies any function on $\mathbb{R}_+$, that is, a function of $\sum_i (x_i + x'_i)$ only. Thus, it is the quadratic Casimir-Laplace operator on $M_B^{(s)}$. The eigenvalue of this operator on $\phi_{\lambda, \lambda'}$ is
\begin{align}
	z^{(s)}_{\lambda, \lambda'} = z_{\lambda, \lambda'} - \frac{Q_\mu^2}{2n}.
\end{align}
It is easy to see that this eigenvalue is the same for any two equivalent weights, that is, it depends only on the equivalence classes $[\mu]$ of weights, as it should.

%(The check:
%\begin{align}
%z^{(s)}_{\mu + c(1^n)} &= \sum_i (\mu_i + c)^2 - \frac{1}{n} \Big(\sum_i \mu_i + n c \Big)^2
%= \sum_i \mu_i^2  + 2c \sum_i \mu_i + n c^2
%\nonumber \\
%& - \frac{1}{n} \Big(\sum_i \mu_i \Big)^2 - 2c \sum_i \mu_i - n c^2
%=  z_\mu - \frac{Q_\mu^2}{n} = z^{(s)}_\mu,
%\end{align}
%Q.E.D.)

Now, the splitting of the scaling dimensions~\eqref{x-mu-splitting} at the level of a one-loop RG looks like
\begin{align}
	x^{\text{1 loop}}_{\lambda, \lambda'} &= - b z^{(s)}_{\lambda, \lambda'} + \alpha(n) Q_\mu^2
	\nonumber \\
	&= - b z_{\lambda, \lambda'} + \Big(\alpha(n) + \frac{b}{2n}\Big) Q_\mu^2.
\end{align}
It is the cancellation of the $1/n$ terms in the brackets in the right-hand side that leads to a finite result
\begin{align}
	x^{\text{1 loop}}_{\lambda, \lambda'} = - b (z^{\text{A}}_{\lambda} + z^{\text{A}}_{\lambda'}) + \tilde{\alpha} (|\lambda| - |\lambda'|)^2,
	\label{x-mu-1-loop}
\end{align}
in agreement with the RG derivation in Appendix~\ref{sec:renorm-comp-op} that leads to Eq.~\eqref{eq:one-loop-x-AIII}, see also Eq.~\eqref{one-loop-dimensions} in the main text. This result is consistent with Eqs.~(17) and~(22) in GW. The special case $\lambda = (1)$, $\lambda' = 0$ corresponds to the scaling of the average local density of states,
so that
\begin{align}
	\tilde{\alpha} = x_\nu \,,
	\label{tilde-alpha}
\end{align}
where $x_\nu$ is the scaling dimension of the average local density of states.

\subsubsection{The Weyl and sublattice symmetry}

Let us now discuss Weyl symmetries of the scaling dimensions $x_{\lambda, \lambda'}$. We begin with the neutral observables. The charge of the exponential function $\phi_{\lambda, \lambda'}$~\eqref{phi-lambda-bar-lambda} is given by Eq.~\eqref{Q-lambda-bar-lambda}, the function is neutral if
\begin{align}
	|\lambda'| = |\lambda|.
	\label{neutrality}
\end{align}
Neutral observables have equal total powers of matrix elements of the $Q$ matrix (or the wave functions) from both sublattices. This includes the cases when one of the weights is zero (say $\lambda' = 0$), while the other has to satisfy $|\lambda| = 0$. An important subclass of neutral observables consists of ``balanced'' observables with $\lambda' = \lambda$.
%When the weights $\lambda$ and $\lambda'$ are integer partitions (or Young diagrams), another class of neutral observables is obtained $\lambda' = \lambda^T$ (conjugate or transposed Young diagram).

The scaling dimensions of neutral observables  are given by the ``semisimple'' term $x_{\lambda, \lambda'}^{(s)}$ in the decomposition~\eqref{x-mu-splitting}. Their perturbative one-loop RG values are given by the first (proportional to $b$) term in Eq.~\eqref{x-mu-1-loop} and thus satisfy Weyl symmetries of class A, including the sign changes of $\mu_i$ or $\mu'_i$ that lead to
\begin{align}
	q_i &\to \tilde{q}_i = - c_i - q_i,
	&
	q'_i &\to \tilde{q}'_i = - c_i - q'_i.
	\label{reflections}
\end{align}
This is suprising in view of the fact that the relevant Weyl group is that of the root system $A_{2n-1}$, that is, the permutation group $S_{2n}$ of all $2n$ radial coordinates $x_i$  and $x'_i$. This group does not contain sign changes of $x_i$'s. Thus, we expect that in higher orders of perturbation theory the symmetries~\eqref{reflections} should be violated. This is also expected because the commutative algebra of invariant differential operators on $M_B^{(s)}$ is generated via the Harish-Chandra isomorphism by the polynomial invariants
\begin{align}
	D_k = e^\rho \sum_{i=1}^{n} (\partial^k_i + \partial^{\prime k}_i) \circ e^{-\rho}
\end{align}
for all $k = 1, 2, \dots, 2n$. In particular, there are invariant operators of odd orders that are not invariant under sign changes $x_i \to - x_i$ and $x'_i \to - x'_i$. If these operators appear in the RG transformation (and they generically are not forbidden), they will spoil the symmetries~\eqref{reflections}. Thus, the class-A-type Weyl symmetry of the perturbative dimensions~\eqref{x-mu-1-loop} for generic neutral observables is likely accidental.

Even without an explicit knowledge of the non-perturbative scaling dimensions $x_{\lambda, \lambda'}^{(s)}$, we can use the general results on the Iwasawa decomposition and the Harish-Chandra isomorphism, which tell us that $x_{\lambda, \lambda'}^{(s)}$ should be symmetric with respect to exchanges of any two components of $\mu = (\lambda, \lambda')$. We can distinguish three possibilities for such exchanges:
\begin{align}
	q_i &\to \tilde{q}_i = q_j + \frac{c_j - c_i}{2},
	&
	q_j &\to \tilde{q}_j = q_i + \frac{c_i - c_j}{2},
	%\label{exchange-1}
	\nonumber \\
	q'_i &\to \tilde{q}'_i = q'_j + \frac{c_j - c_i}{2},
	&
	q'_j &\to \tilde{q}'_j = q'_i + \frac{c_i - c_j}{2},
	%\label{exchange-2}
	\nonumber \\
	q_i &\to \tilde{q}_i = - q'_j - \frac{c_i + c_j}{2},
	&
	q'_j &\to \tilde{q}'_j = - q_i - \frac{c_i + c_j}{2}.
	%\label{exchange-3}
	\label{exchanges}
\end{align}
The first of them corresponds to a permutation of components pertinent to the sublattice A, and the second one corresponds to a permutation of components pertinent to the sublattice B. Importantly, there is also the third type of permutations, with an exchange of components corresponding to different sublattices.
The first two types of the permutations separately preserve $|\lambda|$ and $|\lambda'|$, while the third type preserves the total charge $Q_\mu = 2(|\lambda'| - |\lambda|)$. Thus, in fact, all Weyl group permutations preserve also the U(1) part of thedimension~\eqref{x-mu-splitting}. This allows us to predict symmetries of the most general observables: their scaling dimensions should be symmetric under any ``legitimate'' Weyl transformations, that is, permutations~\eqref{exchanges}:
\begin{align}
	x_{\lambda, \lambda'} = x_{w^\text{AIII}(\lambda, \lambda')}.
	\label{Weyl-symmetry-AIII}
\end{align}

Let us discuss implications of this Weyl symmetry for two important subclasses of observables (which are studied numerically in the present version). First, let us restrict our attention to one-sublattice (say, sublattice-A) observables, i.e., exponents $x_{\lambda,(0)}$.  If we want to stay within this subclass, only permutations within the sublattice A are relevant, which are represented by first line of Eq.~\eqref{exchanges}. This symmetry subgroup is strongly reduced in comparison to the Weyl symmetry of exponents $x_\lambda$ in class A and does not provide particularly useful relations. Second, consider the subclass of balanced observables (i.e., those with $\lambda' = \lambda$) and the corresponding exponents $x_{\lambda,\lambda}$. Here, we make the following important observation: if the weights $\lambda$ and $\lambda'$ contain equal entries at the position $i$, $q_i = q'_i$, then the last type of exchange in Eq.~\eqref{exchanges} with $j=i$ is equivalent to simultaneous reflections~\eqref{reflections} of both $q_i$ and $q'_i$!  In combination with the first two types of transformations in Eq.~\eqref{exchanges}, this leads to the following result: within the subclass of balanced observables ($\lambda' = \lambda$), the dimensions $x_{\lambda, \lambda}$ enjoy the Weyl symmetries $w^\text{A}$ of class-A type (permutations and reflections):
\begin{align}
	x_{\lambda, \lambda} = x_{w^\text{A}(\lambda), w^\text{A}(\lambda)} \,.
	\label{Weyl-symmetry-A}
\end{align}

%\subsection{Sublattice symmetry}

Additional constraints on the scaling dimensions and their symmetry properties come from physical considerations of the sublattice symmetry. It is clear that in a microscopic model the two sublattices should be equivalent, and their interchange should be a symmetry of the system. For example, the scaling dimensions of moments of the local density of states on the two sublattices should be the same, and a similar statement should be valid for generalized MF observables. Translated into the language of the sigma model, the sublattice symmetry is seen as the invariance of the gigma-model actions under the exchange of the fields $U$ and $U^{-1}$.
%$V$ and $V^{-1}$ in Gade ($U$ and $U^\dagger$ in Gade and Wegner).
In our formulation these correspond to the diagonal sub-blocks $\tilde{\mathcal{Q}}_{RR}$ and $\tilde{\mathcal{Q}}_{AA}$ of the rotated $Q$-matrix. Permuting these two sub-blocks is equivalent to the exchange $x_i \leftrightarrow - x_{2n+1-i} = -x'_i$. In turn, this exchange leads to the exchange $q_i \leftrightarrow q'_i$ (same as $\lambda \leftrightarrow \lambda'$) in the radial functions~\eqref{phi-lambda-bar-lambda}. This means that the scaling dimensions of the generalized-multifractality observables should be symmetric under the exchange of the two weights that label them:
\begin{align}
	x_{\lambda, \lambda'} = x_{\lambda', \lambda}.
	\label{sublattice-symmetry}
\end{align}
This symmetry imposes constraints on how Casimir eigenvalues can contribute to the scaling dimensions.

Let us recall that the basic Casimir operators are obtained from the basic Weyl group invariants. For the root system of type $A_{2n-1}$ relevant in our case, these are the symmetric power sums
\begin{align}
	C_k(\lambda, \lambda') &= \sum_{i=1}^{n} ( \mu_i^k + \mu^{\prime k}_i)
	%= \sum_i [(- 2q_i - c_i)^k + (2q'_i + c_i)^k]
	\nonumber \\
	&= \sum_i [(-1)^k (2q_i + c_i)^k + (2q'_i + c_i)^k],
	%\nonumber \\
	%C_1(\lambda, \lambda') &= 2 \sum_i (q'_i - q_i) = 2(|\lambda'| - |\lambda|).
\end{align}
(The first of these, $C_1(\lambda, \lambda')$ is the same as the charge $Q_\mu$.)
It follows that if we interchange the two weights, we get
\begin{align}
	C_k(\lambda, \lambda') = (-1)^k C_k(\lambda', \lambda).
	\label{C-k-symmetry}
\end{align}

The scaling dimensions can be written as linear combinations of products of the Casimir eigenvalues:
\begin{align}
	x_{\lambda, \lambda'}
	&= \sum_k \sum_{\pi \vdash k} \alpha_\pi \prod_{i=1}^{l(\pi)} C_{\pi_i}(\lambda, \lambda'),
\end{align}
where $\pi$ are partitions of the integer $k$, and $l(\pi) = \sum_i \pi_i$ are their lengths. The U(1) term is included as a part of the $k=2$ term with $\pi = (1,1)$, since
\begin{align}
	x_{\lambda, \lambda'}^{\text{U}(1)}
	&= x_\nu (|\lambda'| - |\lambda|)^2
	= \frac{1}{4} x_\nu C_1^2(\lambda, \lambda').
	\label{x-Casimirs}
\end{align}
Now we use Eq.~\eqref{C-k-symmetry} here, and get
\begin{align}
	x_{\lambda', \lambda}
	&
	%= \sum_k \sum_{\pi \vdash k} \alpha_\pi \prod_{i=1}^{l(\pi)} C_{\pi_i}(\lambda', \lambda)
	%= \sum_k \sum_{\pi \vdash k} \alpha_\pi \prod_{i=1}^{l(\pi)} (-1)^{\pi_i} C_{\pi_i}(\lambda, \lambda')
	= \sum_k (-1)^k \sum_{\pi \vdash k} \alpha_\pi \prod_{i=1}^{l(\pi)} C_{\pi_i}(\lambda, \lambda').
\end{align}
This form is consistent with the sublattice symmetry~\eqref{sublattice-symmetry} only if $\alpha_\pi = 0$ for all odd values of $|\pi| = k$. Thus, the sublattice symmetry imposes a constraint on the possible terms in the expansion~\eqref{x-Casimirs}: only partitions of even integers $k$ are allowed. Notice that this does not exclude odd-order Casimirs from showing up. For example, at the order $k=4$ we can have a term with the product $C_1 C_3$. If such terms are present in the scaling dimensions, they spoil the class-A type symmetries. Thus, in general, we only have the class-AIII symmetries~\eqref{Weyl-symmetry-AIII}. However, if the two weights are the same, $\lambda' = \lambda$, then all odd Casimirs simply vanish, and we get back our result~\eqref{Weyl-symmetry-A} for the scaling dimensions of balanced observables.

\subsection{Class BDI}

In this section we present details of the Iwasawa construction for class BDI. In this case
\begin{align}
	M_B &= \frac{\text{GL}(n, \mathbb{R})}{\text{O}(n)}
	= \frac{\text{GL}(1, \mathbb{R})}{\text{O}(1)}
	\times \frac{\text{SL}(n, \mathbb{R})}{\text{SO}(n)}
	= \mathbb{R}_+ \times M_B^{(s)}.
	\label{MB-R-MBs-BDI}
\end{align}
We write the elements of $\mfg = \text{gl}(n, \mathbb{R})$ as real matrices in the space~\eqref{tensor-product-space-2}
\begin{align}
	Z = \sigma_0 \otimes X + \sigma_2 \otimes Y
	= \begin{pmatrix*}[r] X & -iY \\ iY & X \end{pmatrix*},
\end{align}
where $X$ and $Y$ are $n \times n$ blocks in the RA space satisfying $X^T = - X$ and $Y^T = Y$, and in the semisimple case we also need to impose the condition
\begin{align}
	\text{tr} \, Y &= 0.
	\label{trace-condition-BDI}
\end{align}
The elements $Z \in \mfg$ satisfy the conditions
\begin{align}
	Z^\dagger \Sigma_{3} + \Sigma_{3} Z &= 0,
	&
	Z \Sigma_{2} - \Sigma_{2} Z = 0,
	\label{Z-conditions-BDI}
\end{align}
and their combinations. The subalgebra o$(n)$ is the one with $Y = 0$.

The Cartan involution is
\begin{align}
	\theta(Z) = - Z^\dagger = \Sigma_{3} Z \Sigma_{3},
\end{align}
$Z \in \mfk$ if $Y = 0$, and $Z \in \mfp$ if $X = 0$. The generators for $\mfk$ and $\mfp$ are
\begin{align}
	X_{ij} &= \sigma_0 \otimes E^-_{ij},
	&
	Y_{ij} &= \sigma_1 \otimes E^+_{ij}.
\end{align}
These are the same as $X^{(0)}_{ij}$ and $Y^{(0)}_{ij}$ in Eq.~\eqref{0-generators-AIII}.

We choose the maximal Abelian subspace $\mfa \subset \mfp$ as
\begin{align}
	\mfa &= \text{span}\left\{ H_k = \sigma_{1} \otimes E_{kk}
	= Y_{kk}/2 \right\}.
\end{align}
The system of restricted roots is $A_{n-1}$ given by Eq.~\eqref{roots} with $m_o = 1$. The positive restricted root vectors are the same as $E_{\alpha_{ij}}^{(0)}$ in Eq.~\eqref{positive-RRV-AIII}:
\begin{align}
	E_{\alpha_{ij}} &= X_{ij} + Y_{ij},
	\label{positive-RRV-BDI}
\end{align}
and the Weyl vector in the replica limit is
\begin{align}
	\rho &= \sum_{i}^n c_i x_i,
	&
	c_i &= \frac{1}{2} - i.
	\label{Weyl-vector-BDI}
\end{align}
The rest proceeds exactly as in class AIII.

\subsection{Class CII}

In this section we present details of the Iwasawa construction for class CII. In this case
\begin{align}
	M_B = \frac{\text{U}^*(2n)}{\text{Sp}(2n)}
	= \mathbb{R}_+ \times \frac{\text{SU}^*(2n)}{\text{Sp}(2n)}
	= \mathbb{R}_+ \times M_B^{(s)}.
\end{align}

We realize the elements of $\mfg = \text{u}^*(2n)$ as matrices in the space~\eqref{tensor-product-space-3}
\begin{align}
	Z &= \sigma_0 \otimes
	\begin{pmatrix*}[r] X_1 & X_2 \\ -X_2^* & X_1^* \end{pmatrix*}
	+ \sigma_2 \otimes
	\begin{pmatrix*}[r] Y_1 & Y_2 \\ -Y_2^* & Y_1^* \end{pmatrix*},
\end{align}
where the $n \times n$ blocks satisfy
\begin{align}
	X_1^\dagger &= -X_1, & X_2^T &= X_2, & Y_1^\dagger &= Y_1, & Y_2^T &= -Y_2,
\end{align}
and in the semisimple case we also need to impose the condition $\tr Y_1 = 0$.

In this realization the elements $Z \in \mfg$ of $\text{u}^*(2n)$ satisfy
\begin{align}
	Z^\dagger \Sigma_{30} + \Sigma_{30} Z &= 0,
	&
	Z \Sigma_{20} - \Sigma_{20} Z &= 0,
	\nonumber \\
	Z^T \Sigma_{02} + \Sigma_{02} Z &= 0.
	\label{Z-conditions-CII-k=2}
\end{align}
The Cartan involution is
\begin{align}
	\theta(Z) = - Z^\dagger = \Sigma_{30} Z \Sigma_{30},
\end{align}
and its eigenspaces are characterised as follows: $Z \in \mfk$ if $Y_1 = Y_2 = 0$, and $Z \in \mfp$ if $X_1 = X_2 = 0$. We have four groups of generators in both $\mfk$ and $\mfp$:
\begin{align}
	X^{(0)}_{ij} &= \sigma_{00} \otimes E^-_{ij},
	&
	Y^{(0)}_{ij} &= \sigma_{20} \otimes E^+_{ij},
	\nonumber \\
	X^{(1)}_{ij} &= i \sigma_{01} \otimes E^+_{ij},
	&
	Y^{(1)}_{ij} &= i \sigma_{21} \otimes E^-_{ij},
	\nonumber \\
	X^{(2)}_{ij} &= i \sigma_{02} \otimes E^+_{ij},
	&
	Y^{(2)}_{ij} &= i \sigma_{22} \otimes E^-_{ij},
	\nonumber \\
	X^{(3)}_{ij} &= i \sigma_{03} \otimes E^+_{ij},
	&
	Y^{(3)}_{ij} &= i \sigma_{23} \otimes E^-_{ij}.
	\label{generators-CII-k=2}
\end{align}

We choose the maximal Abelian subspace $\mfa \subset \mfp$ as
\begin{align}
	\mfa &= \text{span}\left\{ H_k = \sigma_{20} \otimes E_{kk}
	= Y^{(0)}_{kk}/2 \right\}.
\end{align}
The system of restricted roots is $A_{n-1}$ given by Eq.~\eqref{roots} with $m_o = 4$. The positive restricted root vectors are
\begin{align}
	E_{\alpha_{ij}}^{(0)} &= X^{(0)}_{ij} + Y^{(0)}_{ij},
	&
	E_{\alpha_{ij}}^{(1)} &= X^{(1)}_{ij} + Y^{(1)}_{ij},
	\nonumber \\
	E_{\alpha_{ij}}^{(2)} &= X^{(2)}_{ij} + Y^{(2)}_{ij},
	&
	E_{\alpha_{ij}}^{(3)} &= X^{(3)}_{ij} + Y^{(3)}_{ij},
	\label{positive-RRV-CII-k=2}
\end{align}
and the Weyl vector in the replica limit is
\begin{align}
	\rho &= \sum_{i}^n c_i x_i,
	&
	c_i &= 2 - 4i.
	\label{Weyl-vector-CII}
\end{align}

The unitary transformation that makes the generators of $\mfa$ diagonal and the generators of $\mfn$ strictly upper-triangular is accomplished with the help of the matrix
\begin{align}
	R_\text{CII} &= R \otimes \sigma_0 \otimes I_n,
	\label{U-CII}
\end{align}
where the matrix $R$ was defined in Eq.~\eqref{U-Iwasawa-chiral}. We also need the permutation matrix $\Pi_1$ with elements $(\Pi_1)_{ij} = \delta_{\pi_1(i), j}$ where the permutation $\pi_1$ of the basis of the space~\eqref{tensor-product-space-3} can be described as follows: for $i \in 1, \ldots, n$, we have
\begin{align}
	\pi_1(i) &= 2n + 2 - 2i,
	&
	\pi_1(n+i) &= 2n + 1 - 2i,
	\nonumber \\
	\pi_1(2n+i) &= 2n + 2i-1,
	&
	\pi_1(3n+i) &= 2n + 2i.
	\label{permutation-pi-1-CII-k=2}
\end{align}
The unitary transformation
\begin{align}
	\tilde{M} &= \Pi_1^{-1} R_\text{CII} M R_\text{CII}^{-1} \Pi_1
	\label{tilde-M-CII}
\end{align}
rotates the $\Lambda$ matrix to
\begin{align}
	\tilde{\Lambda} &= \sigma_{2} \otimes \mathcal{I}_{2n},
\end{align}
makes the elements of $\mfa$ diagonal, and the positive restricted root vectors $\tilde{E}$ strictly upper-triangular.

The subsequent construction is almost verbatim as in class AIII, except that each entry in the diagonal matrix $\tilde{a} \in A$ is now repeated twice:
\begin{align}
	\tilde{a} = \text{diag} (e^{x_1} \sigma_0, \ldots, e^{x_n} \sigma_0,
	e^{- x_n} \sigma_0, e^{-x_1} \sigma_0).
	\label{tilde-a-CII}
\end{align}
This doubling of the elements of $\tilde{a}$ is characteristic for all ``spinful'' symmetry classes (AII, C, CI, DIII, and CII) which possess either time reversal symmetry with $T^2 = -1$, or particle-hole symmetry with $P^2 = -1$, or both.

The structure of the matrix $\tilde{\mathcal{Q}}$ is again lock-daigonal as in Eq.~\eqref{Q-tilde-blocks}. The lower-right $2m \times 2m$ submatrices $\tilde{\mathcal{Q}}_{RR}^{(2m)}$ and $\tilde{\mathcal{Q}}_{AA}^{(2m)}$ of the blocks $\tilde{\mathcal{Q}}_{RR}$ and $\tilde{\mathcal{Q}}_{AA}$ are the same as in Eq.~\eqref{Q-tilde-blocks-m}, except that now all entries are understood as $2 \times 2$ matrices, with the blocks on the diagonals proportional to the identity matrix $\sigma_0$. Determinants of $\tilde{\mathcal{Q}}_{RR}^{(2m)}$ give the basic positive $N$-radial eigenfunctions
\begin{align}
	d_{2m}(x) = \exp \Big(- 4 \sum_{i=1}^{m} x_i \Big).
	\label{basic-d-2m}
\end{align}
We can form the most general $N$-radial eigenfunctions for the sublattice A as products
\begin{align}
	%\phi_{(q_1,\ldots,q_n)} = d_2^{(q_1 - q_2)/2} d_4^{(q_2 - q_3)/2}
	%\ldots d_{2n}^{q_n/2}.
	\phi_\lambda(x) = \prod_{k=1}^{n} d_{2k}^{(q_k - q_{k+1})/2}(x).
	\label{d-product-CII}
\end{align}
It is easy to see that the product~\eqref{d-product-CII} is the same as the exponential eigenfunction~\eqref{plane-wave-2}, while the basic function $d_{2m}(x)$ is $\phi_{(2,2,\ldots)}(x)$ with $m$ twos in the subscript. The construction of the other sublattice is identical but uses the determinants of $\tilde{\mathcal{Q}}_{AA}^{(2m)}$.

Notice that the doubling of the diagonal entries $e^{-2x_i}$ for each $i$ in Eq.~\eqref{tilde-a-CII} compelled us to take determinants of sub-matrices of even size and raise the resulting functions to powers written as $(q_i - q_{i+1})/2$. In the Iwasawa formalism it is also possible to obtain directly the basic solutions $\phi_{(1^m)}$. Using definitions above, it is straightforward to show that the matrices $\tilde{\mathcal{Q}}_{RR}^{(2m)} (I_{m} \otimes i\sigma_2)$ and $\tilde{\mathcal{Q}}_{AA}^{(2m)} (I_{m} \otimes i\sigma_2)$ are anti-symmetric, and that their Pfaffians gives the basic $N$-radial eigenfunctions
\begin{align}
	p_m(x) &= \text{Pf} \, [\tilde{\mathcal{Q}}_{RR}^{(2m)} (I_{m} \otimes i\sigma_2)]
	= \exp \Big(-2 \sum_{i=1}^{m} x_i \Big),
	\nonumber\\
	p_m'(x') &= \text{Pf} \, [\tilde{\mathcal{Q}}_{AA}^{(2m)} (I_{m} \otimes i\sigma_2)]
	= \exp \Big(2 \sum_{i=1}^{m} x_i' \Big).
\end{align}
The general $N$-radial functions are then obtained as products of powers of $p_m$:
\begin{align}
	\phi_{\lambda, \lambda'}(x,x') =
	\prod_{k=1}^{n} p_k^{q_k - q_{k+1}}(x)
	\prod_{k=1}^{n} p_k^{\prime \,q'_k - q'_{k+1}}(x').
	\label{p-product-CII}
\end{align}

The resulting form of the Iwasawa construction for class CII as given by Eqs.~\eqref{tilde-a-CII}--\eqref{p-product-CII} is fully analogous to that in other ``spinful'' classes Ref.~\cite{karcher2022generalized-2}.

\pagebreak

\bibliography{gener-MF}

\end{document}